\documentclass[a4paper]{JHEP3}
\usepackage[centertags]{amsmath}
\usepackage{amssymb}
\usepackage{graphicx}

\title{Small Hairy Black Holes in Global AdS  Spacetime}

\author{Pallab Basu$^a$\footnote{pallab@phas.ubc.ca}, \
Jyotirmoy Bhattacharya$^b$\footnote{jyotirmoy@theory.tifr.res.in}, \
Sayantani Bhattacharyya$^b$\footnote{sayanta@theory.tifr.res.in}, \
R.Loganayagam$^b$\footnote{nayagam@theory.tifr.res.in}, \
Shiraz Minwalla$^b$\footnote{minwalla@theory.tifr.res.in}, \
V Umesh $^b$\footnote{vumesh.physics@gmail.com}\\
\small{\emph{$^{a}$University of British Columbia,}} \\
\small{\emph{Vancouver, Canada, V6T 1Z1.}}\\
\small{\emph{$^{b}$ Tata Institute of Fundamental Research,}}\\
\small{\emph{Homi Bhabha Rd, Mumbai 400005.}}  }

\abstract{
We study small charged black holes in global AdS spacetime in 
the presence of a charged massless minimally coupled scalar field.
In a certain parameter range these black holes suffer from 
well known superradiant instabilities. We demonstrate that 
the end point of the resultant tachyon condensation process
is a hairy black hole which we construct analytically in 
a perturbative expansion in the black hole radius. At leading order
our solution is a small undeformed RNAdS black hole immersed 
into a charged scalar condensate that fills the AdS `box'. These
hairy black hole solutions appear in a two parameter family 
labelled by their mass and charge. Their mass is bounded 
from below by a function of their charge; at  the lower bound 
a hairy black hole reduces to a regular horizon free soliton  
which can also be thought of as a nonlinear Bose condensate.
We compute the microcanonical phase diagram of our system at 
small mass, and demonstrate that it  exhibits a 
second order `phase transition' between the RNAdS black hole 
and the hairy black hole phases. }

\preprint{TIFR/TH/10-07}

\begin{document}

\section{Introduction}\label{sec:intro}

There has been much recent interest in the physics of charged black 
brane solutions of the Lagrangian\footnote{We have chosen to work 
in 5 spacetime dimensions, and chosen the scalar field
below to be massless merely for simplicity. The analysis of this paper
carries over, without qualitative modification, in arbitrary spacetime
and for arbitrary scalar potential.}
\begin{equation} \label{lagrangian} 
\begin{split}
S&=\frac{1}{8\pi G_5}\int d^5x  \sqrt{g} \left[ \frac{1}{2} \left(\mathcal{R}[g] +12\right) 
-\frac{1}{4} \mathcal{F}_{\mu\nu}\mathcal{F}^{\mu\nu} 
- |D_\mu \phi|^2 \right]\\
D_\mu \phi&= \nabla_\mu\phi -i e A_\mu \phi 
\end{split} 
\end{equation}
where $G_5$ is the Newtons's constant and the radius of AdS$_5$ is set to unity\footnote{
See Appendix \ref{app:notation} for a summary of notation employed in this paper.}.
This system (sometimes called the massless Abelian Higgs model) admits
a well known set of charged black brane solutions which are asymptotically
Poincare AdS. Recent interest in this system is due to Gubser's observation \cite{Gubser:2008px} that, at large $e$ and when they are near enough to
extremality, these black branes are unstable . The end point of the
tachyon condensation sparked by this instability is a so called
hairy black brane - a solution with a planar horizon immersed 
in a charged scalar condensate. Black branes interacting with 
such matter condensates are novel and interesting, and have been 
studied intensively over the last few years (see \cite{Sachdev:2010ch,Horowitz:2010gk,
Hartnoll:2009qx, Hartnoll:2009sz,Herzog:2009xv} and references therein).
Unfortunately almost all constructions of these solutions have been
numerical\footnote{See however \cite{Basu:2008bh,Herzog:2009ci,Gregory:2009fj}
for analytical studies in a related context.}. 

Under the AdS/CFT correspondence, these planar AdS$_5$  solutions are
dual to  the states of a conformal field theory living on the flat
spacetime $R^{3,1}$. Another natural arena to study $3+1$ dimensional
conformal field theories is to work on $S^{3} \times R_{\text{time}}$
instead. States of such a boundary field theory living on $S^3$ 
are dual to gravitational solutions that asymptote to global AdS$_5$
instead of planar AdS$_5$. The corresponding charged black holes 
in global AdS$_5$ spacetime are characterised by their radius in units of the 
AdS$_5$  radius and their charge. At large horizon radius,
these black holes are locally well approximated by black branes and
we expect their physics to be qualitatively similar to the Poincare AdS
charged branes. It is natural to enquire about the opposite limit:
do small hairy black holes exist, and what are their qualitative 
properties? In this paper we  answer this question by 
explicitly constructing a set of spherically symmetric hairy charged black 
holes whose radii are small compared to the AdS$_5$  radius\footnote{
See \cite{Sonner:2009fk} for earlier work on scalar condensation
in black hole backgrounds in global AdS$_5$.}. Our construction is 
perturbative in the radii of our solutions,
but is otherwise analytic. It permits an analytic construction 
of the microcanonical phase diagram of our system at small mass 
and charge. In the rest of this introduction, we will describe in
detail our construction of small hairy black holes, their properties,
and the phase diagram of our system.

To begin with, we start our discussion with a consideration that may,
at first, seem unrelated to the study of AdS  black holes.  Consider a spherically 
symmetric shell of a scalar field of frequency $\omega$ incident on a 
charged black hole in flat spacetime. One might naively expect a part 
of this wave to be absorbed by the black hole while the rest is
reflected back to infinity. It is, however, a well known fact 
that the reflection coefficient for this process actually exceeds unity 
when $\omega < e \mu$ ($\mu$ is the chemical potential of the black hole). 
Under these conditions more of the incident wave comes out than 
was sent in. This phenomenon, called superradiance \cite{Bekenstein:1973mi},
has immediate and well known implications for the stability of 
small RNAdS black holes, as we now explain.

Consider a superradiant wave incident on a small charged black hole 
sitting at the centre of global AdS spacetime. Such a wave 
reflects off the black hole, propagates out to large $r$, 
but unlike the flat spacetime case, bounces back from 
the boundary of AdS$_5$ and then finds itself re-incident 
on the black hole. This process continues indefinitely. As 
every reflection increases the amplitude of this wave by 
a fixed factor, this process constitutes an instability of 
the charged black hole. A closely related instability, the
so called black hole bomb, was discussed (in the context of
a flat spacetime black hole surrounded by mirrors) as early as the 1970s 
\cite{1972Natur.238..211P}. 

As the spectrum of frequencies of a minimally coupled charged
scalar field (in a gauge where $A_t^{(r=\infty)}=0$) in
AdS$_5$ is bounded from below $\omega\geq\Delta_0\equiv 4$  ,
we expect small charged black holes in AdS$_5$  space to 
exhibit superradiant instabilities whenever the condition
$e \mu \geq \omega\geq \Delta_0$ is satisfied\footnote{In 
Appendix \ref{app:qnm}, we verify this expectation by 
direct computation of the lowest quasi normal mode of
this system. We find that for small R, the imaginary part of the 
frequency of this mode is given by $3R^3(e\mu-4)=
3R^3(e\mu-\Delta_0) $, where $R$ is the Schwarzschild radius
of the black hole. This imaginary part changes sign precisely where
we expect the instability.}. Now the chemical potential
$\mu$ of a small black hole is bounded from above by 
the chemical potential of the extremal black hole; i.e. 
$\mu^2 \leq \mu_c^2=\frac{3}{2}$. It follows that small
charged AdS  black holes are always stable when 
$e^2\leq \frac{\Delta_0^2}{\mu_c^2}\equiv e_c^2= \frac{32}{3}$. 
When $e^2 \geq  e_c^2$, however, small black holes that 
are near enough to extremality suffer from a superradiant
instability.

The superradiant instability described above admits a very 
simple thermodynamical interpretation. Notice that the Boltzmann factor 
for a mode of energy $\Delta_0$ and charge $e$ is given 
by $e^{-T^{-1}(\Delta_0-e\mu)}$ where $T$ is the temperature 
of the black hole. Now this factor leads to an exponential
enhancement (rather than the more usual suppression) 
whenever $\mu e \geq \Delta_0$. In other words, a small 
charged black hole with $\mu e \geq \Delta_0$ is unstable 
against Bose condensation of the lightest scalar mode. 
Indeed the leading unstable mode of a small charged 
black hole with $\mu \geq \frac{\Delta_0}{e}$ is a 
small deformation of the lightest scalar mode in
global AdS$_{5}$ space. 

The considerations outlined above suggests that superradiant tachyon 
condensation proceeds in the following manner. The black hole emits into 
a scalar condensate, thereby losing mass and charge itself. As the 
charge to mass ratio of the condensate (i.e superradiant mode), 
$\frac{e}{\Delta_0}$, exceeds $\frac{1}{\mu}$, the chemical potential 
of the black hole also decreases as this emission proceeds.
Now the decay rate of the black hole is proportional to 
$(\Delta_0-\mu e)$ and so slows down as $\mu$ approaches $\frac{\Delta_0}{e}$.
It seems intuitively plausible that the system asymptotes to a configuration 
consisting of a $\mu \approx \frac{\Delta_0}{e}$ 
stationary charged black hole core surrounded by a diffuse AdS scale charge
condensate, i.e. a hairy black hole. We will provide substantial 
quantitative evidence for the correctness of this picture in this paper.

In the discussion of the previous paragraph we have ignored 
both the backreaction
of the scalar field on the geometry as well as the effect of 
the charged black hole core on the scalar condensate. However 
these effects turn out to be small whenever the starting black hole is 
small enough. In other words the end point of the superradiant
instability of a small charged black hole is given 
approximately by a non-interacting mix of the black hole
core and the condensate cloud at leading order. We will now 
pause to explain why this is the case.

First note that the charge and energy density of the superradiant mode 
is contained in an AdS radius scale cloud. As the charge and mass the 
 initial unstable black hole is small, the same is true charge and mass 
of the eventual  the  scalar condensate. Consequently, the scalar 
condensate is of low density and so backreacts
only weakly on the geometry everywhere.\footnote{ Note that,
in contrast, for the small charged black hole at the core has
its mass and charge concentrated within a small Schwarzschild
radius. Consequently even a black hole of very small mass and charge 
is a large perturbation about the AdS vacuum at length scales 
comparable to its Schwarzschild radius.} For this reason the 
metric  of the final solution is a small deformation
of the RNAdS black hole with $\mu=\frac{\Delta_0}{e}$, and the scalar 
condensate does not significantly affect the properties of the RNAdS black 
hole. On the other hand the condensate cloud is very large compared 
to the RNAdS black hole at its core. This difference in scales ensures 
that the charged black hole also does not significantly affect 
the properties of the scalar condensate.

Motivated by these considerations, in this paper we construct 
the hairy black hole  that marks the end point of the 
superradiant  tachyon condensation process in a perturbative
expansion around a small RNAdS black hole with $\mu=\frac{\Delta_0}{e}$
and small but arbitrary radius. The perturbative procedure we employ in our 
construction is completely 
standard except for one twist, which we now explain. As is usual in 
perturbation theory, we expand out the metric, gauge field and scalar 
field in a power series in $\epsilon$ which is the small parameter
of our expansion\footnote{As we explain in Section \ref{sec:hairybh}
below, we find it convenient to choose $\epsilon$ to be the coefficient of the $\frac{1}{r^4}$ decay of the scalar field at infinity, i.e. 
the vacuum expectation value of the operator dual to the scalar field.} 
\begin{equation}\begin{split} \label{pertexp}
g_{\mu \nu}&= g_{0\mu\nu} + \epsilon^2 g_{2\mu\nu} + \ldots\\
A_t&=A_{0t} + \epsilon^2 A_{2t} + \ldots\\
\phi&= \epsilon \phi_1 + \epsilon^3 \phi_3 + \ldots\\
\end{split}
\end{equation}
Here $g_{0\mu\nu}$ and $A_{0\mu}$ are the metric and gauge field 
of our starting RNAdS black hole solution. We then plug this expansion 
into the equations of motion, expand the latter in a power series 
in $\epsilon$, and attempt to solve the resultant equations 
recursively. Unfortunately,  the linear ordinary differential 
equations that appear in this process do not appear to be analytically 
solvable in full generality.
However it turns out to be easy to solve these equations separately 
in two regimes: at large $r$ (in an expansion in $\frac{R}{r}$ which 
we call as the far-field expansion and mark by a superscript `out') 
and at small $r$ (in an expansion in $r$ which we call the 
near-field expansion and mark by a superscript 'in'). Here $r$ is
the radial coordinate (that is zero at the black hole singularity
 and infinity at the boundary of AdS ) and $R$ is the
Schwarzschild radius of the unperturbed RNAdS black hole solution.
The first expansion is valid when $r \gg R$, while the second expansion 
works when $r \ll 1$. As we are interested in $R \ll 1$, the validity domains
of these two approximations overlap. Consequently, we are  able to solve 
the resultant linear equations everywhere, in a power series expansion in $R^2$. 

\begin{figure}[ht]
 \begin{center}
 \includegraphics{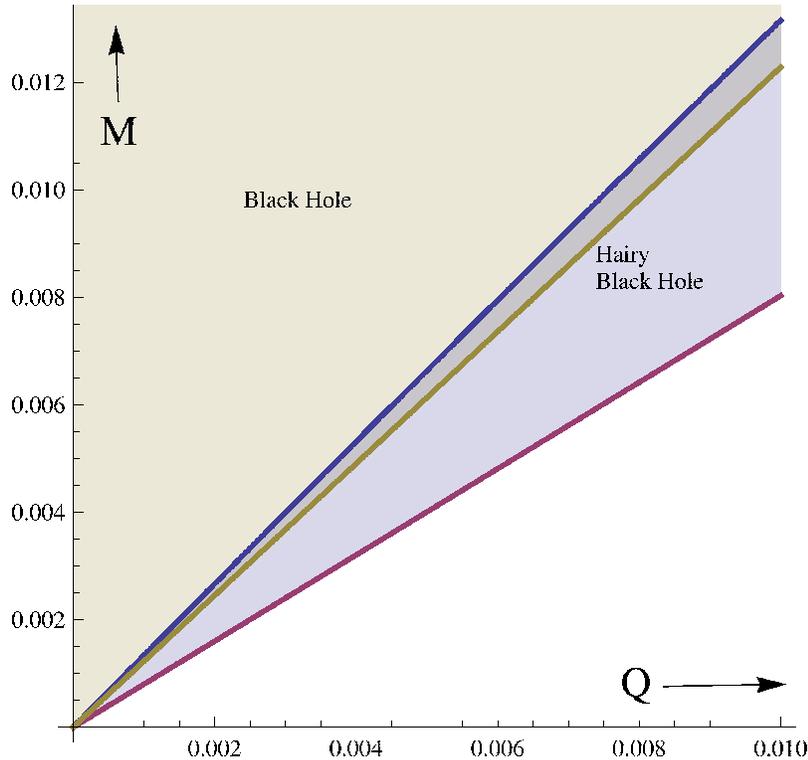}
\end{center}
\caption{\label{fig10} Micro Canonical Phase Diagram at small mass and charge. The 
overlapping region is dominated by the hairy blackhole.}
\end{figure}

When the dust has settled we are thus able to solve for the hairy black 
holes only in a double expansion in $\epsilon^2$ and $R^2$. This expansion
is sufficient to understand small hairy black holes \footnote{The
technical obstruction to solving the equations at arbitrary $R$ has a physical interpretation. Even at arbitrarily small $\epsilon$, the black hole
reacts significantly on the condensate  at finite $R$. Our system 
can be regarded as a non interacting mix of the black hole and the
condensate only at very small $R$. Effectively, in this paper, 
we perturb around this non interacting limit.}. In section \ref{sec:hairybh}
below we  have explicitly implemented this expansion to 
${\cal O}(\epsilon^m R^2)$ for $m \leq 5$. Our calculations allow us 
to determine the microcanonical phase diagram of our system, as a function 
of mass and charge at small values of these parameters\footnote{The 
microcanonical ensemble is well suited to our purposes. We discuss
the phase diagram in other ensembles, in particular the canonical 
and grand canonical ensemble, in Appendix \ref{app:Can} and \ref{app:GC}
below. We are able to make less definite statements in these 
ensembles because it turns out that the system at a given 
fixed chemical potential and temperature often receives contributions 
both from small as well as big black holes. As the approximation techniques
of this paper do not apply to big black holes, we are unable to quantitatively
assess the relative importance of these saddle points.}; our results are 
plotted for $e=5$ in Fig.\ref{fig10} below (the results are qualitatively similar 
for every $e$ provided $e^2\geq \frac{32}{3}$, and may also be simply 
generalised to the study of \eqref{lagrangian} with a mass term added for the 
scalar field- see \ref{ssec:mphi}).

As summarised in Fig.\ref{fig10}, hairy black holes exist only in the mass range 
\begin{equation} \label{range} 
\frac{4}{e} Q+\frac{\left(9 e^2-64\right) }{7 \pi
   e^2}Q^2+{\cal O}(Q^3) \leq M   \leq 
\frac{3 e}{16}\left(1+\frac{32}{3e^2}\right) Q-\frac{3 \left(5 e^4+64
   e^2-1024\right)}{64 \pi
   e^2} Q^2 +{\cal O}(Q^3),
\end{equation}
where $M$ and $Q$ are the mass and the charge of the hairy black hole.

Above the upper bound in \eqref{range}  (i.e. in the shaded grey region 
above the blue line in Fig.\ref{fig10}.), 
RNAdS black holes are stable and are the only known stationary solutions. 
The upper end of \eqref{range} coincides with the onset of superradiant
instabilities  for RNAdS black holes. The lower bound in \eqref{range} 
is marked by the lowest (i.e. red) line in Fig.\ref{fig10}.
The extremality line for RNAdS black holes (the yellow line in Fig.\ref{fig10}.) 
lies in the middle of this range in \eqref{range}. At masses below lower 
bound of \eqref{range} (red line in Fig.\ref{fig10}.), the system presumably has 
no states. 

As we have emphasised, a hairy black hole may approximately be thought of as a 
non interacting superposition of a RNAdS black hole and a 
scalar Bose condensate. At the upper bound of 
\eqref{range} the condensate vanishes, and hairy black holes
reduce to a  RNAdS black holes. As we decrease the mass of a hairy 
black hole at fixed charge (or increase the charge at fixed mass), the 
fraction of the condensate increases. Eventually, 
at the lower end of \eqref{range} - the red line 
in Fig.\ref{fig10}. - the black hole shrinks to zero. 
In this limit \footnote{We take the $R \to 0$ limit purely within classical 
relativity. Of course stringy and quantum gravity effects (including the 
one loop energy density of the gas outside the black hole) become important 
when the black hole becomes parametrically small. Such effects are important 
only in an infinitesimal wedge above the red line in Fig. \ref{fig10} and 
depend on the detailed microphysics of the system (they would be different, 
for instance, in string theory and M theory). We ignore all such effects
in this paper. We thank K. Papadodimas for a discussion on this issue.}
the solution reduces to a regular horizon 
free soliton (see \cite{Nishioka:2009zj} for the appearance of a similar 
solitonic solution in a qualitatively similar context). The solitonic solution,
on the red line of Fig.\ref{fig10} is simpler than the hairy black hole solution. As $R=0$
on this solution, it may be generated as an expansion in a single parameter $\epsilon$; in 
Appendix \ref{app:HigherOrdSoliton} we have carried out this expansion to 17th order. 
Hairy black holes may be thought of as solitons with small RNAdS solutions in their
centre (see the related general analysis\footnote{We thank G. Horowitz and H. Reall for drawing our attention to this reference.} of \cite{Kastor:1992qy} ).

In summary, the hairy black hole interpolates between pure black hole and pure condensate as we scan from the upper to the lower bound of \eqref{range} (or down from the blue line 
to the red line in Fig.\ref{fig10}.).Throughout the range of its existence, the hairy black hole is the only known stable solution\footnote{Except for excited solitons, see 
below for details.}. It is also the thermodynamically dominant solution, 
as the entropy of the hairy black hole exceeds that of the RNAdS 
black hole of the same mass and charge, whenever both solutions exist.

Note that the solitonic solutions described above have some similarities to so called boson stars\footnote{We thank G. Horowitz, M. Rangamani and K. Papadodimas  for drawing our attention to the literature on boson stars and explaining their properties to us. K. Papadodimas has further drawn our attention to the 
fact that our soliton reduces precisely to a traditional boson star in the 
limit $e \to 0$. Of course hairy black holes exist only when $e >e_{c}$ 
and so do not exist in the small $e$  limit.}, 
which have been extensively discussed in the General Relativity literature, mainly in asymptotically flat space (see e.g. \cite{Schunck:2003kk}) but also in asymptotically AdS spaces (see e.g. \cite{Astefanesei:2003qy}). However boson stars usually have scalar fields with a harmonic time dependence, which obstructs placing black holes in their centre (the scalar field would oscillate an infinite number of times as it approaches the horizon). Our solitonic solutions are genuinely stationary\footnote{The presence of the 
 gauge field allows us to evade Derrick's theorem. We thank K. Papadodimas 
pointing this out.} in a particular gauge. In this gauge the temporal 
component of the gauge field takes a particular non zero value at the origin of the soliton. This allows us to construct stationary hairy black hole solutions by placing charged black holes at the centre of the soliton (roughly via the procedure of \cite{Kastor:1992qy}.) if and only if the black holes are 
chosen so that their chemical potential matches the gauge field at the 
centre of the soliton. The last condition has a simple and intuitive
 thermodynamical interpretation; solitons and black holes can be put into 
an approximately non interacting mix only when their chemical potentials 
match!

The hairy black hole we have described so far in this introduction is a 
weakly interacting mix of the RNAdS black hole and a condensate
of the ground state of the scalar field. The reader may wonder whether it is 
possible to construct an excited hairy black hole solution that is a weakly
interacting mix of a RNAdS black hole and an excited state of the
 scalar field. This is indeed the case. The set 
of spherically symmetric linearised excitations of a massless scalar field 
appear in a one parameter family\footnote{These modes
are dual, under the state operator map, to the operator $\partial^{2n} O$, 
where $O$ is the dimension 4 operator that corresponds to the bulk field
$\phi$ according to the rules of the AdS/CFT dictionary.}
labelled by an integer $n$. The energy of the $n^{th}$ state is
 $\Delta_n\equiv 4+2n$ where $n= 0 \ldots \infty$.  It turns out to be 
possible to mimic the construction described above to construct excited 
hairy black holes that reduce, at small masses, to the superposition of 
a RNAdS black hole with $\mu= \frac{\Delta_n}{e}= \frac{4+2n}{e}$ with a
condensate of the $n^{th}$ scalar excited state.  It turns out that these excited hairy black holes are all unstable to the superradiant decay of the scalar mode with energy $\Delta_0=4$. They presumably decay to the ground state hairy black hole, in comparison to which they are all turn out to be entropically sub dominant.

Each excited hairy black hole exists in a limited mass range which turns out 
to be a subset of the mass range  \eqref{range}. At the lower end of this range, each excited hairy black hole
reduces to a horizon free scalar condensate, i.e. an excited soliton. 
Here we find a surprise. Recall that the instability of excited state hairy 
black holes is superradiant in nature. Superradiant instability rates 
scale like $R^3$, and so go to zero as $R \to 0$. It thus appears that excited 
state scalars are actually stable to small fluctuations, and so cannot 
decay classically. The discussion of this paragraph actually leaves open the 
possibility that excited solitons may have an independent non super radiant 
instability mode. It is, however, possible to demonstrate that the spectrum 
of small oscillations about arbitrary unstable solitons has no exponentially growing 
eigenmode, at least within the assumption of spherical symmetry (see \ref{ssec:dynstabl}).
While this result does not rigorously prove the stability of excited solitons
\cite{Wald:1992bd,Seifert:2006kv}, it at least suggests that they are stable.
We find this surprising and do not have a clear sense for its implications.

It may be worth pausing to consider the relative merits and 
disadvantages of the perturbative procedure of this paper versus the 
numerical approach more usually used to study hairy black branes. On the 
negative side our perturbative procedure gives us no information about 
the regime of large masses and charges (where the perturbative expansion 
breaks down). Within its regime of validity, however, our perturbative 
procedure is very powerful. It allows us, once and for all,  to compute the 
phase diagram and 
thermodynamics of all relevant solutions - including each of the infinite 
number of excited state hairy 
black holes - as analytic function of the parameters of the problem 
(e.g. the mass and charge of the scalar field). Perhaps more importantly our 
procedure gives us qualitative intuitive insight into the nature of hairy 
black holes.
For instance, as we have explained many times, the hairy black hole is 
an approximately non interacting mix of a RN-AdS black hole and the 
scalar condensate. This picture together with a few lines of algebra, 
immediately yields a formula for the entropy of the hairy black hole, 
to leading order in its mass and charge (see \ref{ssec:mphi}). In other 
words the perturbative approach employed in this paper gives more than 
numerical answers; it helps us to understand why small hairy black holes 
behave the way they do.

In this paper we have focussed on the black holes in global AdS  at small mass and charge. 
Almost all previous studies of the Lagrangian \eqref{lagrangian} have  
studied the system in the Poincare patch\footnote{See however \cite{Sonner:2009fk}
for a study in global AdS. }.The local dynamics of black holes in global AdS
reduces to the dynamics of black branes of Poincare AdS at large mass and charge. It follows that the phase diagram displayed in Fig.\ref{fig10} should make contact with the results of previous analyses at large mass and charge. In section \ref{sec:discussion} we present a conjectural phase diagram that interpolates between the small mass and charge behaviour 
derived in this paper and the large mass and charge behaviour determined 
in previous work. The analysis of that section makes it clear that something
special happens to the phase diagram of hairy at charges, in units of the 
inverse Newton constant, of order unity. We leave the detailed study of this
special behaviour to future work.

The reader who is interested in asymptotically
AdS gravitational dynamics principally because of the AdS/CFT correspondence
might legitimately complain that our choice of the Lagrangian 
\eqref{lagrangian} was arbitrary; the dynamics of charged scalar
fields in any given example of the AdS/CFT correspondence is 
unlikely to be given by \eqref{lagrangian}. Our attitude to this is the 
following: we regard \eqref{lagrangian} as a toy model which we have 
chosen to study (in common with much earlier work on the subject), 
largely because it is a simple system that possesses several 
of  the ingredients that are qualitatively important for hairy black hole 
dynamics. Our exhaustive study in this paper of the toy model 
sets the stage for a similar analysis of small hairy black holes in 
`realistic' theories IIB theory on AdS$_5 \times S^5$. We expect small of 
black holes in this special bulk theory to share many of the qualitative
features discussed in this paper, but also to have some properties 
that result from dynamical features special to it. We have already 
commenced this study and hope to report on it in the near future. 
\footnote{In this context it is however important to keep in mind that 
small black holes in IIB theory on $AdS_5\times S^5$ sometimes suffer from 
Gregory Laflamme instabilities \cite{Hubeny:2002xn} (in addition to 
potential superradiant instabilities), an additional feature that
complicates (but enriches) the dynamics of small black holes in a `realistic'
theory like IIB SUGRA on $AdS_5 \times S^5$.}

\textbf{Note added in v2 :} After the first version of this paper 
appeared, we became aware of a recent work by Maeda.et.al.\cite{Maeda:2010hf} studying
a similar subject. 
 
\section{The basic setup for the Hairy BHs}\label{sec:hairybh}

\subsection{Basic equations of motion}

As mentioned in the introduction, in this paper we study the Lagrangian 
\eqref{lagrangian}. This action describes the interaction of a massless minimally coupled scalar field, of charge $e$, interacting with a negative cosmological constant Einstein Maxwell system. Through most of this paper we will be interested in stationary, spherically 
symmetric solutions of this system . However, in Appendix.\ref{app:qnm} we will 
generalise to the study of time-dependent configurations to investigate the 
stability against small fluctuations. We adopt a Schwarzschild like gauge and set 
\begin{equation} \label{ansatz}
\begin{split}
ds^2&=-f(r) dt^2+ g(r) dr^2+ r^2 d \Omega_3^2\\
A_t&=A(r)\\
A_r&=A_i=0\\
\phi&=\phi(r)\\
\end{split} 
\end{equation}
The four unknown functions $f(r)$, $g(r)$, $A(r)$ and $\phi(r)$ are constrained by Einstein's equations, the Maxwell equations and the minimally coupled scalar equations. It is possible to demonstrate that $f, g, A , \phi$ are solutions to the equations of motion 
if and only if
\begin{equation} \label{maineqs}
\begin{split}
& r \left(3 f'(r)-2 e^2 r g(r) A(r)^2 \phi (r)^2+r
   A'(r)^2\right)-2 f(r) \left(\left(6 r^2+3\right) g(r)+r^2
   \phi '(r)^2-3\right)=0\\
& f(r) \left(3 r g'(r)-2 g(r) \left(r^2 \phi '(r)^2+3\right)  \right. \\ & \qquad \qquad \qquad \left. +6
   \left(2 r^2+1\right) g(r)^2\right)-r^2 g(r) \left(2 e^2 g(r)
   A(r)^2 \phi (r)^2+A'(r)^2\right)=0\\
& r g(r) f'(r) A'(r)+f(r) \left(r g'(r) A'(r)+4 e^2 r g(r)^2 A(r)
   \phi (r)^2 \right.
   \\ & \qquad \qquad \qquad \qquad \qquad \qquad \qquad \qquad \qquad \qquad
   \left.-2 g(r) \left(r A''(r)+3 A'(r)\right)\right)=0\\
& g(r) \left(\left(r f'(r)+6 f(r)\right) \phi '(r)+2 r f(r) \phi
   ''(r)\right)-r f(r) g'(r) \phi '(r)+2 e^2 r g(r)^2 A(r)^2
   \phi (r)=0.
\end{split}
\end{equation}
The four equations listed in \eqref{maineqs} are the $rr$ and $tt$ 
components of Einstein's equations, the Maxwell
equation and the minimally coupled scalar equation, in that order.

The equations \eqref{maineqs} contain only first derivatives of $f$ and $g$, 
but depend on derivatives upto the second order for 
$\phi$ and $A$. It follows  that \eqref{maineqs} admit a 6 parameter set 
of solutions. One of these solutions is empty AdS$_5$  space, given by $f(r)=r^2+1$, 
$g(r)=\frac{1}{1+r^2}$, $A(r)=\phi(r)=0$. We are interested in those solutions to \eqref{maineqs} that asymptote to AdS spacetime, i.e. solutions whose large $r$ behaviour 
is given by  
\begin{equation}\label{gCondns} \begin{split}
f(r)&=r^2+1+{\cal O}(1/r^2)\\
g(r)&=\frac{1}{1+r^2} + {\cal O}(1/r^6)\\
A(r)&={\cal O}(1) + {\cal O}(1/r^2) \\
\phi(r)&={\cal O}(1/r^4)\\
\end{split} 
\end{equation}
It turns out that these conditions effectively impose two conditions on the 
solutions of \eqref{maineqs}, so that the system of equations admits a 
four parameter set of asymptotically AdS  solutions\footnote{
For example, the equations above are easily solved in linearisation 
about AdS$_5$ ;the six dimensional  solution space is given by  
\begin{equation}\label{linerasolns} \begin{split}
\delta f(r)&= a_1 (1+r^2)-\frac{a_2}{r^2}\\
\delta g(r)&=\frac{a_2}{r^2(1+r^2)^2}\\
\delta A(r)&=a_3+\frac{a_4}{r^2}\\
\delta \phi(r)&=a_5+ a_6 \int \frac{dr}{r^3(1+r^2)}
\end{split}
\end{equation}
The asymptotically AdS  condition set $a_1=a_5=0$. }.
We will also be interested in solutions that are regular (in a suitable sense)
in the interior. As we will see below, this requirement will cut down solution 
space to distinct classes of two parameter space of solutions 
(the parameters may be thought of as the mass and charge of the solutions).
In particular, we are seeking the hairy black hole solutions of the above 
equations that constitute the endpoint of the superradiant instability 
of small RNAdS black holes. To set the stage and notations for our
computation we first briefly review the charged RNAdS black hole 
solutions in global AdS spacetime.

\subsection{RNAdS Black Holes and their superradiance}\label{subsec:RNBH}

The AdS-Reissner-Nordstrom black holes constitute a very well known 
two parameter set of solutions to the equations \eqref{maineqs}.
These solutions are given by 
\begin{equation}\label{bhsol}
\begin{split}
ds^2&= -V(r) dt^2+ \frac{dr^2}{V(r)}+ r^2 d \Omega_3^2 \\
V(r)&\equiv 1+r^2-\frac{R^2}{r^2}\left[1+R^2+\frac{2}{3}\mu^2\right] + 
\frac{2}{3}\mu^2\frac{R^4}{r^4} \\
&=  \left[1-\frac{R^2}{r^2}\right]\left[1+r^2+R^2-\frac{2}{3}\frac{\mu^2 R^2}{r^2}\right]\\
A(r)&= \mu \left[1-\frac{R^2}{r^2}\right]\\
\phi(r)&=0 \\
\end{split}
\end{equation}
where $\mu$ is the chemical potential of the RNAdS black hole. The
function $V(r)$ in \eqref{bhsol} vanishes at $r=R$ and 
consequently this solution has a horizon at $r=R$. In fact, it 
can be shown that $R$ is the outer event horizon provided 
\begin{equation}\label{extbound}
\mu^2 \leq \frac{3}{2}(1+2 R^2).
\end{equation}

We will review later the thermodynamics of these solutions in more detail
with a particular focus on small charged black holes whose $R\ll 1$. 
Consider the small RNAdS black hole solutions of the system described
by the Lagrangian in \eqref{lagrangian}. As we have explained in the
introduction, in the limit $R \ll 1$ we expect the solution in
\eqref{bhsol} to be unstable to superradiant decay provided
$e\mu \geq \Delta_0=4$.
In appendix \ref{app:qnm}, we verify this intuitive expectation 
by determining the lowest quasinormal mode of this system in a
power series in $R$. In a gauge where $A_t^{(r=R)}=0$, we
find that the time dependence of this lowest mode is given by
$e^{-i \omega t}$ where 
\begin{equation}\label{lowestqnm}
\begin{split}
\omega &= (\Delta_0-e\mu) + R^2\ (-6 + 3 e \mu -4 \mu^2) - i\  3R^3(\Delta_0-e\mu)  +{\cal O}(R^4)\\
&= (4-e\mu) + R^2\ (-6 + 3 e \mu -4 \mu^2) - i\  3R^3(4-e \mu)  +{\cal O}(R^4).
 \end{split}
\end{equation}

Note in particular that 
\[{\rm Im} \left(\omega \right)= -3R^3(\Delta_0-e\mu) + {\cal O}(R^4)\]
it follows that the time dependence $e^{-i \omega t}$ of this mode 
represents an exponential damping when $\mu e < \Delta_0$ but an 
exponential growth  when $\mu e >\Delta_0$. Consequently, small 
charged black holes are unstable when $\mu e >\Delta_0$, in agreement
with the intuitive expectations outlined  in the introduction. Further, 
note that the decay (or growth) constant of the lowest quasi normal mode
is given by $3 R^3 |\Delta- e \mu|$, and goes to zero either when $R$ goes
to zero or as $\mu$ goes near $\frac{\Delta_0}{e}$. As we have argued in the
introduction, this motivates us to seek a hairy black hole solution which
is constructed in a perturbation theory about these RNAdS blackholes.

\subsection{Setting up the perturbation theory}\label{subsec:pertheo}
The starting point of our construction is a small RNAdS black hole 
\begin{equation}\label{bhsol1}
\begin{split}
ds^2&= -V(r) dt^2+ \frac{dr^2}{V(r)}+ r^2 d \Omega_3^2 \\
V(r)&= \left[1-\frac{R^2}{r^2}\right]\left[1+r^2+R^2-\frac{2}{3}\frac{\mu^2 R^2}{r^2}\right]\\
A(r)&= \mu \left[1-\frac{R^2}{r^2}\right]\\
\end{split}
\end{equation}
at arbitrary but small $R$, and 
\begin{equation} \label{muexp} \begin{split}
\mu &= \mu(\epsilon, R)=\sum_{n=0} \epsilon^{2n} \mu_{2n}(R)\\
\mu_{2n}(R)&= \sum_{k=0}^\infty \mu_{(2n,2k)} R^{2k}\\
\mu_{(0,0)}&=\frac{4}{e}
\end{split}
\end{equation}  
Here $\mu= \mu(R, \epsilon)$ is the as yet unknown chemical potential of 
our final solution. Note that, at the leading order in the perturbative 
expansion, $\mu=\frac{4}{e}$. 

To proceed we simply expand every unknown function 
\begin{equation} \label{fieldexp}
 \begin{split}
  f(r, R, \epsilon) &= \sum_{n=0}^\infty\epsilon^{2n}f_{2n}(r, R)\\
  g(r, R, \epsilon) &= \sum_{n=0}^\infty\epsilon^{2n}g_{2n}(r, R)\\
  A(r, R, \epsilon) &= \sum_{n=0}^\infty\epsilon^{2n}A_{2n}(r, R)\\
  \phi(r, R, \epsilon) &= \sum_{n=0}^\infty\epsilon^{2n+1}\phi_{2n+1}(r, R)
 \end{split}
\end{equation}
Here $f_0$, $g_0$ and $A_0$ are the values of the functions $f$, $g$ and 
$A$ for a RNAdS black hole with radius $R$ and chemical potential $\mu = \mu_0(R)$.
given in \eqref{muexp}. We expand our equations in a power series in 
$\epsilon$. At each order in $\epsilon$ we have a set of linear differential 
equations (see below for the explicit form of the equations),
 which we solve subject to the requirements of the normalisability 
of $\phi(r)$ and $f(r)$ at infinity together with the regularity of 
$\phi(r)$ and the metric at the horizon. These four physical requirements
turn out to automatically imply that $A(r=R)=0$ i.e. the gauge field 
vanishes at the horizon, as we would expect of a stationary solution.
These four physical requirements determine 4 of the six integration 
constants in the differential equation, yielding a two parameter 
set of solutions. We fix the remaining two integration constants by adopting 
the following conventions to label our solutions: we require that $\phi(r)$ 
fall off at infinity like $\frac{\epsilon}{r^4}$ (definition of $\epsilon$), 
that the horizon area of our solution is $2 \pi^2 R^3$ (definition of $R$). 
This procedure completely determines our solution as a function of $R$ and $\epsilon$.
We can then read of the value of $\mu$ in \eqref{muexp} on our solution from the 
value of the gauge field at infinity.

As we have explained in the introduction, the linear differential equations 
that arise in perturbation theory are difficult to solve exactly, but 
are easily solved in a power series expansion in $R$, by matching near field 
and far field solutions. At every order in $\epsilon$ we thus have a solution 
as an expansion in $R$. Our final solutions are, then presented in a 
double power series expansion in $\epsilon$ and $R$. 

In the next few sections, we present a detailed description of the 
implementation of this perturbation expansion at order $\epsilon$ 
and $\epsilon^2$. In Appendix \ref{app:PertExp} we present explicit 
results for this perturbation expansion at higher orders.

\section{Perturbation theory at ${\cal O}(\epsilon)$}\label{ssec:ordreps}

We will now present a detailed description of the implementation of our 
perturbative expansion at ${\cal O}(\epsilon)$. The procedure described 
in this subsection applies, with minor modifications, to 
the perturbative construction at ${\cal O}(\epsilon^{2m+1})$ for all 
$m$.

In this section we wish to construct the first order correction around 
the black 
hole 
\begin{equation}\label{bhsolm}
\begin{split}
f_0(r,R) &= V(r),~~g_0(r,R) = \frac{1}{V(r)}\\
A_0(r,R)&=\mu_{0} (1-\frac{R^2}{r^2} )\\
V(r)&=1+r^2\left(1-\frac{\frac{2 R^4 \mu_{0}^2}{3}+R^4+R^6}{r^4 R^2} + 
\frac{2 R^4 \mu_0^2}{3 r^6}\right)\\
\end{split}
\end{equation}
Plugging in \eqref{fieldexp}, 
we expand the equations of motion in a power series in $\epsilon$ to 
${\cal O}(\epsilon)$. Of course all equations are automatically obeyed at 
${\cal O}(\epsilon^0)$. 
The only nontrivial equation at ${\cal O}(\epsilon)$ is 
$D^2 \phi=0$  where $D_\mu = \nabla_\mu - i e A_\mu$ is  the linearised 
gauge covariantised Laplace equation about the background \eqref{bhsolm}. 
We will now solve this equation subject to the 
constraints of normalisability at infinity, regularity at the horizon, 
and the requirement that $\phi(r) \sim \frac{\epsilon}{r^4}$ at large $r$.  

\subsection{Far Field Region}\label{sssec:farfield}

Let us first focus on the region $r \gg R$. In this region the background 
\eqref{bhsolm} is a small perturbation about global AdS space. 
For this reason we expand
\begin{equation}\label{phiexpout}
\phi_1^{out}(r)=\sum_{k=0}^\infty R^{2k} \phi^{out}_{(1,2k)}(r) ,
\end{equation}
where the superscript {\it out} emphasises that this expansion is good at large $r$.
In the limit $R \to 0$, \eqref{bhsolm} reduces to global AdS spacetime 
with  $A_t=\frac{4}{e}$. A stationary linearised fluctuation about this background 
is gauge equivalent to a linearised fluctuation with time dependence 
$e^{-4 i t}$ about global AdS space with $A_t=0$ ($A_t$ is the temporal 
component of the gauge field). The required solution is 
simply the ground state excitation of a massless minimally coupled scalar 
field about global AdS

\begin{equation}\label{normalisable}
\phi^{out}_{(1,0)}(r)=\frac{1}{(1+r^2)^2}.
\end{equation} 
The overall normalisation of the mode is set by the requirement 
$$\phi^{out}_{(1,0)}(r)=\frac{1}{r^4} +{\cal O}(1/r^6).$$

We now plug \eqref{phiexpout} into the equations of motion $D^2 \phi=0$ and 
expand to ${\cal O}(R^2)$ to solve for $\phi^{out}_{1,2}$. Here $D^2$ is the gauge
covariant Laplacian about the background \eqref{bhsol1}. Now  
$$(D^2)^{out}= (D_0^2)^{out}+R^2 (D_2^2)^{out} + \ldots$$
where $(D_0^2)^{out}$ is the gauge covariant Laplacian about global AdS spacetime with 
background gauge field $A_t=\frac{4}{e}$. It follows that, at ${\cal O}(R^2)$, 
$$(D_0^2)^{out}\phi^{out}_{(1,2)}= -(D_2^2)^{out} \phi_{(1,0)}^{out}=-(D_2^2)^{out} \left[\frac{1}{(1+r^2)^2} \right]$$
This equation is easily integrated and we find 
\begin{equation}\label{phout}
\begin{split}
\phi^{out}_{(1,2)}(r)=&\frac{2 \left(-3 e^2+6 \left(e^2-32\right) \left(r^2+1\right) \log (r)-3
   \left(e^2-32\right) \left(r^2+1\right) \log
   \left(r^2+1\right)-32\right)}{3 e^2 \left(r^2+1\right)^3}\\
&+\left(\mu_{0,2} - \frac{6 e^2-64}{e^3}\right)\left(\frac{e \left(6 \log (r) r^2-3 \log \left(r^2+1\right) r^2-1\right)}{6
   \left(r^3+r\right)^2}\right)
\end{split}
\end{equation}

We could iterate this process to generate $\phi^{out}_{(1,2k)}$ till any 
desired $k$. As in \eqref{phout}, it turns out that the expressions 
$\phi^{out}_{(1,2k)}$ are increasingly singular as $r \to 0$.  In fact it may 
be shown that the most singular piece of $\phi^{out}_{(1,2k)}$ scales like 
$\frac{1}{r^{2k}}$, upto logarithmic corrections. In other words the 
expansion of $\phi^{out}$ in powers of $R^2$ is really an expansion
in $\frac{R^2}{r^2}$ (upto log corrections)  and breaks down at $r \sim R$. 

In summary we have found that, to ${\cal O}(R^2)$ 
\begin{equation}\label{fint}
\begin{split}
 \phi^{out}_1(r) =&\frac{1}{(r^2 +1)^2}\\
&+R^2\bigg[\frac{2 \left(-3 e^2+3 \left(e^2-32\right) \left(r^2+1\right) \log (r^2/(r^2+1))-32\right)}{3 e^2 \left(r^2+1\right)^3}\\
&+\left(\mu_{(0,2)} - \frac{6 e^2-64}{e^3}\right)\left(\frac{e \left(6 \log (r) r^2-3 \log \left(r^2+1\right) r^2-1\right)}{6
   \left(r^3+r\right)^2}\right)\bigg] \\
& +{\cal O}({R^4/r^4})
\end{split}
\end{equation}
The small $r$ expansion of this result is given by
\begin{equation}\label{outrexp1}
 \begin{split}
\phi^{out}_1(r)=& \left[1 - 2 r^2 + {\cal O}(r^4)\right] + R^2\bigg[\frac{4}{e^2}(e^2 - 32)\log(r)
-2 \left(1 + \frac{32}{3 e^2}\right)+{\cal O}(r^2)\bigg]\\
& -R^2\left(\mu_{(0,2)} - \frac{6 e^2-64}{e^3}\right)\left[-\frac{e}{6 r^2} + e\log(r) + \frac{e}{3} + {\cal O}(r^2)\right] + {\cal O}(R^4)
\end{split}
\end{equation}
Note that this result depends on the as yet unknown parameter $\mu_{(0,2)}$. 
This quantity will be determined below by matching with the near field 
solution.

\subsection{Near Field Region}\label{sssec:nearfield}

Let us now turn to inner region $r \ll 1$. Over these length scales the 
small black hole is far from a small perturbation about AdS$_5$  space. 
Instead the simplification in this region arises from the fact that 
background gauge field, which is of order unity, is negligible compared 
to the mass scale set by the horizon radius $\frac{1}{R}$. In other 
words the gauge field is a small perturbation about the black hole
background in this region. To display this fact it is convenient to work in 
a rescaled radial coordinate $y=\frac{r}{R}$ and a rescaled time coordinate 
$\tau=\frac{t}{R}$. Note that the near field region consists of spacetime 
points with $y$ of order unity. In these coordinates the 
background black hole solution takes the form 
\begin{equation}\label{bhsolmy}
\begin{split}
ds^2&= R^2 \left( -V(y) d\tau^2+ \frac{dy^2}{V(y)}+ y^2 d \Omega_3^2 \right) \\
V(r)&=\left[1-\frac{1}{y^2}\right]\left[1-\frac{2}{3} \frac{\mu^2}{y^2} +R^2\left(1+y^2\right)\right]\\
A_\tau&= R \mu_0 (1-\frac{1}{y^2} )\\
\end{split}
\end{equation}
The explicit factor of $R$ in $A_\tau$ in \eqref{bhsolmy} demonstrates 
the effective weakness of the gauge field. This justifies an expansion of the near 
field solution in a power series in $R$
\begin{equation}\label{nearexp}
\phi^{in}_1(y)=\sum_{k=0}^\infty R^{2k} \phi^{in}_{(1,2k)}(y)
\end{equation}
To determine the unknown functions in this expansion, we must solve the 
equation $D^2 \phi^{in}=0$, where $D^2$ is the gauge covariant Laplacian 
about the background \eqref{bhsolmy}. Our solutions are subject to 
the constraint of regularity at the horizon. Further, they must match with 
the far field expansion in equations \eqref{fint} and \eqref{outrexp1} above.  

Note 
$$(D^2)^{in}=\frac{1}{R^2}(D_0^2)^{in} +  (D_2^2)^{in} + \ldots$$
where $(D_0^2)^{in}$ in is the leading part of $(D^2)^{in}$ in an
R expansion. At leading order we find 
$D_0^2 \phi^{in}_{(1,0)}(y)=0$ 
The two linearly independent solutions of this equation are easily obtained 
by integration. The only solution regular at the horizon is the constant 
$$\phi^{in}_{(1,0)}(y)= 1$$
where we have determined the value of the constant by matching with 
equations \eqref{fint} and \eqref{outrexp1} (more below about the matching).

At next order in $R^2$ we obtain an equation of the form 
$$D_0^2 \phi^{in}_{(1,2)}= -D_2^2 \phi^{in}_{(1, 0)}(y).$$
Even though  $\phi^{in}_{(1, 0)}(y)$ is a constant, the RHS of this equation 
is nonzero because of the gauge field in \eqref{bhsolm}. The equation is 
easily solved by integration; imposing regularity of the 
solution at the horizon we find\footnote{The apparent logarithmic singularities
at $y=1$, in two of the terms of \eqref{PhiIn12}, actually cancel.} 
\begin{equation}\label{PhiIn12}
\begin{split}
\phi^{in}_{(1,2)}(y)=&\alpha + \frac{1}{3 e^2}\bigg[-6 e^2 y^2-128 \log \left(3 e^2-32\right) \log \left(\frac{y^2-1}{3 e^2
   y^2-32}\right)-192 \log \left(3 e^2 y^2-32\right)\\
&+6 \log \left(3 e^2 y^2-32\right) e^2+128 \log \left(-\frac{3 e^2
   \left(y^2-1\right)}{3 e^2-32}\right) \log \left(3 e^2 y^2-32\right)\\
&+64 \log ^2\left(3 e^2 y^2-32\right)+128 \text{Li}_2\left(\frac{32-3 e^2
   y^2}{32-3 e^2}\right)\bigg]
\end{split}
\end{equation}
where $\text{Li}_2(x)$ is the polylog function as defined in Mathematica 6  
$$\text{Li}_2(z)= \sum_{k=1}^\infty \frac{z^k}{k^2}$$
The single unknown parameter $\alpha$ in this solution will be determined by 
matching below.  

The perturbative procedure described above may be iterated to arbitrary order.
It turns out that the  fields $\phi^{in}_{(1,2m)}$ at high $m$ 
are increasingly singular at large $y$. In fact it may be shown that the 
dominant growth of $\phi^{in}_{(1,2m)}$ is generically $y^{2m}$. It follows 
that the near field perturbative expansion is an expansion in $(yR)^{2}=r^2$. 

In summary
\begin{equation} \label{phin}
\begin{split}
\phi^{in}_1(y) =& 1+ R^2 \alpha\\
&+\frac{R^2}{3 e^2}\bigg[-6 e^2 y^2-128 \log \left(3 e^2-32\right) \log \left(\frac{y^2-1}{3 e^2
   y^2-32}\right)-192 \log \left(3 e^2 y^2-32\right)\\
 &+6 \log \left(3 e^2 y^2-32\right) e^2+128 \log \left(-\frac{3 e^2
   \left(y^2-1\right)}{3 e^2-32}\right) \log \left(3 e^2 y^2-32\right)\\
&+64 \log ^2\left(3 e^2 y^2-32\right)+128 \text{Li}_2\left(\frac{32-3 e^2
   y^2}{32-3 e^2}\right)\bigg]+{\cal O}(yR)^4
\end{split}
\end{equation}
The large $y$ expansion of this result is given by 
\begin{equation}\label{inrexp1}
\begin{split}
\phi^{in}_1(y) =& 1 + R^2\bigg[\left(-2y^2 + \frac{4}{e^2}(e^2 - 32)\log(y)\right)
+\alpha - \frac{64\pi^2}{3 e^2} + 6\left(1 - \frac{32}{e^2}\right)\log(3)\\
&+\frac{1}{3 e^2}\bigg(-192 \log ^2\left(\frac{1}{32-3 e^2}\right)+384 \log (3) \log \left(3
   e^2-32\right)\\
&+12 \log (e) \left(3 e^2+64 \log \left(3
   e^2-32\right)-96\right)\bigg)+{\cal O}\left(\frac{1}{y^2}\right)\bigg]+{\cal O}(Ry)^4
\end{split}
\end{equation}
In \eqref{inrexp1} we have determined $\phi^{in}_1(y)$ in terms of the as yet unknown 
parameter $\alpha$ which will be determined by matching below.

\subsection{Matching}\label{sssec:matching}

In order to match the near and far field results, we substitute 
$y=\frac{r}{R}$ in \eqref{inrexp1} (the large $y$ expansion 
of $\phi^{in}_1$) and view the resultant expression as an expansion
 about small $r$ and small 
$R$. As we have explained above, 
the resultant expression is reliable to all order in $R$  but only 
to order ${\cal O}(r^2)$; all terms of order ${\cal O}(r^4)$ or higher 
receive contributions from as yet undetermined fourth order 
 terms in the perturbation expansion of $\phi_1^{in}$ \eqref{inrexp1}.

We then compare this expression with the small $r$  expansion of 
$\phi^{out}$, \eqref{outrexp1}. We can generate this expansion to any order 
in $r$ that we desire (merely by Taylor expanding \eqref{outrexp1});  however 
the resultant expression is clearly valid only to  ${\cal O}(R^2)$
in $R$ (terms of ${\cal O}(R^4)$ obviously receive contributions from as yet 
undetermined fourth order terms in the expansion of $\phi^{out}$). 
Terms of the form $r^0 R^0$, $r^{2} R^{0}$ and $r^0 R^2$ (together with logarithmic corrections)
are reliably computed by both expansions and so must agree. 
The unknown parameters $\alpha$ and $\mu_{(0,2)}$ are determined to 
ensure this (as we have more conditions than variables we also obtain 
valuable consistency checks). We find 
\begin{equation*}
 \begin{split}
  \mu_{(0,2)}=& \frac{6 e^2-64}{e^3} \\
\alpha =& \frac{2 \left(-9 e^2-192 \log \left(3 e^2-32\right)+288\right)}{9 e^2}\log(3)
-\frac{2 \left(3 e^2-32 \log ^2\left(32-3 e^2\right)+32\right)}{3 e^2}\\
&+\frac{64 \pi ^2}{9 e^2}-18 \left(e^2-32\right) \log (R) +6 \log (e) \left(3 e^2+64 \log \left(3 e^2-32\right)-96\right)
 \end{split}
\end{equation*}
This completes our determination of our solution to ${\cal O}(R^2)$. 
The procedure described in this subsubsection can be iterated
to obtain the solution at higher orders.

\section{Perturbation theory at ${\cal O}(\epsilon^2)$}\label{ssec:ordepssq}

We now briefly outline the procedure used to evaluate the solution at 
${\cal O}(\epsilon^2)$. We proceed in close imitation to the previous 
subsection. The main difference is that at this (and all even orders) in 
the $\epsilon$ expansion, perturbation theory serves to determine the 
corrections to the functions $f$, $g$ and $A$ rather than the function $\phi$. 
The procedure described here applies, with minor modifications, to 
the perturbative construction at ${\cal O}(\epsilon^{2m})$ for all 
$m$. 

\subsection{Far Field Region}

In the far field region $r \gg R$ we expand
\begin{equation}\label{othexpout} \begin{split}
f_2^{out}(r)&=\sum_{m=0}^\infty R^{2m} f^{out}_{(2,2m)}(r)\\
g_2^{out}(r)&=\sum_{m=0}^\infty R^{2m} g^{out}_{(2,2m)}(r)\\
A_2^{out}(r)&=\sum_{m=0}^\infty R^{2m} A^{out}_{(2,2m)}(r)\\
\end{split}
\end{equation}
Plugging this expansion into the equations of motion and expanding to 
${\cal O}(R^{2m})$ we find equations of the form
\begin{equation}\label{bahir}
 \begin{split}
  \frac{d}{dr}\bigg(r^2(1 + r^2)^2 g^{out}_{(2,2m)}(r)\bigg) &= P^g_{(2,2m)}(r)\\
\frac{d}{dr}\left(\frac{f^{out}_{(2,2m)}(r)}{1+r^2}\right) &= \frac{2(1 + 2 r^2)}{r}g^{out}_{(2,2m)}(r) + P^f_{(2,2m)}(r)\\
\frac{d}{dr}\left(r^3\frac{dA^{out}_{(2,2m)}(r,R)}{dr}\right) &= P^A_{(2,2m)}(r)\\
 \end{split}
\end{equation}

Where $P^g_{(2,2m)}(r),~P^f_{(2,2m)}(r),~P^A_{(2,2m)}(r)$ are the source terms 
at order $\epsilon^2 R^{2m}$ which are obtained from terms quadratic in $\phi_1^{out}$. The most general normalisable 
solution to these equations takes the form 
\begin{equation}\label{bahirsama}
 \begin{split}
g^{out}_{(2,2m)}(r) =& \frac{b_{(2,2m)}}{r^2(1 + r^2)^2}-\frac{1}{r^2(1 + r^2)^2 }
\left(\int_r^\infty dx~P^g_{(2,2m)}(x)\right)\\
f^{out}_{(2,2m)}(r) =& - (1 + r^2)\left(\int_x^\infty  dx~ \left[ \frac{2(1 + 2 x^2)}{r}g^{out}_{(2,2m)}(x)+ P^f_{(2,2m)}(x) \right] \right)\\
A^{out}_{(2,2m)}(r) =& \frac{ h_{(2,2m)}}{r^2} +  k_{(2,2m)} + 
\int_r^\infty \frac{dx}{x^3}\left[\int_x^\infty dw~P_{(2,2m)}^A(w) \right]\\
\end{split}
\end{equation}
Note that this solution has three undetermined integration constants
$b_{(2,2m)},h_{(2,2m)}$ and $k_{(2,2m)}$. 

Let us first focus on ${\cal O}(R^0)$. The constants $b_{(2,0)}$ and 
$h_{(2,0)}$ are determined by the requirement that the expansion of 
 $g_{(2,0)}^{out}$ and $A_{(2,0)}^{out}$ at small $r$ starts out regular (i.e. 
has no term that goes like $\frac{1}{r^2}$). This requirement follows from 
matching with the near field solution. For example, a term in 
$g_{(2,0)}^{out} \propto \frac{1}{r^2}$ results would match onto a term 
in $g_{2}^{in}$ that scales like $\frac{1}{y^2 R^2}$. However  
$g_{2}^{in}(y, R)$ has a regular power series expansion in $R$ and 
so does not have such a term  \footnote{At higher orders in the expansion in 
$R$, similar reasoning will not set $b_{(2,2m)}$ and 
$h_{(2,2m)}^{(2)}$ to zero but will instead determine them by matching 
with $g_{(2, 2m-2)}^{in}$. }. 
The constant $k_{(2,2m)}=\mu_{(2,2m)}$, is as yet undetermined. 

\subsection{Near Field Region}

We now turn to the solution in the inner region $r \ll 1$. As in the previous 
section we find it convenient to solve the equations here in the rescaled 
$y$ and $\tau$ coordinates. We expand
\begin{equation}\label{othexpout1} \begin{split}
f_2^{in}(y)&=\sum_{m=0}^\infty R^{2m} f^{in}_{(2,2m)}(y)\\
g_2^{in}(y)&=\sum_{m=0}^\infty R^{2m} g^{in}_{(2,2m)}(y)\\
A_2^{in}(y)&=\sum_{m=0}^\infty R^{2m} A^{in}_{(2,2m)}(y)\\
\end{split}
\end{equation}

The equations are slightly simpler when rewritten in terms of a new function 
$$K_{(2,2m)}(y) = V_0(y) g^{in}_{(2,2m)}(y) + \frac{f^{in}_{(2,2m)}(y)}{V_0(y)}$$ where 
$$V_0(y) = \frac{(y^2-1)(y^2 - \frac{2}{3}\mu_{(0,0)}^2)}{y^4}$$
 In terms of this function the final set of equations take the following form.
\begin{equation}\label{samikaran}
 \begin{split}
\frac{dK_{(2,2m)}(y)}{dy} &= S^{(K)}_{(2,2m)}(y)\\
\frac{d}{dy}\left(y^3\frac{dA^{in}_{(2,2m)}(y)}{dy}\right) &= S^{(A)}_{(2,2m)}(y)  + \mu_{(0,0)}\left(\frac{d K_{(2,2m)}(y)}{dy}\right)\\
\frac{d}{dy}\bigg(y^2V_0(y)f^{in}_{(2,2m)}(y)\bigg) &= S^{(f)}_{(2,2m)}(y) + 2 y K_{(2,2m)}(y) -\frac{4\mu_{(0,0)}}{3}\left( \frac{dA^{in}_{(2,2m)}(y)}{dy}\right)\\
\end{split}
\end{equation}
Where $S^{(K)}_{(2,2m)}(y),~S^{(A)}_{(2,2m)}(y)~\text{and}~S^{(f)}_{(2,2m)}(y)$ are the source terms which depend on the solutions at all previous orders.
The most general solution to these equations is given by 
\begin{equation}\label{samadhan}
 \begin{split}
K_{(2,2m)}(y) =&f_{(2,2m)} + \int_1^y dx~S^{(K)}_{(2,2m)}(x)\\
A^{in}_{(2,2m)}(y) =&  \tilde h_{(2,2m)} (1-\frac{1}{y^2} )+ \int_1^y \frac{dx}{x^3}\left[\int_1^x dw~\left(S_{(2,2m)}^{(A)}(w) + \mu_{(0,0)} S_{(2,2m)}^{(K)}(w)\right)\right]\\  
f^{in}_{(2,2m)}(y) =& \frac{4\mu_{(0,0)}}{3}A^{in}_{(2, 2m)}(y) + \int_1^y dx\left(S_{(2,2m)}^{(f)}(x) + 2 x K_{(2,2m)}(x)\right)\\
\end{split}
\end{equation}
In the above solution two of the four integration constants are chosen 
such that the solution obeys  the requirement that $A(1)=0$ vanishes at the 
horizon (regularity of the gauge field at 
the horizon in Euclidean space) and that $f(1)=0$ (this is the requirement 
that the horizon is located at $y=1$, which follows from our definition 
of $R$). It may also be shown that the remaining two constants in the 
inner solution at ${\cal O}(R^{2k})$ may be determined 
by matching to the outer solution at the same order (${\cal O}(R^{2k})$).

In particular, the inner solution at order $R^0$ 
is completely determined by matching with the ${\cal O}(R^0)$ outer solution 
that we have already determined above, in terms of a single unknown $\mu_{(2,0)}$. 
This yields the complete solution at order ${\cal O}(R^0)$ in terms of this one 
unknown number.

\subsection{Iteration}

This process may now be iterated. Our determination of the inner solution at 
${\cal O}(R^0)$ permits an unambiguous determination of the  
integration constants in the outer solution at ${\cal O}(R^2)$. This then 
 allows the complete determination of the inner solution at ${\cal O}(R^2)$, 
which, in turn, permits the determination of the outer solution at 
${\cal O}(R^4)$ and so on. This procedure may be iterated indefinitely. 

In summary the procedure described in this subsection permits the complete 
determination of the ${\cal O}(\epsilon^2)$ correction to our solution (order 
by order in $R^2$), as a function of the shift in the chemical potential 
$\mu$  at ${\cal O}(\epsilon^2)$, i.e. in terms of the as yet unknown numbers 
$\mu_{(2, 2m)}$. These numbers are left undetermined by ${\cal O}(\epsilon^2)$ 
analysis, but turn out to be fixed by the requirement that there exist 
regular solutions of the scalar equation at ${\cal O}(\epsilon^3)$. This is completely analogous to the fact  that the ${\cal O}(\epsilon^0)$ shift in the chemical potential was determined from the analysis of the scalar equation at ${\cal O}(\epsilon)$. 

It is relatively straightforward (though increasingly tedious) to carry out 
our perturbative expansion to higher orders in perturbation theory. 
The equations at odd orders in the $\epsilon$ expansion serve to determine 
scalar field corrections, while equations at even orders serve 
to determine corrections to the metric and gauge field. In Appendix \ref{app:PertExp}
we list explicit results for the correction to the metric, gauge field and 
scalar field at low orders in perturbation theory. We will analyse the
thermodynamics of these hairy black holes in more detail in later sections.
%
%

\section{The Soliton}\label{sec:soliton}

The hairy black hole solitons of the previous section appear in a two 
parameter family labelled by $R$ and $\epsilon$. In general our solutions 
may be thought of as a small RNAdS black hole surrounded by a cloud of 
scalar condensate. As we have described in the introduction, the limit  
 $\epsilon \to 0$ switches off the condensate cloud. In this limit  (the 
blue line of Fig. \ref{fig10}) hairy black holes reduce to RNAdS black holes. 
On the other hand, in the limit $R \to 0$ the black hole at the centre of 
the condensate cloud shrinks to zero size, apparently leaving behind 
a horizon free scalar cloud.  This is indeed the case. Indeed the solitonic 
solutions so obtained are considerably simpler than hairy black holes, as 
they may be generated as a single expansion in $\epsilon$. The linear
 differential equations that we encounter at every order in this process 
are exactly solvable without recourse to the elaborate matching procedure 
described in the previous section.

 In this section we 
will directly construct the hairy black hole at $R=0$ in a perturbation 
expansion in $\epsilon$. We refer to the solution of this section as 
the `soliton'. In order to construct the soliton, we search for all 
stationary charged solutions 
that are everywhere completely singularity (and horizon) free. We use 
global AdS  as a starting point for these solutions, which we construct 
in a perturbative expansion in the scalar amplitude. As in the previous 
section we will only study rotationally invariant solutions, i.e. solutions 
that preserve the full $SO(4)$ symmetry group of AdS$_5$ .

At linear order the complete set of regular, asymptotically 
AdS$_5$ , $SO(4)$ symmetric
fluctuations about global AdS  is given by\footnote{We remind the reader that the Gauss's hypergeometric function ${}_2F_1\left[a,\ b,\ c,\ z\right]$ is a 
solution to the equation
\[
\left[\left(z\frac{d}{dz}+a\right)\left(z\frac{d}{dz}+b\right) - \left(z\frac{d}{dz}+c\right)\frac{d}{dz}\right] {}_2F_1\left[a,\ b,\ c,\ z\right]=0
\]
defined by the series
\[ {}_2F_1\left[a,\ b,\ c,\ z\right] \equiv \sum_{k=0}^{k=\infty}
\frac{(a)_k (b)_k}{(c)_k} \frac{z^k}{k!} \]
where the `Pochhammer symbol' $(a)_k$ is defined by the raising factorial
\[ (a)_k \equiv a(a+1)(a+2)\ldots(a+k-1) \]
Note also that for n an integer, the function ${}_2F_1\left[-n,-n-2,2,\ z\right]$ is actually an n-th degree polynomial in the variable $z$.} 

\begin{equation} \label{scl}
\begin{split}
\delta \phi& = \sum_{n}\frac{a_n e^{-i \omega_n t}}{(1+r^2)^{n+2}}\  {}_2F_1\left[-n,\ -(n+2),\ 2,\ -r^2\right] ,\qquad\text{with}\ \omega_n\equiv 4+2n -e \mathfrak{a}\\
A_t&=\mathfrak{a} ,\qquad\text{and}\qquad \delta g_{\mu\nu}=\delta A_i=0 .
\end{split}
\end{equation}

The equation \eqref{scl} is simply the most general rotationally invariant normalisable 
and regular solution to the equation $\partial^2 \phi$. The constant $\mathfrak{a}$ in 
\eqref{scl} can be set to any desired value by a choice of gauge. While \eqref{scl} is generically time dependent, it reduces to a stationary solution when only
one of the modes is turned on and $\mathfrak{a}$ is chosen accordingly.i.e.,
\[ \mathfrak{a}=\frac{2k+4}{e} \qquad\text{and}\qquad a_k\propto\delta_{nk} \]
for some non-negative integer k. This yields a discrete set of stationary, 
one parameter, solutions to the equations of motion, labelled by their amplitude. In the rest of this section we describe the construction of the nonlinear counterparts of these solutions in a power series expansion in the amplitude. We refer to these stationary 
solutions as nonlinear `solitons'. 

Unlike the two parameter set of black hole solutions described in the 
previous section, the solitonic solutions of this subsection appear in 
a one parameter family, labelled by their charge; the soliton mass is 
a determined function of its charge. The fact that their are fewer solitonic 
than black hole solutions is related to the fact that the solitons we 
construct in this subsection have no horizons, and therefore carry no 
macroscopic entropy. 

The ground state soliton (i.e. the soliton at $n=0$) has a specially simple 
interpretation. It may be thought of as the nonlinear version of the Bose 
condensate, that forms when a macroscopic number of scalar photons each 
occupies the scalar ground state `wave function'. It also represents the 
$R \to 0$ limit of the hairy black hole solution of the previous section. 
We will construct this solution in this section, postponing discussion 
of excited solitons to the next section. 

\subsection{Perturbation theory for soliton}
To initiate the perturbative construction of the ground state soliton we 
set
\begin{equation}
 \begin{split}
f(r)&=1+r^2 + \sum_{n} \epsilon^{2n} f_{2n}(r)\\
g(r)&=\frac{1}{1+r^2}+ \sum_{n=1}^\infty \epsilon^2n g_{2n}(r)\\
A(r)&=\frac{4}{e} + \sum_{n=1}^\infty \epsilon^{2n} A_{2n}(r)\\
\phi(r)&=\frac{\epsilon}{(1+r^2)^2} + \sum_{n=1}^\infty
\phi_{2n+1}(r) \epsilon^{2n+1}\\
\end{split}
\end{equation}
and plug these expansions into \eqref{maineqs}. We then expand out 
and solve these equations order by order in $\epsilon$. All equations are 
automatically solved upto ${\cal O}(\epsilon)$. At order 
$\epsilon^{2n}$ the last equation in \eqref{maineqs} is trivial while the first 
three take the form 
\begin{equation}\label{bahirs}
 \begin{split}
  \frac{d}{dr}\bigg(r^2(1 + r^2)^2 g_{2n}(r)\bigg) &= P^{(g)}_{2n}(r)\\
\frac{d}{dr}\left(\frac{f_{2n}(r)}{1+r^2}\right) &= \frac{2(1 + 2 r^2)}{r}
g_{2n}(r) + P^{(f)}_{2n}(r)\\
\frac{d}{dr}\left(r^3\frac{dA_{2n}(r)}{dr}\right) &= P^{(A)}_{2n}(r).\\
 \end{split}
\end{equation}
On the other hand, at order $\epsilon^{2n+1}$ the first three equations in 
\eqref{maineqs}
is trivial while the last equation reduces to 
\begin{equation} \label{phs}
\frac{d}{dr}\bigg(\frac{r^3}{(1 + r^2)^3}\frac{d}{dr}
\left[(1 + r^2)^2\phi_{2n +1}(r)\right]\bigg) = P^{(\phi)}_{2n+1}(r)
\end{equation}
Here the source terms $P^{(g)}_{2n}(r),~P^{(f)}_{2n}(r),~P^{(A)}_{2n}(r)~~ \text{and}~~P^{(\phi)}_{2n+1}(r)$ are the source terms which are completely 
determined by the solution to lower orders in perturbation theory, and 
so should be thought of as known functions, in terms of which we wish to 
determine the unknowns $f_{2n}$, $g_{2n}$, $A_{2n}$ and $\phi_{2n+1}$. 

Note that \eqref{bahirs} are identical to the equations that appear in 
the far field expansion of the hairy black hole solution of the previous 
section. This is intuitive; in the limit $R \to 0$ all of the hairy black 
hole spacetime lies within the far field region. The soliton is simpler 
to construct than the hairy black hole precisely because it has no 
separate near field region. The differential equations that arise, at any 
given order of perturbation theory, may simply be solved once and for all, 
with no need for an elaborate matching procedure. 

The equations \eqref{bahirs} are all easily integrated. It also turns out that 
all the integration constants in these equations are uniquely determined by 
the requirements of regularity, normalisability and our definition of 
$\epsilon$, as we now explain. 

The integration constant in the first equation of \eqref{bahirs} is 
determined by the requirement that $g(r)$ 
is regular at the origin. The integration constant in the second equation 
is fixed by requirement of normalisability for $f_{2n}(r)$. The constant for
the first of the two integrals needed to solve the third equation 
is fixed by the regularity of $A_{2n}(r)$ at the origin. The constant in the 
second integral (an additive shift in $A_{2n}$) is left unfixed at 
${\cal O}(\epsilon^{2n})$ but is fixed at ${\cal O}(\epsilon^{2n+1})$ 
(see below).

The equation \eqref{phs} is also easily solved by integration. The constant in the first 
integration needed to solve this equation is determined by the requirement 
of regularity of $\phi_{2n+1}(r)$ at the origin. Once we have fixed this 
constant, it turns out that the solution for $\phi_{2n+1}$ is generically 
non normalisable for every value of the second integration constant. 
In fact normalisability is achieved only when  the previously undetermined 
constant shift of $A_{2n}$ takes a specific value, a condition 
that determines this quantity. The constant in 
the last integral that determines $\phi$ from \eqref{phs} is determined by 
our definition of $\epsilon$ which implies that
$$\phi_{2n+1} \sim {\cal O}(1/r^6)$$ 
for $n \geq 1$. 

\subsection{Soliton upto ${\cal O}(\epsilon^4)$}

In summary, the perturbative procedure outlined in this subsection is very easily implemented
to arbitrary order in perturbation theory. In fact, by automating the 
procedure described above, we have implemented this perturbative series to 17th order in a Mathematica programme.

We present some of the results, to this order, in Appendix \ref{app:HigherOrdSoliton}. 
In the rest of this subsection we content ourselves with a presentation of our 
results to ${\cal O}(\epsilon^4)$.

\begin{equation}
 \begin{split}
  \phi(r) &=\frac{\epsilon }{\left(r^2+1\right)^2} + \frac{\epsilon ^3}{63
   \left(r^2+1\right)^6}
   \Big(-e^2 \left(9 r^6+30 r^4+34 r^2+13\right) +64 r^6+260
   r^4 \\ & +360 r^2+150\Big) + \mathcal{O}(\epsilon^5)\\
 \end{split}
\end{equation}

\begin{equation}
\begin{split}
   f(r) &= \left(r^2+1\right)-\frac{8 \left(r^4+3 r^2+3\right) \epsilon
   ^2}{9 \left(r^2+1\right)^3} +\frac{\epsilon ^4}{39690
   \left(r^2+1\right)^7} \left. \Big(e^2
   \left(6767 r^{12}+48104 r^{10}  
   \right. \right. \\ & \left. \left.+147252 r^8+256816 r^6+271348
   r^4+163008 r^2+42426\right)-32 \left(2448 r^{12}+17136
   r^{10}\right. \right. \\ & \left. \left. +51408 r^8+86688 r^6+87794 r^4+50014
   r^2+11213\right)\right. \Big) + \mathcal{O}(\epsilon^5)\\
\end{split}
\end{equation}

\begin{equation}
\begin{split}
g(r) &= \frac{1}{r^2+1}+\frac{8 r^2 \left(r^2+3\right) \epsilon ^2}{9
   \left(r^2+1\right)^5} -\frac{\epsilon ^4}{39690
   \left(r^2+1\right)^9} \left. \Big(r^2 \left(e^2
   \left(6767 r^{10}+48104 r^8+147252 r^6 \right. \right. \right.\\ & \left. \left. \left. +229600 r^4+180460
   r^2+58800\right)-64 \left(1224 r^{10}+8568 r^8+26194
   r^6+43260 r^4 \right. \right. \right.\\ & \left. \left. \left.
   +37065 r^2+11025\right)\right)\right. \Big)+ \mathcal{O}(\epsilon^5)\\ 
\end{split}
\end{equation}

\begin{equation}
 \begin{split}
  A(r) &= \frac{4}{e}+\epsilon ^2 \left(-\frac{e}{6 r^2}+\frac{e}{6 r^2
   \left(r^2+1\right)^3} +\frac{3 e}{14}-\frac{32}{21
   e}\right) +\epsilon ^4 \left(\frac{1}{105840 r^2
   \left(r^2+1\right)^7} \Big(e^3 \left(945 r^8 \right. \right. \\ &\left. \left.+315
   r^6-5691 r^4-8917 r^2-3856\right) +16 e \left(241
   e^2-2658\right) \left(r^2+1\right)^7\right. \\ &\left.-32 e \left(210 r^8+21
   r^6-1967 r^4-3527 r^2-1329\right) \Big) \right. \\ &\left.-\frac{6383817 e^4-122400480
   e^2+574944256}{97796160 e}\right)+ \mathcal{O}(\epsilon^5)\\
\end{split}
\end{equation}

We will postpone the discussion of the thermodynamics of these solution for later.

\subsection{Excited state solitons and hairy BHs}\label{ssec:soliton}

As explained around \eqref{scl}, the stationary solitonic solution
constructed in the previous section is simply one (albeit 
a special one, as it has the smallest mass to charge ratio) of an 
infinite class of stationary solitonic solutions, each of which may be 
constructed in a perturbative expansion in $\epsilon$, exactly as in the 
previous section. We label solitonic solutions by an integer $n$; 
the $n^{th}$ stationary soliton has chemical potential $\mu = 4+2n$ at small 
amplitude. 

We have explicitly constructed the excited solitons with $n=1$ and $n=2$ 
upto a high order in perturbation theory. In Appendix \ref{sec:excitedstate}
below we present some of the details of our results at low orders.  Further,
we can further construct a large class of excited hairy black holes which 
reduce to these excited state solitons  as their horizon size goes to zero. 
These black holes may be thought of as a mixture of the excited solitons 
and a small RNAdS black holes with $\mu \approx \frac{4+2n}{e}$.
In Appendix.\ref{app:ExcHairy1}, we construct this excited state hairy 
black hole at $n=1$. We have a simple program in  Mathematica  that
may be used to generate the excited hairy black hole solution at any 
fixed value of $n$. It should prove possible to generalise this 
construction once and for all at arbitrary $n$, but we have not
attempted this generalisation. We will present a detailed analysis of the
thermodynamics (and stability) of the excited solitons ad black holes
later in the paper.

\section{Thermodynamics in the Micro Canonical Ensemble}\label{sec:thermoMicro}

In this section we compare the entropies of the various solutions constructed
in this paper as a function of their mass and charge. We find it convenient
to present all formulae in terms of the rescaled mass $m$ and the rescaled 
charge $q$. The physical mass and charge of the system, $M$ and $Q$, differ from 
$m$ and $q$ by the  rescaling 
\begin{equation} \label{masschargescal} \begin{split}
Q&=\frac{\pi q}{2}\\
M&=\frac{3 \pi}{8} m\\
\end{split}
\end{equation}
The grand canonical partition function for the system is defined by the 
formula 
$$Z_{GC}= {\rm Tr} \exp\left[ -T^{-1} \left\{ M-\mu Q \right\} \right].$$
where $T$ is the temperature and $\mu$ is the chemical potential.

\subsection{RNAdS Black Hole}\label{ssec:RNAdSBH}
The basic thermodynamics for an RNAdS blackhole is summarised by the following
formulae\footnote{Throughout this paper, we find it convenient to consistently omit a
factor of $G_5^{-1}$ from all our extensive quantities.}
\begin{equation} \label{basictherm} \begin{split}
M&\equiv \frac{3 \pi}{8} m = \frac{3 \pi}{8} R^2\left[1+R^2+\frac{2}{3}\frac{q^2}{R^4}\right] \\
& = \frac{3 \pi}{8} R^2\left[1+R^2+\frac{2}{3} \left(\frac{2Q}{\pi R^2}\right)^2 \right]
 = \frac{3 \pi}{8} R^2\left[1+R^2+\frac{2}{3} \mu^2 \right]\\
Q&\equiv \frac{\pi}{2}q = \frac{\pi}{2} \mu R^2\\
S&= \frac{A_H}{4}= \frac{1}{4} (2\pi^2 R^3)=\frac{\pi^2}{2}R^3\\
T&= \frac{V'(R)}{4\pi} =\frac{1}{2\pi R}\left[1+2R^2-\frac{2}{3}\frac{q^2}{R^4}\right] \\
&=\frac{1}{2\pi R}\left[1+2R^2-\frac{2}{3}\left(\frac{2Q}{\pi R^2}\right)^2\right]  = \frac{1}{2\pi R}\left[1+2R^2-\frac{2}{3}\mu^2\right]\\
\mu&=A_t^{(r=\infty)}-A_t^{(r=R)}=\frac{q}{R^2}= \frac{2Q}{\pi R^2}.\\
\end{split}
\end{equation}
were $Q$ is the charge, $M$ is the mass of the black hole, $S$ is its 
entropy, $T$ its temperature and $\mu$ its chemical potential.We use
the symbol $A_H$ to denote the area of the outer horizon. The condition
for $R$ to be the outer horizon radius is 
\begin{equation}\label{extboundA}
\frac{q^2}{R^4}=\mu^2\leq \frac{3}{2}(1+2 R^2).
\end{equation}

We are mainly interested in small RNAdS black holes with $R\ll 1$. 
and the thermodynamic expressions can be simplified in this limit. 
The mass of RNAdS black holes at fixed charge is bounded from below; 
at small $m$ and $q$, we have  
\[ m \geq 2 \sqrt{\frac{2}{3}}q + \left(\sqrt{\frac{2}{3}} q\right)^2 - 
\left(\sqrt{\frac{2}{3}} q\right)^3 + {\cal O}(q^4)\] 
For every pair $(m, q)$ that obeys this inequality, there exists a 
unique black hole solution.

At small mass and charge (with mass and charge taken to be of 
the same order) the  entropy and the radius of the black hole
is given by
\begin{equation} \label{entt} \begin{split}
S&=\frac{\pi ^{2}R^3}{2}\\
R^2& =\frac{ m+\sqrt{ m^2-\frac{8}{3} q^2}}{2} + \mathcal{O}(m^2,q^2,mq)
\end{split}
\end{equation}
At leading order in mass and charge, the chemical potentials of these 
black holes are given as solutions to the equation 
\begin{equation}\label{tempchem}
\frac{m}{q}=\frac{1}{\mu}\left(1+\frac{2 \mu ^2}{3}\right)\\
\end{equation}
while the temperature is given by 
\begin{equation}
T = \frac{1}{2\pi R}\left[\frac{m}{R^2} - \frac{4 q^2}{3 R^4}\right]= \frac{1}{2\pi R}\left[1 - \frac{2 q^2}{3 R^4}\right]
\end{equation}
where $R^2$ is given in \eqref{entt}.

\subsection{Soliton - Ground State and Excited states}\label{ssec:GSSoliton}

Using the soliton solution in the previous section, the mass of the ground
state soliton can be easily determined as a function of its charge. We find 
\begin{equation}
 m =  \frac{16 q}{3 e}+\frac{2}{21}\left(9-\frac{64}{e^2}\right) q^2+O\left(q^3\right) 
\end{equation}
In Appendix \ref{app:eps17}, we give the relation of the mass and the charge
implicitly upto higher orders.

Upon continuing to Euclidean space, our soliton yields a regular solution for arbitrary
periodicity of the Euclidean time coordinate (it is similar 
to global AdS spacetime in this respect). It follows that, within the 
classical gravity approximation, this soliton can be in thermodynamical
equilibrium at arbitrary temperature. As the soliton has no horizon,
its entropy vanishes in the classical gravity approximation. This implies 
that the free energy of the soliton is equal to its mass. 

The chemical potential of the soliton is given by the value of the gauge 
potential at infinity
\begin{equation}
\mu  = \frac{4}{e}+\left(\frac{9}{7}-\frac{64}{7 e^2}\right)q+O\left(q^2\right)
\end{equation}
Note that the coefficient of $q$ in the formula above is positive when 
$e^2 > \frac{32}{3}\equiv e_c^2$ so that the chemical potential of the
soliton increases with charge whenever hairy black holes exist. It is
plausible that the only classical gravity state in the system with
$\mu < \frac{4}{e}$ is the vacuum (or more precisely a thermal gas 
about the vacuum; this gas is absent in classical gravity).

The grand free energy of the ground state soliton is given by
\begin{equation}
G(\mu) \equiv M-TS -\mu Q = -\frac{343 \pi  e^2 (e \mu -4)^4}{4 \left(9 e^2-64\right)^3} + \mathcal{O}\left((\mu - 4/e)^3\right).
\end{equation}


The above analysis is easily generalised to excited state solitons.
For the general excited state solitons, we present thermodynamical formulae 
only at leading order. The mass and chemical potential of the
soliton are given by 
\begin{equation} \begin{split}
 m & = \frac{4(4+2n) q}{3 e} +{\cal O}(q^2)\\
\mu &= \frac{4+2n}{e}+ {\cal O}(q)\\
\end{split}
\end{equation}

We have explicitly constructed the excited solitons with $n=1$ and $n=2$ 
upto a high order in perturbation theory. In Appendix \ref{sec:excitedstate}
we present some of the details of our results at low orders. This allows us to 
give the thermodynamic formulae for these cases upto a higher order
At $n=1$ we find 
\begin{equation} \label{oneex}\begin{split}
 M(q)& = \frac{3 \pi  q}{e}+\frac{1}{308} \pi 
   \left(109-\frac{2544}{e^2}\right) q^2 + \mathcal{O}\left(q^3\right),\\
\mu&=\frac{6}{e}+\frac{1}{77} \left(109-\frac{2544}{e^2}\right)
   q+O\left(q^2\right)\\
G(\mu)& = -\frac{77 \pi  (e \mu -6)^2}{436 e^2-10176}
      + \mathcal{O}\left(\mu-\frac{6}{e}\right)^3.
\end{split}
\end{equation}

while at $n=2$
\begin{equation} \label{twoex}\begin{split}
 M(q)& = \frac{4 \pi  q}{e}+\frac{\pi 
   \left(4741-\frac{228352}{e^2}\right)
   q^2}{12012}+O\left(q^3\right)\\
\mu(q)&=\frac{8}{e}+\frac{\left(4741-\frac{228352}{e^2}\right)
   q}{3003}+O\left(q^2\right)\\
G(\mu)& = -\frac{3003 \pi  (e \mu -8)^2}{18964 e^2-913408} 
+ \mathcal{O}\left(\mu-\frac{8}{e}\right)^3
\end{split}
\end{equation}
Note that the coefficient of $q$ in the expansion of $\mu$ in the expansion 
of \eqref{oneex} is positive whenever $e^2 > 24$ so that the first excited 
hairy black hole exists. On the other hand, the coefficient of $q$ in the 
expansion of $\mu$ is negative at $e^2=\frac{128}{3}$, the threshold for 
the existence of the second excited hairy black hole (see below).

\subsection{Dynamical Stability of solitons}\label{ssec:dynstabl}

In this subsection, we comment on the dynamical stability of the
excited solitons. In particular we will prove below 
that the spectrum of small fluctuations about excited
solitons have no $SO(4)$ symmetric exponentially growing modes, 
within the $\epsilon$ perturbative expansion. This result suggests
(but does not strictly prove \cite{Wald:1992bd,Seifert:2006kv})
that small excited state solitons are all dynamically stable 
against small fluctuations.

As we have mentioned above, the normal modes of the scalar field 
constitute the only $SO(4)$ symmetric fluctuations of \eqref{lagrangian} 
about global AdS spacetime. At small $\epsilon$ the solitonic solution 
is everywhere a small perturbation around global AdS spacetime. It follows 
that the $SO(4)$ symmetric perturbations about the soliton, at small 
$\epsilon$, are small perturbations of spherically symmetric scalar 
normal modes about global AdS spacetime. These modes obey the equation 
\begin{equation} \label{stab}
D^2 \phi=0
\end{equation}
where $D$ is the gauge covariant derivative about the soliton background. 
We study perturbations of the form 
$$\phi(r,t)= \psi(r) e^{-i \omega t}$$
and wish to investigate whether the frequencies $\omega$ (which are all real 
about global AdS ) can develop a small imaginary piece about the solitonic 
background. We will now demonstrate that this is impossible in the 
$\epsilon$ expansion. To establish this we multiply the equation 
\eqref{stab} by $\phi^*$ and integrate the resultant scalar over AdS spacetime. 
We find 
$$ \int \sqrt{g} |g^{0 0}| \left( \omega-e A_t(r) \right)^2 |\psi|^2 
= \int \sqrt{g} g^{rr} |\partial_r \psi|^2$$
Here $g_{\mu\nu}$ is the soliton metric and $A_t(r)$ is the gauge field 
of the solitonic solution. 

Now recall that $A_t= \frac{4}{e} +{\cal O}({\epsilon^2})$. It follows that 
$$(\omega-4)^2= \frac{ \int \sqrt{g} g^{rr} |\partial_r \psi|^2}
{\int \sqrt{g} |g^{0 0}|  |\psi|^2 }
+{\cal O}(\epsilon^2).$$
As the leading term on the RHS is ${\cal O}(\epsilon^0)$ and positive, 
it follows that $\omega$ is real within the $\epsilon$ expansion.
Consequently the spectrum of spherically symmetric small fluctuations 
about the soliton background does not have exponentially growing modes 
in the $\epsilon$ expansion. This suggests that all excited solitons 
are classically stable. We find this result surprising, and think that 
it warrants further study.

\subsection{Massive scalar : Hairy black hole thermodynamics}\label{ssec:mphi}

We concluded at the end of the previous subsection that the 
excited solitons seem to be classically stable. In contrast,
our calculations in the appendix \ref{app:qnm} indicate
that the RNAdS blackhole can become superradiantly unstable.
It is an interesting question to ask what is the thermodynamics
of the resultant hairy black hole. Since, we have an explicit
construction of the hairy black hole solutions, we can directly
go ahead to compute the thermodynamic quantities for these
hairy solutions. Before doing that however we wish to present
an argument in this subsection which gives us some intuition about
the kind of thermodynamics we should expect at the leading order.

We will present this argument in a slightly more general framework
than we have been working till now - we wish to consider the effect of
adding a scalar mass term to the Lagrangian\eqref{lagrangian}.,i.e.,
we work with a more general system  
\begin{equation}\begin{split}
S&=\frac{1}{8\pi G_5}\int d^5x  \sqrt{g} \left[ \frac{1}{2} \left(\mathcal{R}[g] +12\right) 
-\frac{1}{4} \mathcal{F}_{\mu\nu}\mathcal{F}^{\mu\nu} 
- |D_\mu \phi|^2 - m_\phi^2|\phi|^2\right]\\
\mathcal{F}_{\mu\nu} &\equiv \nabla_\mu A_\nu-\nabla_\nu A_\mu \qquad\text{and}\qquad
D_\mu \phi \equiv \nabla_\mu\phi -i e A_\mu \phi 
\end{split} 
\end{equation}
This system has a minimally coupled charge scalar with mass $m_\phi$ and charge $e$
in AdS$_5$. By the standard rules of AdS/CFT, the dual boundary operator
$\mathcal{O}_\phi$ has a scaling dimension
\[ \Delta_0 = \left[\frac{d}{2} + \sqrt{ \left(\frac{d}{2}\right)^2 +m^2_\phi }\
 \right]_{d=4} =  2 + \sqrt{ 4 +m^2_\phi}  \]
In a gauge where $A_t^{r=\infty}=0$, this is also the energy of 
the lowest $\phi$ mode in vacuum AdS$_5$. For the case $m_\phi=0$, this 
reduces to $\Delta_0=4$. 

The other spherically symmetric modes of $\phi$ (dual to the descendants $\partial^{2n}\mathcal{O}_\phi$ )have an energy 
\[ \Delta_n \equiv \Delta_0 + 2n =  2 + \sqrt{ 4 +m^2_\phi}  + 2n \]
For the case $m_\phi=0$, $\Delta_n=4+2n$. Hence, in a gauge where
$A_t^{r=\infty}=0$, the energy of the n-th excited state is also 
given by  $\Delta_n$. We can form a non-linear Bose condensate by
dumping a charge $Q_{sol}$ into this n-th excited state - this is
equivalent to populating this excited state with $Q_{sol}/e$ number
of charged Bosons. To the leading order, where we neglect 
self-interaction between these Bosons, the mass of such
a soliton is given by 
\[ M_{sol}=\frac{Q_{sol}}{e}\Delta_n+{\cal O}(Q^2_{sol}) \]
This sets the chemical potential of the soliton to be 
\[ \mu_{sol} \equiv \frac{\partial M_{sol}}{\partial Q_{sol}} = \frac{\Delta_n}{e}+{\cal O}(Q_{sol}) \]
The entropy of such a solution is zero $S_{sol}=0$. This in particular means 
that within this approximation, this solution exists at arbitrary temperatures
$T_{sol}$.

Now, let us form a hairy black-hole by placing at the core of this non-linear
Bose condensate a small ordinary RNAdS black hole with a small outer horizon
radius $R$ and chemical potential $\mu_{BH}$. Such a black hole has a mass
\[ M_{BH} = \frac{3\pi}{8}R^2\left[1+\frac{2}{3}\mu_{BH}^2\right] + {\cal O}(R)^4 \]
a charge 
\[ Q_{BH} = \frac{\pi}{2} \mu_{BH} R^2  \]
an entropy 
\[S_{BH} = \frac{1}{4}(2\pi^2R^3)=\frac{\pi^2}{2} R^3 \]
and  a temperature
\[ T_{BH} =\frac{1}{2\pi R} \left[1-\frac{2}{3}\mu_{BH}^2\right] + {\cal O}(R) \]
 
If the number of Bosons $Q_{sol}/e$ is small, then the condensate outside
is a small perturbation on the  RNAdS black hole. And if the radius $R$ 
of the blackhole is small, then it is a small perturbation on the soliton
on length scales large compared to $R$. Hence, if both these conditions are
met, it is legitimate at the leading order to assume that there is no 
interaction between the core and the condensate parts of the hairy 
black hole. In this regime, since the core and the condensate can 
still exchange charge and energy, all that is needed for a 
stationary solution is that the core and the condensate be at a
thermal and chemical equilibrium,i.e.,
\[ \bar{T}=T_{sol} = T_{BH} = \frac{1}{2\pi R} \left[1-\frac{2}{3}\mu_{BH}^2\right] + {\cal O}(R) \]
and
\[ \bar{\mu}=\mu_{BH} =\mu_{sol} = \frac{\Delta_n}{e}+{\cal O}(Q_{sol})\]

Using these equilibrium conditions we want to figure out the `mole
fractions' of these two phases at equilibrium as a function of total
mass and charge
\[ M = M_{sol}+ M_{BH} \qquad\text{and}\qquad Q= Q_{sol}+Q_{BH}\]
This is easily done and we get the mass fractions of the core and the condensate are given by
\begin{equation}\begin{split}
M_{BH} &=\frac{(1+\frac{2}{3}\bar{\mu}^2)}{(1-\frac{2}{3}\bar{\mu}^2)}(M-\bar{\mu}\ Q)+{\cal O}\left(M^2,Q^2,M Q\right) \\
m_{BH} &\equiv \frac{3}{8\pi} M_{BH}  = \frac{(1+\frac{2}{3}\bar{\mu}^2)}{(1-\frac{2}{3}\bar{\mu}^2)}(m-\frac{4}{3}\bar{\mu}\ q)
+{\cal O}\left(m^2,q^2,m q\right) \\
&= \frac{(1+\frac{2\Delta_n^2}{3e^2})}{(1-\frac{2\Delta_n^2}{3e^2})}(m-\frac{4}{3}\bar{\mu}\ q)
+{\cal O}\left(m^2,q^2,m q\right) \\
M_{sol} &= \frac{(1+\frac{2}{3}\bar{\mu}^2)\bar{\mu}\ Q - \frac{4}{3}\bar{\mu}^2\ M}{(1-\frac{2}{3}\bar{\mu}^2)} +{\cal O}\left(M^2,Q^2,M Q\right) \\
m_{sol} &\equiv \frac{3}{8\pi} M_{sol} = \frac{4\bar{\mu}}{3}\frac{(1+\frac{2}{3}\bar{\mu}^2)\ q - \bar{\mu}\ m}{(1-\frac{2}{3}\bar{\mu}^2)}+{\cal O}\left(m^2,q^2,m q\right) \\
&= \frac{4\Delta_n}{3e}\frac{(1+\frac{2\Delta_n^2}{3e^2})\ q - \frac{\Delta_n}{e}\ m}{(1-\frac{2\Delta_n^2}{3e^2})}+{\cal O}\left(m^2,q^2,m q\right) \\
\end{split}\end{equation}
The charge fractions of the core and the condensate are given by
\begin{equation}\begin{split}
Q_{BH} &=\frac{4\bar{\mu}}{3}\frac{(M-\bar{\mu}\ Q)}{(1-\frac{2}{3}\bar{\mu}^2)}+{\cal O}\left(M^2,Q^2,M Q\right)\\
q_{BH} &\equiv \frac{2}{\pi}Q_{BH}  = \bar{\mu}\frac{(m-\frac{4\bar{\mu}}{3}\ q)}{(1-\frac{2}{3}\bar{\mu}^2)}+{\cal O}\left(m^2,q^2,m q\right) \\
&= \frac{\Delta_n}{e}\frac{(m-\frac{4\Delta_n}{3e}\ q)}{(1-\frac{2\Delta_n^2}{3e^2})}+{\cal O}\left(m^2,q^2,m q\right) \\
Q_{sol} &= \frac{(1+\frac{2}{3}\bar{\mu}^2)\ Q - \frac{4}{3}\bar{\mu}\ M}{(1-\frac{2}{3}\bar{\mu}^2)} +{\cal O}\left(M^2,Q^2,M Q\right) \\
q_{sol} &\equiv \frac{2}{\pi} Q_{sol} = \frac{(1+\frac{2}{3}\bar{\mu}^2)\ q - \bar{\mu}\ m}{(1-\frac{2}{3}\bar{\mu}^2)}+{\cal O}\left(m^2,q^2,m q\right) \\
&=\frac{(1+\frac{2\Delta_n^2}{3e^2})\ q - \frac{\Delta_n}{e}\ m}{(1-\frac{2\Delta_n^2}{3e^2})}+{\cal O}\left(m^2,q^2,m q\right) \\
\end{split}\end{equation}

The radius of the black hole at the core is
\begin{equation}\begin{split}
R &= \left[\frac{8}{3\pi}\frac{(M-\bar{\mu}\ Q)}{(1-\frac{2}{3}\bar{\mu}^2)}+{\cal O}\left(M^2,Q^2,M Q\right)\right]^{1/2}
= \left[\frac{m-\frac{4}{3}\bar{\mu}\ q}{1-\frac{2}{3}\bar{\mu}^2}+{\cal O}\left(m^2,q^2,m q\right)\right]^{1/2}\\
&= \left[\frac{m-\frac{4\Delta_n}{3e}\ q}{1-\frac{2\Delta_n^2}{3e^2}}+{\cal O}\left(m^2,q^2,m q\right)\right]^{1/2}\\
\end{split}\end{equation}
and the entropy of the hairy black hole is given by
\begin{equation}\begin{split} 
S &= \frac{\pi^2}{2} R^3 = \frac{\pi^2}{2}\left[\frac{8}{3\pi}\frac{(M-\bar{\mu}\ Q)}{(1-\frac{2}{3}\bar{\mu}^2)}+{\cal O}\left(M^2,Q^2,M Q\right)\right]^{3/2} \\
&=\frac{\pi^2}{2} \left[\frac{m-\frac{4}{3}\bar{\mu}\ q}{1-\frac{2}{3}\bar{\mu}^2}+{\cal O}\left(m^2,q^2,m q\right)\right]^{3/2}\\
&= \frac{\pi^2}{2}\left[\frac{m-\frac{4\Delta_n}{3e}\ q}{1-\frac{2\Delta_n^2}{3e^2}}+{\cal O}\left(m^2,q^2,m q\right)\right]^{3/2}\\
\end{split}\end{equation}

The existence region of the hairy black holes is in between where 
the hairy black hole coincides with the RNAdS black hole on one side 
and where it coincides with the pure soliton on the other side.
This gives the existence region as 
\begin{equation}\begin{split}
&\frac{3e}{4\Delta_n}\left(1+\frac{2}{3}\frac{\Delta_n^2}{e^2}\right)Q +{\cal O}(Q^2)\geq M \geq \frac{\Delta_n}{e} Q +{\cal O}(Q^2) \\
&\frac{e}{\Delta_n}\left(1+\frac{2}{3}\frac{\Delta_n^2}{e^2}\right)q+{\cal O}(q^2)\geq  m \geq \frac{4}{3}\frac{\Delta_n}{e} q+{\cal O}(q^2)  \\
\end{split}\end{equation}
and this happens only if 
\[ e\geq \sqrt{\frac{2}{3}} \Delta_n = \frac{\Delta_n}{\mu_c}\equiv e_c \]
where $e_c$ is the critical charge above which the pure black hole becomes
superradiantly unstable to radiation in the n-th excited state.

\begin{figure}[ht]
\begin{center}
 \includegraphics[width=0.5\textwidth]{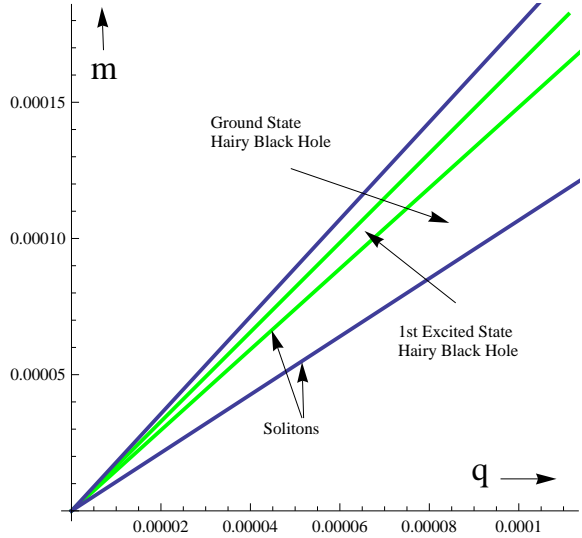}
\end{center}
\caption{\label{fig4} Microcanonical ensemble for $e= 5.4$ : Existence region of 
ground state and excited state hairy black holes.}
\end{figure}

In this regime, the upper bound on $M$ is a decreasing function of 
$\Delta_n$ whereas the lower bound is an increasing function of $\Delta_n$.
This implies that the existence region of n-th excited state hairy
black hole is entirely inside the existence region of (n-1)th excited
state black hole (see Figure.\ref{fig4}). Further, in this regime, 
one can show that the radius $R$ is a decreasing function of $\Delta_n$. 
Hence, the higher excited state hairy black holes have smaller
cores and consequently are entropically subdominant
to the lower excited state hairy black holes.

Armed with the above intuition, in the next few sections, we will
derive the thermodynamics of hairy black holes with $m_\phi=0$
directly from our solutions and show that their leading order 
behaviour is captured by the kind of non-interaction arguments
that we have presented in this subsection. 

\subsection{Ground State Hairy Black Hole}\label{ssec:GSHairyBH}

Once we have our solutions for hairy black holes from Appendix \ref{app:PertExp}, 
the evaluation of their thermodynamic charges and potentials is a 
straight forward exercise. At low orders in the perturbative expansion we 
find\footnote{Throughout this paper, we find it convenient to consistently omit a
factor of $G_5^{-1}$ from all our extensive quantities.}

\begin{equation}\label{thermoHairy}
 \begin{split}
 M =&\frac{3\pi}{8}\bigg(\left[\left(1+\frac{32}{3 e^2}\right) R^2-\left(-1-\frac{32}{e^2}+\frac{1024}{3
   e^4}\right) R^4 + {\cal O}(R)^6\right]\\
&+\epsilon^2 \left[\frac{8}{9}-\left(\frac{1016}{189}-\frac{21760}{189 e^2}\right) R^2 + {\cal O}(R)^4\right] \bigg)+ {\cal O}(\epsilon)^4\\
Q =&\frac{\pi}{2}\bigg(\left[ \frac{4 R^2}{e}-\left(\frac{64}{e^3}-\frac{6}{e}\right) R^4 + {\cal O}(R)^6\right]\\
&+\epsilon^2\left[\frac{e}{6}-\left(\frac{317 e}{252}-\frac{1528}{63 e}\right) R^2+ {\cal O}(R)^4\right]\bigg)+ {\cal O}(\epsilon)^4\\
\mu =& \bigg[\frac{4}{e}+ R^2\left(\frac{6}{e}-\frac{64}{e^3}\right)+R^4\bigg(-\frac{21}{2 e}-\frac{736}{3 e^3}+\frac{40448}{9 e^5}\\
&-\frac{256 \log \left(1-\frac{32}{3 e^2}\right)}{e^3}+\frac{8192 \log
   \left(1-\frac{32}{3 e^2}\right)}{3 e^5}-\frac{512 \log
   (R)}{e^3}+\frac{16384 \log (R)}{3 e^5}\bigg)+ {\cal O}(R)^6\bigg]\\
&+\epsilon^2 \bigg[\left(\frac{9 e^2-64}{42 e}\right) + R^2\left(-\frac{75969 e^4-2256672 e^2+13746176}{26460 e^3}\right) + {\cal O}(R)^4\bigg]
+ {\cal O}(\epsilon)^4\\
 T =&\frac{1}{4\pi R}\bigg(\bigg[\left(2-\frac{64}{3 e^2}\right)+\left(\frac{64 \left(32-3 e^2\right)}{3
   e^4}+4\right) R^2 +  {\cal O}(R)^4\bigg] + \epsilon^2\bigg[\frac{8 \left(e^2-32\right)}{21 e^2}\\
&-R^2\bigg(\frac{256 \left(13357 e^2-157376\right)}{6615 e^4}
+\frac{2048 \left(3 e^2-32\right) \left(\log \left(e^2-\frac{32}{3}\right)+2
   \log \left(\frac{R}{e}\right)\right)}{27 e^4}\bigg)\\
& +{\cal O}(R)^4\bigg]\bigg)+{\cal O}(\epsilon)^4
 \end{split}
\end{equation}

It may be verified that these quantities obey the first law of thermodynamics
$$dM=T d S+ \mu dQ .$$
Equation \eqref{thermoHairy} above lists formulae for the mass and charge of small hairy black holes as a function of $R$ and $\epsilon$. Inverting these relations we 
find
\begin{equation}\label{microsol}
 \begin{split}
 R^2 =& \left(\frac{e}{3 e^2 - 32}\right)\left(3 e m - 16 q\right)+  {\cal O}(m^2,q^2, m q)\\ \epsilon^2 =&\left( \frac{6}{e\left(3 e^2 - 32\right)}\right)\left[\left(3 e^2 + 32\right)q - 12 e m\right]+  {\cal O}(m^2,q^2, m q)\\ 
 \end{split}
\end{equation}

Hairy black holes exist for all positive values of $R$ and $\epsilon$. 
Of course $R^2$ and $\epsilon^2$ are positive; this implies that the mass 
and charge of hairy black holes vary over the range \eqref{range}. 
As we have mentioned in the introduction, it is possible to satisfy 
this inequality only when $e\geq\sqrt{\frac{32}{3}}=e_c$. Assuming this
is the case, we have hairy black hole solutions only within the band 
\eqref{range}. At the upper end of the band the solution reduces to a RNAdS 
black hole $(\epsilon =0)$.  At the lower end of the band the solution 
reduces to the soliton with $R=0$. 

It is now a simple matter to plug \eqref{microsol} into \eqref{thermoHairy}
to determine the entropy, temperature and chemical potential of the black 
hole as a function of its mass and charge. We find  
\begin{equation}\label{entsn} \begin{split}
 S =& \frac{\pi^2}{2} R^3 \\
=& \frac{\pi^2}{2}\left(\frac{e\left(3 e m - 16 q\right)}{3 e^2 - 32}\right)^{\frac{3}{2}} \bigg[\\
&1 +\frac{9}{14 e \left(32-3 e^2\right)^2 (3 e m-16 q)}\bigg(3 e^2 \left(21 e^4-384 e^2+5120\right) m^2\\
&+2 \left(27 e^6-64 e^4-1024 e^2+131072\right) q^2
p K-8 e \left(75 e^4-1152  e^2+17408\right) m q\bigg)\\
&+\mathcal{O}(m^2,q^2,m q)\bigg]\\
T&= \frac{2 \left(3 e^2-32\right)^{3/2}}{3 \sqrt{e^5 (3 e m-16 q)}}\left[1 +\mathcal{O}(m,q)\right]\\
\mu&=\frac{4}{e} +\frac{\left(576 e-18 e^3\right) m+\left(-27 e^4+576 e^2-5120\right) q}{224
   e^2-21 e^4} +\mathcal{O}(m^2,q^2,mq)\\
\end{split}
\end{equation}

As we have explained in the previous subsection, at leading order, these 
formulae have a very simple and intuitive explanation in terms of a
noninteracting mixture of a RNAdS black hole and the ground state soliton.
It is easily checked that the formulae in this section agree with the expressions
that we had derived before if we put $\Delta_n=4$.

\subsection{Excited Hairy black holes}\label{sec:HairyEx}

We begin by reporting some of the basic thermodynamical formulae
that follow from the formulae presented in Appendix.\ref{app:ExcHairy1}.
For the mass, charge and chemical potential we find 
\begin{equation}
\begin{split}
 m =& \left(\left(1+\frac{24}{e^2}\right)
   R^2+\left(1-\frac{304}{e^2}-\frac{1536}{e^4}\right)
   R^4+O\left(R^5\right)\right)\\ & +\left(\frac{2}{9}+\left(\frac{2
   6948}{231 e^2}-\frac{8111}{1386}\right)
   R^2+O\left(R^5\right)\right) \epsilon ^2+O\left(\epsilon
   ^3\right)
\end{split}
\end{equation}
\begin{equation}
\begin{split}
 q = &\left(\frac{6
   R^2}{e}+\left(\frac{12}{e}-\frac{432}{e^3}\right)
   R^4+O\left(R^5\right)\right) \\ &+\left(\frac{e}{36}+\left(\frac{
   7661}{462 e}-\frac{1061 e}{3696}\right)
   R^2+O\left(R^5\right)\right) \epsilon ^2+O\left(\epsilon
   ^3\right)
\end{split}
\end{equation}
\begin{equation}
\begin{split}
 \mu = \left(\frac{6}{e}+\left(\frac{12}{e}-\frac{432}{e^3}\right) 
    R^2+O(R)^4\right)+\epsilon ^2
   \left(\left(\frac{109 e}{2772}-\frac{212}{231
   e}\right)+O(R)^2\right)+O\left(\epsilon ^3\right)
\end{split}
\end{equation}

The $n^{th}$ excited state is a small deformation of a small RNAdS black hole 
at $\mu = \frac{4+2n}{e}$. RNAdS black holes at this value of the chemical 
potential have $n-1$ superradiant instabilities. It follows that the 
$n^{th}$ excited hairy black hole also has $n-1$ unstable linear fluctuation 
modes, which tend to flow the black hole to lower excited (generically ground 
state) hair black holes.

The $n^{th}$ excited hairy black hole exists only when $e^2 \geq \frac{2 (4+n)^2}{e}$. When this condition is fulfilled, the $n^{th}$ excited state black hole exists only when 
$$\frac{8}{3} (n+2) q+\mathcal{O}(q^2)\leq m\leq \frac{\left(3 e^2+8 (n+2)^2\right) q}{6 e
   (n+2)}+\mathcal{O}(q^2)  $$
This is completely in accordance with our non-interaction argument as expected.

\section{Discussion}\label{sec:discussion}

In this paper we have demonstrated that very small charged hairy black holes 
of the Lagrangian \eqref{lagrangian} are extremely simple objects. To leading
order in an expansion of the mass and charge, these objects may be thought 
of as an {\it non interacting} superposition of a small RNAdS black hole 
and a charged soliton. The different components of this mixture 
interact only weakly for two related reasons. The black hole does not affect 
the soliton because it is parametrically smaller than the soliton. 
The soliton does not backreact on the black hole because its energy 
density is parametrically small.

We have constructed an infinite class of hairy black hole 
solutions labelled by a single parameter $n$. Excepting the ground state 
hairy black hole, each of these solutions is classically unstable. 
The time scale for this instability is proportional to the area of the 
RNAdS black hole that sits inside the hairy solution, and goes to zero 
in the limit that this area goes to zero. In particular all excited
state solitons could well be stable configurations (see \ref{ssec:dynstabl}). We find the 
likely existence of an infinite number of classically stable classical solutions 
surprising, and do not have a good feeling for the implications of this 
observation. Of course these excited solitons will all eventually decay 
to the ground state hairy black hole via quantum tunnelling, but the rate 
for this decay will be exponentially suppressed. It would be interesting to 
construct the instanton that mediates this decay process. 

\begin{figure}[ht]
 \begin{center}
 \includegraphics[scale=0.44]{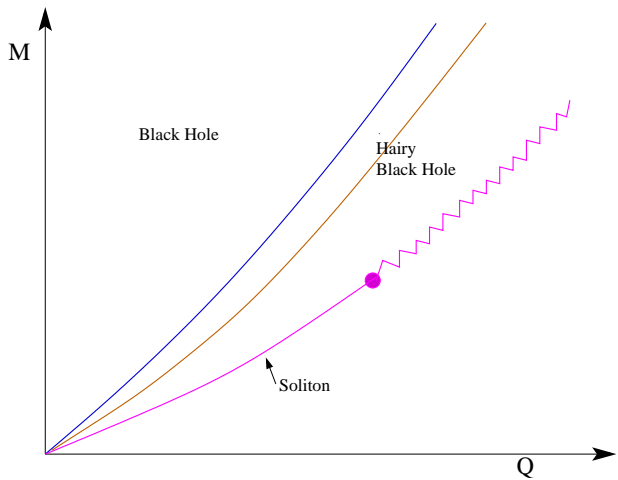}
\end{center}
\caption{\label{fig11} Proposed Microcanonical phase diagram for large $M$ and $Q$.}
\end{figure}

We have constructed hairy black holes in a perturbative 
expansion in their mass and charge. As we increase the mass and charge, the 
soliton and the black hole begin 
to interact with each other. At large mass and charge (where this system 
has been intensively previously investigated) there is probably 
no sense in which the hairy black hole can usefully be regarded as a mix of 
two independent entities. In fact we suspect that the soliton does not even 
exist as an independent object at large enough charge. It is very natural 
to wonder how the phase diagram of Figure 1 continues to large mass and 
charge. We sketch one possibility for this continuation in Fig. \ref{fig11}

The main point of interest of the conjectured phase diagram of Fig. 
\ref{fig11} is 
the lower edge of the diagram. As we have explained below, the hairy black 
hole phase is bounded from below by the solitonic solution at small mass 
and charge. The temperature of the hairy black hole also tends to infinity 
as we approach this line. On the other hand, it seems plausible that 
the solitonic solution goes singular past its `Chandrasekhar' critical 
charge $q=q_c$ (the dot in Fig. \ref{fig11}). If this is indeed the case, it 
is of great interest to know the nature of the lower bound of Fig. \ref{fig11}
at higher values of the charge (the jagged line in Fig. \ref{fig11}). We get
some information from studies of the system in the Poincare patch limit (see for instance \cite{Gubser:2008wz,Gubser:2009cg,Horowitz:2009ij}),so asymptotically large charges in Fig. \ref{fig11}. In this limit prior studies appear to suggest that the jagged line is the zero temperature limit 
of hairy black brane solution\footnote{It is interesting that the zero temperature limit of 
hairy black branes appears to depend qualitatively on the mass of the charged scalar field, 
and $m=0$ case is rather special. In the small black hole limit, on the other hand, we do 
not expect a qualitative dependence of our phase diagram on the mass of the scalar field. 
Indeed the leading order thermodynamics of hairy black holes in the Lagrangian \eqref{lagrangian} supplemented by a mass term for the scalar field can be easily  obtained based on general non-interaction arguments (see \ref{ssec:mphi}) and it 
gives results that are qualitatively similar to the $m=0$ case. We thank A. Yarom 
for a discussion about this point.}.
If the phase diagram of Fig. \ref{fig11} is indeed correct, it is natural to 
suppose that the jagged line everywhere represents the zero temperature 
limit of a hairy black hole. This suggests that the neighbourhood of the 
limiting soliton solution (the dot in Fig. \ref{fig11}) is extremely 
interesting. Hairy black holes on the solid lower line in the neighbourhood 
of this dot are at infinite temperature. On the other hand, points on the jagged  
line in the neighbourhood of this dot are at zero temperature. 
As we circle the dot from the solid to the jagged line we presumably 
pass through all temperatures in between. While all this is very 
speculative, would be very interesting to investigate it further.

Although all the analysis of this paper focussed on the concrete and especially
simple context of charged black holes in the system \eqref{lagrangian}, 
the basic physical picture of small hairy black holes 
as a linear combination of approximately non interacting pieces is 
based on very general considerations, should apply equally to the study of 
any bulk gravitational asymptotically AdS system that hosts RNAdS black 
holes which suffer from a superradiant instability. It should be 
straightforward to generalise the calculations of this paper to systems 
in which \eqref{lagrangian} is modified by the addition of a potential for
the scalar field, and / or is studied in different dimensions. It may 
also be possible to generalise the constructions of this paper to 
superradiant instabilities in  charged rotating black holes\cite{Kunduri:2006qa,
Cardoso:2006wa}.

As we have explained in section \ref{sec:hairybh}, at leading 
order in perturbation theory, small hairy black holes exist 
for $e^2 >\frac{32}{3}$, but do not exist when $e^2 < \frac{32}{3}$.
The case $e^2=\frac{32}{3}$ lies on the edge, and is 
particularly interesting. The question of whether hairy black holes
exist at this critical value is determined by a second rather 
than first order calculation. In the system under study in this paper, 
it turns out that hairy black holes do exist for $e^2 <\frac{32}{3}$.
One way of understanding this statement goes as follows.

At any fixed value of the charge $q$, hairy black holes exist if and only 
if $e^2 \geq e_c^2(q)$ where  
$$e_c^2(q)=\frac{32}{3}-\frac{16}{21}q +{\cal O}(q^2)$$ 
(one may derive this result by comparing the mass of the soliton with 
the extremal RNAdS black hole at equal charge). It follows that
small charge hairy black holes do exist at  $e^2=\frac{32}{3}$, but we have to go beyond leading order in perturbation theory to see this\footnote{On the other 
hand, the study of the near horizon BF bound in highly charged
extremal black hole backgrounds (which are locally well-approximated
by extremal black-branes) indicates that as $q\rightarrow \infty$, 
\[ e_c^2(q)\approx 3+ \frac{3^{1/3}}{2q^{1/3}}+ {\cal O}(q^{-4/3}) \]
Note that the leading deviation from the black-brane result of 
$e_c^2=3$ is positive and in the large q limit $e_c^2$ continues to be 
monotonically decreasing. In other words, all available data is consistent
with the conjecture that $e_c^2(q)$ is a monotonically decreasing
function that interpolates between $\frac{32}{3}$ and $3$ as $q$ ranges
from 0 to $\infty$.} .  Although we will not 
elaborate in detail in this paper, it turns out that the properties of small charged black holes at $e^2=\frac{32}{3}$ differ qualitatively from the properties of the same 
objects at larger $e^2$. This observation is particularly relevant as it turns
out that charged scalar fields are forced to sit at this critical value of 
charge in some natural supersymmetric bulk theories. We postpone the 
elaboration of these remarks and their consequences to future work.

A very interesting arena in which the ideas of this paper might find 
application is to the study of charged and rotating black holes in IIB theory 
on AdS$_5 \times S^5$. Small charged rotating black holes in 
IIB theory on AdS$_5 \times S^5$ sometimes suffer 
from superradiant instabilities. It should be possible to 
apply the ideas of this  to the study small black holes on 
AdS$_5 \times S^5$ (we have started work in this direction). 
In this context one should, however, also keep in mind that small black holes 
in this system sometimes also 
suffer from Gregory Laflamme type instabilities\cite{Hubeny:2002xn}
 even when they are uncharged, a feature that complicates (but enriches) 
the study of this system. 

Recall also that IIB supergravity on AdS$_5 \times S^5$ hosts a 4 parameter 
set of one sixteenth BPS supersymmetric black hole solutions\cite{Gauntlett:2002nw,
Gutowski:2004ez,Gutowski:2004yv}. It is possible
that there exist new BPS hairy black holes consisting of a non interacting 
mix  of these SUSY black holes with a SUSY graviton condensate. It would 
be very interesting to investigate this further.

\subsection*{Acknowledgements}

We would like to thank R. Gopakumar, S. Hartnoll, H. Liu, G. Mandal, 
K. Papadodimas, W. Song, A. Strominger, T. Takayanagi, S. Trivedi, S. Wadia and 
A. Yarom for useful discussions. We would also like to thank S.Hartnoll,
G.Horowitz, K. Papadodimas,  M.Rangamani, H.Reall , A.Strominger, 
S. Trivedi and A.Yarom for useful comments on a draft version of this 
manuscript. SM would like to thank IPMU for 
hospitality while this work was being completed. The work of SM was 
supported in part by a Swarnajayanti Fellowship. We must also acknowledge our debt to the steady and generous support of the people of India for research in basic science.

\appendix

\section{Superradiant Instability of small black holes}\label{app:qnm}

In this section we analyse the dynamical stability of RNAdS black holes in 
AdS$_5$  to superradiant emission in the presence of a massless charged 
minimally coupled scalar field. As we have explained in the introduction, 
we intuitively expect very small black hole to be unstable whenever $\mu \geq 
\frac{4}{e}$. In this Appendix we verify this expectation by computing 
the frequency of the lowest quasi normal mode of the black hole in 
a perturbative expansion $R$, the radius of the black hole. We find that 
the imaginary part of this frequency flips sign (from stable to unstable) as 
$\mu$ increases past $\frac{4}{e}$, exactly as we expected on intuitive 
grounds.

We wish to compute the lowest quasi normal mode of \eqref{bhsol1} at small 
$R$. 
By definition, quasi normal modes are regular at the future horizon, 
so it is useful to work in coordinates that are good at the future horizon. 
We choose to work in ingoing Eddington Finkelstein coordinate; in other words 
we replace the Schwarzschild time $t$ with the new time variable 
$$ v = t + \int \frac{1}{V(r)} dr, $$
where $V(r)$ given by 
\begin{equation}
 \begin{split}
V(r)&=  \left(1-\frac{R^2}{r^2}\right)\left(1+r^2+R^2-\frac{2}{3}\mu^2\right)\\
\end{split}
\end{equation}
In these new coordinates the background \eqref{bhsol1} takes the form 
\begin{equation}\label{metricED}
 ds^2= 2 dv dr - V(r) dv^2 + r^2 (d\theta^2 + \sin^2(\theta) d\psi^2 + \sin^2(\theta) \sin^2(\psi) d\lambda^2 ).
\end{equation}
with 
\begin{equation}\label{gaugefldEF}
\begin{split}
  A_v &= \mu \left(1-\frac{R^2}{r^2}\right).\\
  A_r &= -\frac{\mu}{V(r)}\left(1-\frac{R^2}{r^2}\right).
\end{split}
\end{equation}

A linearised scalar fluctuation about this background takes the form 
\begin{equation}\label{scaleq}
 D^{\mu} D_{\mu} \phi(v,r) = 0 ,
\end{equation}
where 
$$ D_{\mu} \equiv \nabla_{\mu} - i e A_{\mu}, $$
with $A_{\mu}$ being the background gauge field \eqref{gaugefldEF} 

In the rest of this appendix we will solve \eqref{scaleq} separately 
in a far field region, $r \gg R$ and a near field region $r \ll 1$. 
In the limit $R \ll 1$, of interest here, the solution may then be 
determined everwhere by matching the two solutions in their overlapping 
domain of validity. The matching procedure may be carried out systematically
in a power series in $R$, and turns out to determine the frequency of 
the quasi normal mode in a power series in $R$. We carry out this procedure 
to order $R^3$, the first order at which the quasi normal frequency 
develops an imaginary component. \footnote{The fact that the 
quasi normal frequency first develops an imaginary piece at 
${\cal O}(R^3)$ is simply related to the fact that the area - and so low 
frequency absorption cross section of a black hole in 5 dimensions - scales 
like $R^3$.}

\subsection{Solution in near field region }\label{app:nearfield} 

When $r \ll 1$ it is useful to work with the rescaled coordinate 
$y$ given by $$r = R y.$$ 
Let the scalar fluctuation take the form 
\begin{equation}
 \Phi^{in}(v,y) = \exp\left(- i \omega v\right) \Phi^{in}(y).
\end{equation}
where 
\begin{equation}
 \Phi^{in}(y) = \Phi^{in}_0(y) +\Phi^{in}_1(y) R + \Phi^{in}_2(y) R^2 + \mathcal{O}(R^3).
\end{equation}
and 
\begin{equation}
 \omega = 4 - \mu e  + \omega^{(1)} R + \omega^{(2)} R^2 + \omega^{(3)} R^3 + \mathcal{O}(R^4)
\end{equation}
Note we have chosen to study the quasi normal mode with the frequency 
$4 -\mu e  +{\cal O}(R)$; here we have used the physical expectation that 
the lowest quasinormal mode should reduce to the lowest normal mode in 
the limit $R \to 0$. The energy of the lowest normal mode is $4$. 

It is a simple matter to solve \eqref{scaleq} perturbatively in $R$. Imposing 
the physical requirement of regularity at the horizon we find 
\begin{equation}
 \begin{split}
  \Phi^{in}_0(y) &= d_0,\\
  \Phi^{in}_1(y) &= \frac{1}{6
   \left(2 \mu ^2-3\right)} \left( 6 \left(2 \mu ^2-3\right) (d_1-i d_0 y (e \mu
   -4)) \right. \\ &\left. +i d_0 (e \mu -4) \left(9 \left(\log \left(3 y^2-2
   \mu ^2\right)-2 \log (y+1)\right)+4 \sqrt{6} \mu ^3 \tanh
   ^{-1}\left(\frac{\sqrt{\frac{3}{2}} y}{\mu }\right)\right) \right),\\
  \Phi^{in}_2(y) &= -\frac{1}{2} d_0 y^2 \left(e^2 \mu ^2-8 e
   \mu +20\right)\\ &+\frac{y (4-e \mu ) \left(d_0 (e \mu -4) \left(2 \sqrt{6}
   \pi  i \mu ^3-9 \log (3)\right)+6 i
   d_1 \left(2 \mu ^2-3\right)\right)}{6 \left(2 \mu
   ^2-3\right)} \\ & -4 d_0 \left(e \mu -2 \mu ^2-3\right)
   \log \left(\frac{1}{y}\right)+ d_2 + \mathcal{O}\left(\frac{1}{y}\right).
 \end{split}
\end{equation}
where $d_0,d_1$ and $d_2$ are as yet undetermined integration constants 
(they will be determined below by matching). For brevity we have also 
presented the result assuming $\omega^{(1)}=0$, a result that turns out to be 
forced on us by matching with the far field expansion below.

\subsection{Solution in the far field region}\label{app:outer}

In the outer region the fluctuation takes the form 
\begin{equation}
 \Phi^{out}(v,r) = \exp\left(- i \omega v\right) \Phi^{out}(r).
\end{equation}
where
\begin{equation}
 \Phi^{out}(r) = \Phi^{out}_0(r) +\Phi^{out}_1(r) R + \Phi^{out}_2(r) R^2 + 
\Phi^{out}_3(r) R^3+ \mathcal{O}(R^4).
\end{equation}
Solving the equation of motion subject to the requirement of normalisability 
at large $r$ we find 
\begin{equation}
 \begin{split}
  \Phi^{out}_0(r) &= \frac{e^{-i (e \mu -4) \tan ^{-1}(r)}}{\left(r^2+1\right)^2},\\
  \Phi^{out}_1(r) &= 0,\\
  \Phi^{out}_2(r) &= \frac{e^{-i (e \mu -4) \tan ^{-1}(r)}}{6 r \left(r^2+1\right)^3} 
    \Big( \left. 3 i r^2 \left(2 \mu
   ^2+3\right) (e \mu -4)+\frac{3}{2} \left(r^2+1\right) r
   \left(-8 e \mu  \log \left(r^2+1\right) \right. \right. \\ &\left. \left. +16 \left(e \mu -2 \mu
   ^2-3\right) \log (r)+2 i \left(e \mu  \left(2 \mu
   ^2+9\right)-8 \left(2 \mu ^2+3\right)\right) \tan ^{-1}(r)-2
   i \pi  e \mu ^3 \right. \right. \\ &\left. \left. -9 i \pi  e \mu +16 \mu ^2 \log
   \left(r^2+1\right)+24 \log \left(r^2+1\right)+16 i \pi  \mu
   ^2+24 i \pi \right) \right. \\ &\left. +2 i \left(2 \mu ^2+3\right) (e \mu -4)-4
   r \left(2 \mu ^2+3\right) \right. \Big),\\ 
\end{split}
\end{equation}
and,
\begin{equation}
\begin{split}
  \Phi^{out}_3(r) &= -\frac{e^{-i (e \mu -4) \tan ^{-1}(r)}}{6
   \left(r^3+r\right)^2} \omega_3 \Big(3 i \pi  r^2
    + 3 r^2 \left(\log \left(r^2+1\right)-2
   \log (r)-2 i \tan ^{-1}(r)\right)+1 \Big)\\
 \end{split}
\end{equation}
where once again we have presented the results only for $\omega^{(1)}=0$. 
We have also plugged $\omega^{(2)} = -6 + 3 e \mu -4 \mu^2 $ in the 
expression for $\Phi_2(r)$ and $\Phi_3(r)$ (this is forced on us 
by the matching condition below). Here besides imposing normalisability
at infinity, we have also demanded that the coefficient of the 
leading normalisable piece is one.

\subsection{Conditions for patch up}\label{app:patchup}

In order to complete our determination of the solution, we must now match 
the near and far field solutions. The logic for this matching procedure 
is exactly as described in subsection \ref{sssec:matching}. 
Implementing this procedure we find 
\begin{equation}
 \begin{split}
  d_0 &= 1,\\
  d_1 &=\frac{i (e \mu -4) \left(2 \sqrt{6} i \pi  
 \mu ^3-9 \log (3)\right)}{6 \left(2 \mu ^2-3\right)},\\
  d_2&=\frac{1}{12} \left. \Big( 4 \left(2 \mu ^2+3\right) \left(e^2 \mu ^2-8
   e \mu +14\right)+48 \left(e \mu -2 \mu ^2-3\right) \log (R) \right. \\ & \left.-3
   i \pi  \left(2 e \mu ^3+9 e \mu -16 \mu ^2-24\right) \right. \Big).
 \end{split}
\end{equation}
Also the quasi-normal frequency is determined to be
\begin{equation}\label{qnf}
 \begin{split}
 \omega = 4 - e \mu  - R^2 (6 - 3 e \mu +4 \mu^2) - R^3 \left(3 i (4- e \mu) \right)
          +\mathcal{O}\left(R^4\right).
 \end{split}
\end{equation}
Once these matching conditions are imposed, the large $y$ expansion of the 
near field solution (with $y$ substituted by $\frac{r}{R}$) 
and the small $r$ expansion of the far field solution both share the common 
expansions 
\begin{equation}
\begin{split}
 \Phi^{out}(r) &= \Big( 1-i r (e \mu -4)+r^2 \left(-\frac{1}{2} e^2 \mu ^2+4 e \mu
   -10\right)+O\left(r^3\right) \Big)  \\ 
   &+ \left( \frac{i \left(2 \mu ^2+3\right) (e \mu -4)}{3 r}+\left(4 (\mu 
   (e-2 \mu )-3) \log (r)\right. \right. \\ &\left. \left.+\frac{1}{3} \left(2 \mu ^2+3\right) (e
   \mu  (e \mu -8)+14)\right. \right. \\ & \left. \left. -\frac{1}{4} i \pi  \left(2 e \mu ^3+9 e
   \mu -16 \mu ^2-24\right)\right) +O\left(r\right)\right) R^2 \\
   & - \Big( \frac{i (e \mu -4)}{2 r^2} 
   + \mathcal{O}\left(\frac{1}{r}\right) \Big) R^3 + \mathcal{O}\left( R^4 \right). \\
\Phi^{in}(r) &= \Big( r^2 \left(-\frac{1}{2} e^2 \mu ^2+4 e \mu
   -10\right)-i r (e \mu -4)+1+O\left(\frac{1}{r}\right) \Big)  \\ 
   &+ \left( \left(4 (\mu 
   (e-2 \mu )-3) \log (r)+\frac{1}{3} \left(2 \mu ^2+3\right) (e
   \mu  (e \mu -8)+14)\right. \right. \\ & \left. \left. -\frac{1}{4} i \pi  \left(2 e \mu ^3+9 e
   \mu -16 \mu ^2-24\right)\right)+\frac{i \left(2 \mu ^2+3\right) (e \mu -4)}{3 r} +O\left(\frac{1}{r^2}\right)\right) R^2 \\
   & - \Big( \frac{i (e \mu -4)}{2 r^2} 
   + \mathcal{O}\left(\frac{1}{r^3}\right) \Big) R^3 + \mathcal{O}\left( R^4 \right). \\
\end{split}
\end{equation}

Equation \eqref{qnf} is the main result of this Appendix. Note that the imaginary 
part of $\omega$ turns positive as $\mu$ exceeds $\frac{4}{e}$, demonstrating 
that RNAdS black holes with $\mu \geq \frac{4}{e}$ suffer from a super 
radiant instability. 

\section{Results of the Low Order Perturbative Expansion of the Hairy 
Black Hole}\label{app:PertExp}

In this Appendix we present explicit results for the perturbative expansion 
of the Hairy black hole solution at low orders in perturbation theory. 
See section \ref{sec:hairybh} for explanation of the notation etc.

\subsection{Near Field Expansion}\label{app:nearFieldExp}
\begin{equation}\label{uttorfin}
 \begin{split}
  f^{in}_{(0,0)}(y)=&\frac{\left(y^2-1\right) \left(3 e^2 y^2-32\right)}{3 e^2 y^4} \\
f^{in}_{(0,2)}(y) =& \frac{\left(y^2-1\right) \left(3 \left(y^4+y^2\right) e^4-96 e^2+1024\right)}{3
   e^4 y^4}\\
f^{in}_{(0,4)}(y)=&\frac{32 \left(y^2-1\right) \left[27 e^4+1536 e^2+384 \left(3 e^2-32\right)
   \log \left[\left(1-\frac{32}{3 e^2}\right) R^2\right]-22528\right]}{27 e^6
   y^4}\\
f^{in}_{(2,0)}(y) =&-\frac{8 \left(y^2-1\right) \left(3 e^2 \left(7 y^2-4\right)-64\right)}{63 e^2
   y^4}\\
 \end{split}
\end{equation}

\begin{equation}\label{uttorgin}
 \begin{split}
  g^{in}_{(0,0)}(y)=&\frac{3 e^2 y^4}{\left(y^2-1\right) \left(3 e^2 y^2-32\right)} \\
g^{in}_{(0,2)}(y) =& -\frac{3 y^4 \left(3 \left(y^4+y^2\right) e^4-96
   e^2+1024\right)}{\left(y^2-1\right) \left(32-3 e^2 y^2\right)^2}\\
g^{in}_{(0,4)}(y)=&-\frac{2048 y^4 \left(9 \left(5 y^2-6\right) e^4+96 \left(1-11 y^2\right)
   e^2+6656\right)}{3 e^2 \left(y^2-1\right) \left(3 e^2 y^2-32\right)^3}\\
&+\frac{9 e^2 y^6 \left(3 \left(y^3+y\right)^2 e^4-96 \left(2 y^2+3\right)
   e^2+2048 y^2\right)}{\left(y^2-1\right) \left(3 e^2 y^2-32\right)^3}\\
&+\frac{y^4 \left(y^8-12288 \left(3 e^2-32\right) \left(3 e^2
   y^2-32\right)\right)}{3 e^2 \left(y^2-1\right) \left(3 e^2 y^2-32\right)^3}\log \left[\left(1-\frac{32}{3 e^2}\right) R^2\right]\\
g^{in}_{(2,0)}(y) =&\frac{32 e^2 \left(40-3 e^2\right) y^4}{7 \left(y^2-1\right) \left(32-3 e^2
   y^2\right)^2}\\
 \end{split}
\end{equation}

\begin{equation}\label{uttorAin}
 \begin{split}
 A^{in}_{(0,0)}(y)=&\frac{4}{e}\left(1 - \frac{1}{y^2}\right) \\
A^{in}_{(0,2)}(y) =& \left(\frac{\left(6 e^2-64\right)}{e^3}\right) \left(1-\frac{1}{y^2}\right)\\
A^{in}_{(0,4)}(y)=&\left(\frac{189 e^4+4416 e^2+1536 \left(3 e^2-32\right) \log
   \left(\left(1-\frac{32}{3 e^2}\right) R^2\right)-80896}{18 e^5}\right)\left(1-\frac{1}{y^2}\right)\\
A^{in}_{(2,0)}(y) =&-\frac{2 \left(3 e^2+16\right)}{21 e} \left(1-\frac{1}{y^2}\right)\\
 \end{split}
\end{equation}

\begin{equation}\label{uttorphin}
 \begin{split}
 \phi^{in}_{(1,0)}(y)=& 1 \\
\phi^{in}_{(1,2)}(y) =& \alpha + \frac{1}{3 e^2}\bigg[-6 e^2 y^2-128 \log \left(3 e^2-32\right) \log \left(\frac{y^2-1}{3 e^2
   y^2-32}\right)-192 \log \left(3 e^2 y^2-32\right)\\
&+6 \log \left(3 e^2 y^2-32\right) e^2+128 \log \left(-\frac{3 e^2
   \left(y^2-1\right)}{3 e^2-32}\right) \log \left(3 e^2 y^2-32\right)\\
&+64 \log ^2\left(3 e^2 y^2-32\right)+128 \text{Li}_2\left(\frac{32-3 e^2
   y^2}{32-3 e^2}\right)\bigg]\\
\phi^{in}_{(3,0)}(y)=&\frac{1}{63} \left(150-13 e^2\right)
 \end{split}
\end{equation}
where
\begin{equation}
\begin{split} 
 \alpha =& \frac{2 \left(-9 e^2-192 \log \left(3 e^2-32\right)+288\right)}{9 e^2}\log(3)
-\frac{2 \left(3 e^2-32 \log ^2\left(32-3 e^2\right)+32\right)}{3 e^2}\\
&+\frac{64 \pi ^2}{9 e^2}-18 \left(e^2-32\right) \log (R) +6 \log (e) \left(3 e^2+64 \log \left(3 e^2-32\right)-96\right)
\end{split}
\end{equation}
\subsection{Far Field Expansion}\label{app:outside}
\begin{equation}\label{uttorf}
 \begin{split}
  f^{out}_{(0,0)}(r)=& 1 + r^2\\
f^{out}_{(0,2)}(r) =& \left(1 +\frac{32}{3 e^2}\right)\frac{1}{r^2}\\
f^{out}_{(0,4)}(r)=&\frac{32 e^2+\left(1024-3 e^2 \left(e^2+32\right)\right) r^2}{3 e^4 r^4}\\
f^{out}_{(2,0)}(r) =& -\frac{8 \left(r^4+3 r^2+3\right)}{9 \left(r^2+1\right)^3}\\
f^{out}_{(2,2)}(r) =&\frac{1}{189e^2 r^2(1+r^2)^4}\left[-256 \left(84 r^{10}+463 r^8+914 r^6+755 r^4+193 r^2-6\right)\right.\\
&+8 e^2 \left(252 r^{10}+1261 r^8+2419 r^6+2169 r^4+921 r^2+99\right)\\
&\left.+84 \bigg[\left(3 e^2-32\right) r^2 \left(r^2+1\right)^4+\left(32-e^2\right) r^2
   \left(r^4+3 r^2+3\right)\bigg] \log \left(\frac{r^2}{r^2+1}\right)\right]\\
 \end{split}
\end{equation}

\begin{equation}\label{uttorg}
 \begin{split}
  g^{out}_{(0,0)}(r)=&\frac{1}{1 + r^2}\\
g^{out}_{(0,2)}(r) =& \left(1 +\frac{32}{3 e^2}\right)\frac{1}{r^2(1 + r^2)^2}\\
g^{out}_{(0,4)}(r)=&\frac{9 \left(r^4+r^2+1\right) e^4+96 \left(3 r^4+2 r^2+1\right) e^2-1024
   \left(3 \left(r^4+r^2\right)-1\right)}{9 e^4 r^4 \left(r^2+1\right)^3}\\
g^{out}_{(2,0)}(r) =& \frac{8 r^2 \left(r^2+3\right)}{9 \left(r^2+1\right)^5}\\
g^{out}_{(2,2)}(r) =&8 \left(-127 e^2-\frac{448 \left(6 r^6+20 r^4+14
   r^2+5\right)}{\left(r^2+1\right)^4}+2720\right)\\
&+\frac{56 e^2 \left(9 r^6+36 r^4+70 r^2+13\right)}{\left(r^2+1\right)^4}\\
&+\frac{2 \left(e^2 \left(12 r^8+57 r^6+72 r^4+70 r^2+13\right)-384 r^4
   \left(r^4+4 r^2+3\right)\right)}{\left(r^2+1\right)^4}\log \left(\frac{r^2}{r^2+1}\right)
 \end{split}
\end{equation}

\begin{equation}\label{uttorA}
 \begin{split}
  A^{out}_{(0,0)}(r)=&\frac{4}{e}\\
A^{out}_{(0,2)}(r) =& \frac{2 \left(e^2 \left(3-\frac{2}{r^2}\right)-32\right)}{e^3}\\
A^{out}_{(0,4)}(r)=&-\frac{189 e^4+4416 e^2+1536 \left(3 e^2-32\right) \log
   \left(\left(1-\frac{32}{3 e^2}\right) R^2\right)-80896}{18 e^5}+ \frac{64-6 e^2}{e^3 r^2}\\
A^{out}_{(2,0)}(r) =& -\frac{e \left(r^4+3 r^2+3\right)}{6 \left(r^2+1\right)^3}+\frac{9 e^2-64}{42 e}\\
A^{out}_{(2,2)}(r) =&\frac{\left(\frac{33285}{r^2}-75969\right) e^4+96
   \left(23507-\frac{6685}{r^2}\right) e^2-13746176}{26460 e^3}\\
&-\frac{8 \left(24 r^8+60 r^6+20 r^4-51 r^2-29\right)}{9 e r^2
   \left(r^2+1\right)^4} +\frac{e \left(72 r^8+228 r^6+228 r^4+55 r^2-35\right)}{36 r^2
   \left(r^2+1\right)^4}\\
&+\frac{2 \left(-32 \left(r^2+2\right) r^4+e^2 \left(3 r^4+8 r^2+6\right)
   r^2+64\right)}{3 e
   \left(r^2+1\right)^3} \log \left(\frac{r^2}{r^2+1}\right)
 \end{split}
\end{equation}

\begin{equation}\label{uttorph}
 \begin{split}
 \phi^{out}_{(1,0)}(r)=& \frac{1}{(r^2 +1)^2}\\
\phi^{out}_{(1,2)}(r) =&\frac{2 \left(-3 e^2+6 \left(e^2-32\right) \left(r^2+1\right) \log (r)-3
   \left(e^2-32\right) \left(r^2+1\right) \log
   \left(r^2+1\right)-32\right)}{3 e^2 \left(r^2+1\right)^3}\\
\phi^{out}_{(3,0)}(r)=&\frac{64 r^6+260 r^4+360 r^2-e^2 \left(9 r^6+30 r^4+34 r^2+13\right)+150}{63
   \left(r^2+1\right)^6}
 \end{split}
\end{equation}

\section{The soliton at high orders in perturbation theory}\label{app:HigherOrdSoliton}

\subsection{Explicit Results to ${\cal O}(\epsilon^{17})$}\label{app:eps17}

As we have mentioned in section \ref{sec:soliton}, the perturbation theory 
that generates 
the soliton solution as a function of $\epsilon$ is straightforward and hence may be automated on Mathematica. We have implemented this automation
and used it to generate the ground state soliton solution to 
${\cal O}(\epsilon^{17})$. For what its worth, we present the resultant 
explicit formulae for all thermodynamical quantities: the mass, the 
charge and the chemical potential to ${\cal O}(\epsilon^{17})$. Later in 
this appendix we will also speculate that our solution develops a singularity 
at the origin at a finite value of $\epsilon$. To aid this discussion we 
also present formulas for $f(r=0)$ and $\phi(r=0)$ to the same order in 
$\epsilon$.

\begin{equation*}
\begin{split}
& \left. m=\right. \\ & \left. 0.888889 \epsilon ^2 + \left(1.9737-0.170496 e^2\right) \epsilon ^4 + \left(10.7168-1.77184 e^2+0.0725209 e^4\right) \epsilon ^6 \right.  \\   
& \left. + \left(76.4861-18.5347 e^2+1.48588 e^4-0.0394005 e^6\right) \epsilon ^8   \right. \\
& \left. + \left(624.015-198.755 e^2+23.5941 e^4-1.23705 e^6+0.0241682 e^8\right) \epsilon ^{10}  \right. \\
& \left. + \left(5511.63-2173.08 e^2+340.947 e^4-26.6063 e^6+1.03262 e^8-0.0159449 e^{10}\right) \epsilon ^{12}  \right. \\
& \left. + \left(51307.1-24103.6 e^2+4697.17 e^4-485.985 e^6+28.1541 e^8 \right. \right. \\
& \left. \left.-0.865871 e^{10}+0.0110442 e^{12}\right) \epsilon ^{14}  + \left(495774.-270273. e^2+62898.2 e^4-8099.8 e^6 \right. \right. \\
& \left. \left. +623.327 e^8-28.6648 e^{10}+0.729356 e^{12}-0.0079209 e^{14}\right) \epsilon ^{16}+ {\cal O}(\epsilon^{18}) \right.
\end{split}
\end{equation*}

\begin{equation*}
 \begin{split}
&\left. q= \right. \\ & \left. 0.166667 e \mbox{ } \epsilon ^2+\left(0.401814 e-0.0364324 e^3\right) \epsilon ^4+\left(2.1931 e-0.373206 e^3+0.0158055 e^5\right) \epsilon ^6 \right. \\ & \left. +\left(15.6491 e-3.87045 e^3+0.317549 e^5-0.00864522 e^7\right) \epsilon ^8 \right. \\ & \left. +\left(127.56 e-41.2748 e^3+4.98583 e^5-0.266507 e^7+0.00531978 e^9\right) \epsilon ^{10} \right. \\ & \left. +\left(1125.66 e-449.55 e^3+71.5228 e^5-5.66677 e^7+0.223607 e^9-0.00351592 e^{11}\right) \epsilon ^{12} \right. \\ & \left. +\left(10470.4 e-4972.41 e^3+980.312 e^5-102.7 e^7+6.03018 e^9\right. \right. \\
& \left. \left.-0.188165 e^{11}+0.00243798 e^{13}\right) \epsilon ^{14} \right. \\ & \left. +\left(101107. e-55636.5 e^3+13077.2 e^5-1701.96 e^7+132.464 e^9-6.16548 e^{11} \right. \right. \\ & \left. +\left. 0.158912 e^{13}-0.00174981 e^{15}\right) \epsilon ^{16}+ {\cal O}(\epsilon^{18}) \right.
 \end{split}
\end{equation*}

\begin{equation*}
 \begin{split}
& \left. \mu = \right. \\& \left. \frac{1}{e}\left[4.+\left(-1.52381+0.214286 e^2\right) \epsilon ^2+\left(-5.87901+1.25159 e^2-0.0652768 e^4\right) \epsilon ^4 \right. \right. \\ & \left. \left. +\left(-37.0661+10.6372 e^2-1.00385 e^4+0.0312297 e^6\right) \epsilon ^6 \right. \right.  \\ & \left. \left.  +\left(-283.701+102.563 e^2-13.7659 e^4+0.813892 e^6-0.0179038 e^8\right) \epsilon ^8 \right. \right. \\ & \left. \left. +\left(-2410.37+1051.44 e^2-182.027 e^4+15.6419 e^6-0.667552 e^8+0.0113255 e^{10}\right) \epsilon ^{10} \right. \right. \\ & \left. \left. +\left(-21860.7+11170.5 e^2-2363.08 e^4+264.989 e^6-16.6181 e^8\right.  \right. \right. \\ & \left. \left.   \left.+0.552789 e^{10}-0.00762253 e^{12}\right) \epsilon ^{12} \right. \right. \\ & \left. \left. +\left(-207326.+121451. e^2-30325.8 e^4+4184.8 e^6-344.747 e^8+16.9581 e^{10}\right. \right. \right.  \\ & \left.  \left. -0.461289 e^{12}+0.00535403 e^{14}\right) \epsilon ^{14} \right.  \\  & \left.  +\left(-2.03127\times 10^6+1.34191\times 10^6 e^2-386028. e^4+63167.2 e^6-6431.58 e^8+417.306 e^{10} \right. \right. \\ & \left. \left.   \left. -16.8524 e^{12}+0.387334 e^{14}-0.00387981 e^{16}\right) \epsilon ^{16} + {\cal O}(\epsilon^{18})\right]\right.
 \end{split}
\end{equation*}

\begin{equation*}
 \begin{split}
& \left. f(r=0) = \right. \\ & \left. 1.-2.66667 \epsilon ^2+\left(-9.04046+1.06893 e^2\right) \epsilon ^4+\left(-55.7996+11.1848 e^2-0.566798 e^4\right) \epsilon ^6 \right. \\ & \left. +\left(-424.503+118.131 e^2-10.9991 e^4+0.34307 e^6\right) \epsilon ^8 \right. \\ & \left. +\left(-3599.49+1276.91 e^2-170.023 e^4+10.0766 e^6-0.224439 e^8\right) \epsilon ^{10}\right. \\ & \left. +\left(-32626.9+14049.7 e^2-2419.13 e^4+208.257 e^6-8.96698 e^8+0.154554 e^{10}\right) \epsilon ^{12} \right. \\ & \left. +\left(-309433.+156626. e^2-33000.6 e^4+3705.4 e^6-233.893 e^8\right. \right. \\ & \left. \left.+7.87144 e^{10}-0.110374 e^{12}\right) \epsilon ^{14} \right. \\ & \left. +\left(-3.03237\times 10^6+1.76343\times 10^6 e^2-438933. e^4+60625.9 e^6 \right. \right. \\ & \left. \left. -5019.01 e^8+249.084 e^{10}-6.86274 e^{12}+0.080996 e^{14}\right) \epsilon ^{16} + {\cal O}(\epsilon^{18})\right.
 \end{split}
\end{equation*}

\begin{equation*}
 \begin{split}
& \left. \phi(r=0)= \right. \\ & \left. \epsilon+\left(2.38095-0.206349 e^2\right) \epsilon ^3+\left(13.5366-2.20766 e^2+0.0892759 e^4\right) \epsilon ^5 \right. \\ & \left. +\left(99.3332-23.6891 e^2+1.86986 e^4-0.0488621 e^6\right) \epsilon ^7 \right. \\ & \left. +\left(825.529-258.875 e^2+30.261 e^4-1.56287 e^6+0.0300946 e^8\right) \epsilon ^9 \right. \\ & \left. +\left(7388.22-2870.91 e^2+443.951 e^4-34.1507 e^6+1.30689 e^8-0.0199063 e^{10}\right) \epsilon ^{11} \right. \\ & \left. +\left(69458.8-32195.6 e^2+6190.33 e^4-631.96 e^6+36.1289 e^8\right. \right. \\ & \left. \left.-1.09675 e^{10}+0.0138126 e^{12}\right) \epsilon ^{13} \right. \\ & \left. +\left(676349.-364168. e^2+83703. e^4-10646.1 e^6+809.226 e^8-36.7616 e^{10} \right. \right. \\ & \left. \left. +0.924182 e^{12}-0.00991936 e^{14}\right) \epsilon ^{15}+ \left(6.76167\times 10^6-4.14712\times 10^6 e^2\right. \right. \\ & \left. \left.+1.10887\times 10^6 e^4-168826. e^6+16007.9 e^8-967.993 e^{10}+36.4553 e^{12}\right. \right. \\ & \left. \left.-0.781807 e^{14}+0.00731015 e^{16}\right) \epsilon ^{17}+ {\cal O}(\epsilon^{19}) \right.
 \end{split}
\end{equation*}
\subsection{Particular case $e=4$}\label{app:eIS4}

Of course, the formulae of the previous subsection are not immediately 
illuminating. In order to extract some (tentative) physical conclusions 
from those formulae, we specialise, in this section, to a particular value 
of $e$, namely $e=4$. At this specific value of $e$ we were able to coax 
Mathematica into producing results upto ${\cal O}(\epsilon^{30})$. We will not
explicitly list our results, but use them to generate some plots that may 
carry qualitative lessons. 

We are principally interested in the following question: does our solitonic 
solution develop a singularity (and so cease to exist) past a particular 
critical value of $\epsilon$ (or charge)? It would seem intuitively that this
should be the case; no solitonic solution should exist at a mass greater 
than the `Chandrasekhar limit' for this configuration.

Of course it is far from clear that perturbation theory can capture any 
phenomenon - particularly one as interesting as singularity formation - at 
finite values of $\epsilon$. Nonetheless, in this subsection we will 
investigate the clues that we can glean from our perturbative analysis.

\begin{figure}[ht]
 \begin{center}
 \includegraphics[scale=0.6]{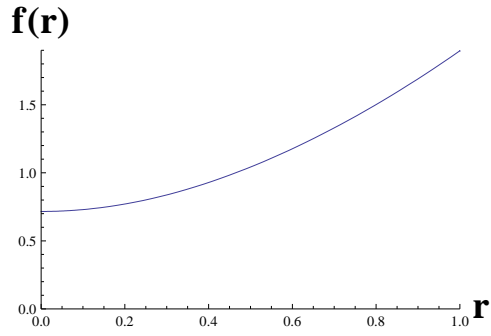}
\end{center}
\caption{\label{figC}$f(r)$ for $\epsilon=0.4$ and $e=4$}
\end{figure}


\begin{figure}[ht]
 \begin{center}
 \includegraphics[scale=0.6]{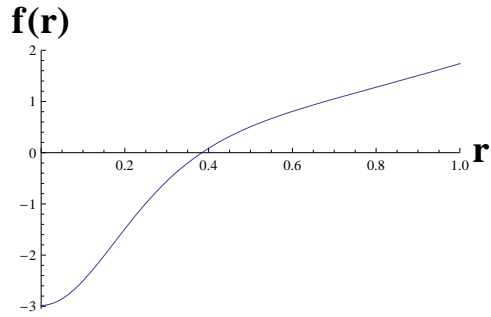}
\end{center}
\caption{\label{figD} $f(r)$ for $\epsilon=0.7$ and $e=4$}
\end{figure}

In Figs \ref{figC} and \ref{figD} we present a plot of $f(r)$ at $\epsilon=0.4$ and at 
$\epsilon=0.7$. We also present a plot of the scalar field $\phi(r)$ at the origin $r=0$ as a function of $\epsilon$. Note that $f(r)$ is everywhere positive at $\epsilon=0.4$ while it goes negative near the origin at $\epsilon=0.7$. Note also that the scalar field behaves quite smoothly at the origin at $\epsilon=0.4$ but shows a pronounced peak near the origin at $\epsilon=0.7$. 



\begin{figure}[ht]
\begin{center}
 \includegraphics[scale=0.6]{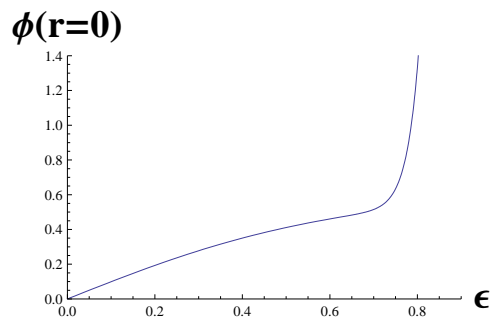}
\end{center}
\caption{\label{figE}$\phi(r=0)$ as a function of $\epsilon$ for $e=4$}
\end{figure}



\begin{figure}[ht]
\begin{center}
 \includegraphics[scale=0.6]{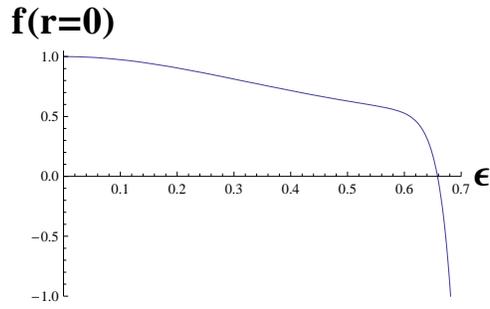}
\end{center}
\caption{\label{figE1}$f(r=0)$ as a function of $\epsilon$ for $e=4$}
\end{figure}

We take these results to indicate that the actual solution develops a 
singularity at some value of $\epsilon$ between $0.4$ and $0.7$. 
Let us use the vanishing of $f(r)$ at the horizon as an estimator of 
the onset of this singularity. In Fig. \ref{figE1} we plot $f(r=0)$ as 
a function 
of $\epsilon$. Note that this graph goes through the origin at 
$\epsilon \approx 0.65$.


\begin{figure}[ht]
\begin{center}
 \includegraphics[scale=0.6]{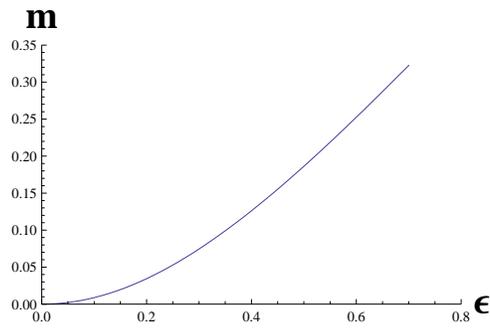}
\end{center}
\caption{\label{figA} Mass of soliton as a function of $\epsilon$ for $e=4$}
\end{figure}


\begin{figure}[ht]
 \begin{center}
 \includegraphics[scale=0.6]{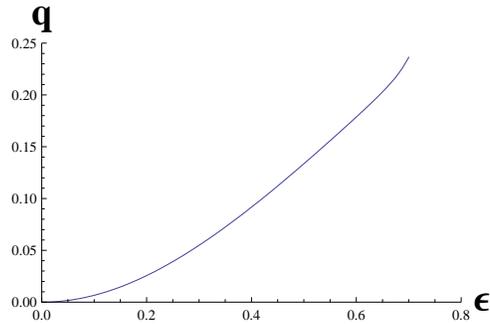}
\end{center}
\caption{\label{figB} Charge of soliton as function of $\epsilon$ for $e=4$}
\end{figure}

In Figs \ref{figA} and \ref{figB} we plot the mass and charge of the solution as a function of 
$\epsilon$. Note that these graphs do not show a pronounced peak near 
$\epsilon=0.65$. We take this result to indicate that the mass and charge 
of the soliton are  finite as we approach the singularity. All of the `conclusions' of this subsection are at best suggestive. A more 
serious analysis of the perturbative expansions described in this appendix 
could plausibly yield more solid indications as to the existence (or otherwise) of a singularity in the solution (rather than simply in the perturbative
expansion) at $\epsilon=0.65$. Numerical solutions to the differential 
equations would also likely yield valuable insights here. We leave all 
such discussions to future work.

\section{Excited State Solitons}\label{sec:excitedstate}
In this section, we will present explicit solution for the 
solitons obtained by populating the first and the second 
excited state. 

\subsection{The first excited state soliton}\label{app:ExcI}

The solution obtained by populating the first excited state takes the form
\begin{equation}
 \begin{split}
  f(r) &= \left(r^2+1\right)-\frac{\left(r^8+5 r^6+10 r^4+6\right)
   \epsilon ^2}{2 \left(r^2+1\right)^5} \\ &+\frac{\epsilon ^4}{34151040
   \left(r^2+1\right)^{11}}
   \left(e^2 \left(939978 r^{20}+10428693 r^{18}+52677075
   r^{16}+165426459 r^{14} \right. \right.\\& \left. \left.+334808793 r^{12}+425064222
   r^{10}+328198398 r^8+171552590 r^6+73128264 r^4\right. \right.\\& \left. \left.+33080568
   r^2+13962524\right)-12 \left(2568757 r^{20}+28256327
   r^{18}+141281635 r^{16}\right. \right.\\& \left. \left.+434422857 r^{14}+871830234
   r^{12}+1127373654 r^{10}+916291332 r^8+520240710
   r^6\right. \right.\\& \left. \left.+186067332 r^4+111081024
   r^2+26865882\right)\right)+O\left(\epsilon ^5\right)\\
 g(r) &= \frac{1}{r^2+1}+\frac{r^2 \left(r^6+5 r^4-2 r^2+12\right)
   \epsilon ^2}{2 \left(r^2+1\right)^7} \\ &-\frac{\epsilon ^4}{11383680
   \left(r^2+1\right)^{13}}
   \left(r^2 \left(e^2 \left(313326 r^{18}+3476231
   r^{16}+17559025 r^{14}\right. \right.\right.\\& \left. \left.\left.+44264825 r^{12}+61767915
   r^{10}+58915626 r^8+58567586 r^6+66442530 r^4\right. \right.\right.\\& \left. \left.\left.+35744940
   r^2+14941080\right)-4 \left(2568757 r^{18}+28256327
   r^{16}+141993115 r^{14}\right. \right.\right.\\& \left. \left.\left.+378069945 r^{12}+580690770
   r^{10}+599020422 r^8+676146702 r^6+535404870 r^4\right. \right.\right.\\& \left. \left.\left.+429618420
   r^2+121663080\right)\right)\right)+O\left(\epsilon ^5\right)\\
 A(r) &= \frac{6}{e}+\epsilon ^2 \left(-\frac{e \left(r^8+5 r^6+10
   r^4+r^2+5\right)}{16 \left(r^2+1\right)^5}+\frac{109
   e}{1232}-\frac{159}{77 e}\right)\\&+\epsilon ^4
   \left(\text{C1}+\frac{1}{273208320 r^2
   \left(r^2+1\right)^{11}} \left( e^3 \left(2266110 r^{16}-3679830
   r^{14}-57389178 r^{12}\right. \right.\right.\\& \left. \left.\left.-149388558 r^{10}-172806150
   r^8-103232855 r^6-29637025 r^4-2330873 r^2-\right. \right.\right.\\& \left. \left.\left.1065873\right)+3 e
   \left(355291 e^2-11254468\right) \left(r^2+1\right)^{11}+12 e
   \left(-4407480 r^{16}+6142752 r^{14}\right. \right.\right.\\& \left. \left.\left.+116586624
   r^{12}+322636512 r^{10}+364337655 r^8+252086890 r^6+50963352
   r^4\right. \right.\right.\\& \left. \left.\left.
  +7639614 r^2+2813617\right)\right) \right)+O\left(\epsilon ^5\right) \\
 \phi(r) &=\frac{\left(2-3 r^2\right) \epsilon }{2
   \left(r^2+1\right)^3}+\frac{\epsilon ^3}{9856
   \left(r^2+1\right)^9} \left(e^2 \left(1308
   r^{12}+6684 r^{10}+13380 r^8+12637 r^6\right.\right.\\& \left. \left.+5460 r^4+117
   r^2-710\right)-4 \left(7632 r^{12}+41946 r^{10}+90252
   r^8+87853 r^6\right.\right.\\& \left. \left.+41412 r^4-1053 r^2-5930\right)\right)+O\left(\epsilon ^4\right)\\
 \end{split}
\end{equation}
Here C1 is a constant that will be determined by the regularity and 
normalisability conditions on $\phi$ at one higher order.

\subsection{The second excited state soliton}\label{app:ExcII}

The solution obtained by populating the second excited state is
\begin{equation}
\begin{split}
 f(r) &= \left(r^2+1\right)-\frac{16 \left(r^{12}+7 r^{10}+21 r^8-11
   r^6+82 r^4-18 r^2+9\right) \epsilon ^2}{45
   \left(r^2+1\right)^7} \\ &+\frac{\epsilon ^4}{182614682250
   \left(r^2+1\right)^{15}} \left(11 e^2
   \left(145257733 r^{28}+2191163280 r^{26}\right.\right.\\& \left. \left.+15436521240
   r^{24}+71111047280 r^{22}+204078916860 r^{20}+363116451984
   r^{18}\right.\right.\\& \left. \left.+447815797000 r^{16}+559745609280 r^{14}+629383835652
   r^{12}+760236964032 r^{10}\right.\right.\\& \left. \left.+508886827176 r^8+129434743200
   r^6+41946040800 r^4-4941526368 r^2\right.\right.\\& \left. \left.+3675474942\right)-128
   \left(809908894 r^{28}+12148633410 r^{26}+85040433870
   r^{24}\right.\right.\\& \left. \left.+383937720530 r^{22}+1113624190770 r^{20}+2086530679662
   r^{18}+2767907054770 r^{16}\right.\right.\\& \left. \left.+3367219159590
   r^{14}+3524605653426 r^{12}+4551299991966
   r^{10}+1947299375658 r^8\right.\right.\\& \left. \left.+1248211079850 r^6+39350124900
   r^4+27778828956 r^2+14011941561\right)\right)+O\left(\epsilon ^5\right) \\
 g(r) &= \frac{1}{r^2+1}+\frac{16 r^2 \left(r^{10}+7 r^8-9 r^6+85 r^4-50
   r^2+30\right) \epsilon ^2}{45 \left(r^2+1\right)^9}\\& +\frac{r^2
   \epsilon ^4}{182614682250
   \left(r^2+1\right)^{17}} \left(256 \left(404954447 r^{26}+6074316705
   r^{24}+42610397025 r^{22}\right.\right.\\& \left. \left.+146943860245 r^{20}+332411328795
   r^{18}+736429945251 r^{16}+1692683167175 r^{14}\right.\right.\\& \left. \left.+2288373934275
   r^{12}+2999779439085 r^{10}+693256048485 r^8+761470058349
   r^6\right.\right.\\& \left. \left.+46619547975 r^4+124153839810 r^2+57083996970\right)-11
   e^2 \left(145257733 r^{26}\right.\right.\\& \left. \left.+2191163280 r^{24}+15436521240
   r^{22}+48745664240 r^{20}+98672175420 r^{18}\right.\right.\\& \left. \left.+226204362384
   r^{16}+539605666600 r^{14}+922637944800 r^{12}+647631553140
   r^{10}\right.\right.\\& \left. \left.+673336503840 r^8-167785201584 r^6+151437686400
   r^4-3202158960 r^2\right.\right.\\& \left. \left.+21249708480\right)\right)+O\left(\epsilon ^5\right)\\
 A(r) &=  \frac{8}{e}+\epsilon ^2 \left(\frac{4741 e^2-228352}{90090
   e}-\frac{e}{30 r^2}+\frac{e \left(8 \left(5 r^4-4
   r^2+4\right) r^4+1\right)}{30 r^2
   \left(r^2+1\right)^7}\right)\\ &+\epsilon ^4
   \left(\text{C1}+\frac{1}{973944972000 r^2
   \left(r^2+1\right)^{15}}\left(11 e^3 \left(776575800
   r^{24}-4643959320 r^{22}\right.\right.\right.\\& \left. \left.\left.-49995986040 r^{20}-153300661800
   r^{18}-226524451725 r^{16}-196951063595 r^{14}\right.\right.\right.\\& \left. \left.\left.-81174009917
   r^{12}-35347997139 r^{10}-47175005605 r^8-22830663835
   r^6\right.\right.\right.\\& \left. \left.\left.-9008399805 r^4-109267667 r^2-82336064\right)
   \right. \right.\\& \left. \left.+64 e
   \left(14151511 e^2-890269174\right)
   \left(r^2+1\right)^{15}-128 e \left(3214411200
   r^{24}\right.\right.\right.\\& \left. \left.\left.-18271242990 r^{22}-210141141210 r^{20}-667533287850
   r^{18}-983744254350 r^{16}\right.\right.\right.\\& \left. \left.\left.-972471806735 r^{14}\right.\right.\right.\\& \left. \left.\left.-123178436381
   r^{12}-380041323207 r^{10}-58304634115 r^8-160254872155
   r^6\right.\right.\right.\\& \left. \left.\left.-31564145265 r^4-979195571
   r^2-445134587\right) \right)\right)+O\left(\epsilon ^5\right)\\
\end{split}
\end{equation}
Here again C1 is determined at one higher order.
Finally the scalar field in this case is given by 
\begin{equation}
\begin{split}
 \phi(r) &= \frac{\left(2 r^4-4 r^2+1\right) \epsilon
   }{\left(r^2+1\right)^4}-\frac{2 \epsilon ^3}{675675
   \left(r^2+1\right)^{12}} \left(11 e^2
   \left(4310 r^{18}+27530 r^{16}+69735 r^{14}\right.\right.\\& \left. \left.+80920
   r^{12}+29848 r^{10}-54222 r^8-68110 r^6-26098 r^4-6504
   r^2+1149\right) \right. \\ &\left. -2 \left(1141760 r^{18}+7624220
   r^{16}+20257200 r^{14}+24337600 r^{12}+10397296
   r^{10}\right.\right.\\& \left. \left.-17335794 r^8-19274920 r^6-9051766 r^4-1800468
   r^2+416883\right)\right)+O\left(\epsilon ^4\right)\\ 
\end{split}
\end{equation}

\section{The First Excited Hairy Black Hole}\label{app:ExcHairy1}

In this Appendix we present the results of our construction of the first 
excited hairy black hole metric. We have written a Mathematica programme
that allows us to generate the corresponding solution for the $n^{th}$ 
excited hairy black hole at any given value of $n$. We use the same 
conventions as in appendix\ref{app:PertExp}. 
\subsection{Near Field Expansion}\label{app:nearFHairy}
\begin{equation}\label{uttorfine1}
 \begin{split}
  f^{in}_{(0,0)}(y)=&\frac{\left(y^2-1\right) \left(e^2 y^2-24\right)}{e^2 y^4} \\
f^{in}_{(0,2)}(y) =& \frac{\left(y^2-1\right) \left(e^4 \left(y^4+y^2\right)+304
   e^2+1536\right)}{e^4 y^4}\\
f^{in}_{(0,4)}(y)=& \frac{8 \left(y^2-1\right) \left(9 \mathcal{C}_1 e^5+361 e^6
   y^2+57 e^4 \left(84 y^2+19\right)+576 e^2 \left(26
   y^2+19\right)+27648\right)}{9 e^6 y^4}\\
f^{in}_{(2,0)}(y) =& -\frac{8 \left(y^2-1\right) \left(e^2 \left(154
   y^2-23\right)-212\right)}{231 e^2 y^4} \\
 \end{split}
\end{equation}

\begin{equation}\label{uttorgine1}
 \begin{split}
  g^{in}_{(0,0)}(y)=&\frac{e^2 y^4}{\left(y^2-1\right) \left(e^2 y^2-24\right)} \\
g^{in}_{(0,2)}(y) =& -\frac{y^4 \left(e^4 \left(y^4+y^2\right)-96 e^2+3456\right)}{\left(y^2-1\right) \left(e^2 y^2-24\right)^2}\\
g^{in}_{(0,4)}(y)=&\frac{y^4}{e^2
   \left(y^2-1\right) \left(e^2 y^2-24\right)^3} \left. \Big(8 \mathcal{C}_1 e^7 y^2-192 \mathcal{C}_1 e^5
   +e^8 y^4 \left(y^2+1\right)^2-96 e^6 \left(2
   y^4+y^2\right)\right. \\ & \left. +6912 e^4 \left(y^4+1\right)+124416 e^2 \left(y^2-4\right)
   +8957952\right. \Big)\\
B^{in}_{(2,0)}(y) =& \frac{8 e^2 \left(712-23 e^2\right) y^4}{231 \left(y^2-1\right) \left(e^2 y^2-24\right)^2}\\
 \end{split}
\end{equation}

\begin{equation}\label{uttorAine1}
 \begin{split}
 A^{in}_{(0,0)}(y)=&\frac{6 \left(y^2-1\right)}{e y^2} \\
A^{in}_{(0,2)}(y) =& \frac{12 \left(e^2-36\right)  \left(y^2-1\right)}{e^3 y^2}\\
A^{in}_{(0,4)}(y)=&  \mathcal{C}_1 \left(1-\frac{1}{y^2}\right)\\
A^{in}_{(2,0)}(y) =&-\frac{\left(23 e^2+212\right) \left(y^2-1\right)}{231 e y^2}\\
 \end{split}
\end{equation}

\begin{equation}\label{uttorphine1}
 \begin{split}
 \phi^{in}_{(1,0)}(y)=& -\frac{2}{3}\\
\phi^{in}_{(1,2)}(y) =& \frac{1}{2 e^2}\Bigg(-288 \text{Li}_2\left(\frac{e^2 y^2-24}{e^2-24}\right)
    -12 \left(e^2-72\right) \log (R)+e^2 \left(6
   y^2-5\right) \\ &-6 \left(48 \log \left(-\frac{e^2 \left(y^2-1\right)}{e^2-24}\right)-24 \log \left(e^2
   y^2-24\right)+e^2-72\right) \log \left(e^2 y^2-24\right) \\ &-144 \log ^2\left(\frac{1}{24-e^2}\right)+12
   \left(e^2-72\right) \log (e)-48 \pi ^2+456 \Bigg)\\
\phi^{in}_{(3,0)}(y)=& \frac{5 \left(71 e^2-2372\right)}{16632}.
 \end{split}
\end{equation}

\subsection{Far Field Solutions}\label{app:outsideHairy}

\begin{equation}\label{uttorf1}
 \begin{split}
  f^{out}_{(0,0)}(r)=& r^2+1\\
f^{out}_{(0,2)}(r) =& -\frac{e^2+24}{e^2 r^2}\\
f^{out}_{(0,4)}(r)=&\frac{1}{6 e^2 r^4}\left(144-6 e^2 \left(\frac{96 
          \left(e^2-36\right)}{e^4}+1\right) r^2 \right)\\
f^{out}_{(2,0)}(r) =&-\frac{2 \left(r^8+5 r^6+10 r^4+6\right)}{9 \left(r^2+1\right)^5} \\
f^{out}_{(2,2)}(r) =&\frac{1}{1386 e^2 r^2
   \left(r^2+1\right)^6} \Big(-2772 r^2 \left(r^2+1\right) 
   \left(2 \left(3 e^2-88\right) r^{12}+12 \left(3 e^2-88\right)
   r^{10} \right. \\ & \left. +\left(89 e^2-2568\right) r^8
   +5 \left(23 e^2-632\right) r^6+80 \left(e^2-24\right) r^4+12 \left(3
   e^2-88\right) r^2 \right. \\ & \left.  +256\right) (\log (1-i r)
   +\log (1+i r)-2 \log (r))+e^2 \left(16632 r^{14}+110675
   r^{12} \right. \\ & \left.  +309542 r^{10}+467241 r^8+416782 r^6
    +179030 r^4+35422 r^2+2952\right) \\ & -24 \left(20328 r^{14}+138869
   r^{12}+394622 r^{10}+595703 r^8+518662 r^6+240194 r^4 \right. \\ & \left. +22250 r^2-424\right) \Big)\\
 \end{split}
\end{equation}

\begin{equation}\label{uttorg1}
 \begin{split}
  g^{out}_{(0,0)}(r)=&\frac{1}{r^2+1}\\
g^{out}_{(0,2)}(r) =& \frac{e^2+24}{e^2 r^2 \left(r^2+1\right)^2}\\
g^{out}_{(0,4)}(r)=&\frac{e^4 \left(r^4+r^2+1\right)+24 e^2 \left(4 r^4+3 r^2+1\right)-576 \left(6 r^4+6 r^2-1\right)}{e^4 r^4
   \left(r^2+1\right)^3}\\
g^{out}_{(2,0)}(r) =&\frac{2 r^2 \left(r^6+5 r^4-2 r^2+12\right)}{9 \left(r^2+1\right)^7} \\
g^{out}_{(2,2)}(r) =& \frac{1}{1386 e^2
   \left(r^3+r\right)^2}\Big( -\frac{77}{\left(r^2+1\right)^6} \left(36 \left(e^2-72\right) 
   \left(r^8+6 r^6+3 r^4+10 r^2 \right. \right. \\ & \left. \left. +12\right) r^4 (\log (1-i r)+\log
   (1+i r)-2 \log (r))-e^2 \left(32 r^{10}+182 r^8+856 r^6
   \right. \right. \\ & \left. \left. +393 r^4+210 r^2+19\right)+24 \left(108 r^{10}+586
   r^8+1292 r^6+1471 r^4+222 r^2+69\right)\right) \\ &-2567 e^2+161688 \Big)\\
 \end{split}
\end{equation}

\begin{equation}\label{uttorA1}
 \begin{split}
  A^{out}_{(0,0)}(r)=&\frac{6}{e}\\
A^{out}_{(0,2)}(r) =& \frac{e^2 \left(12-\frac{6}{r^2}\right)-432}{e^3}\\
A^{out}_{(0,4)}(r)=& \mathcal{C}_1-\frac{12 \left(e^2-36\right)}{e^3 r^2}\\
A^{out}_{(2,0)}(r) =&\frac{1}{2772} \Bigg( -\frac{77 e \left(r^8+5 r^6
       +10 r^4+r^2+5\right)}{\left(r^2+1\right)^5}+109 e-\frac{2544}{e} \Bigg)\\
A^{out}_{(2,2)}(r) =& \mathcal{C}_2 + \frac{1}{11088 e r^2 \left(r^2+1\right)^6} 
   \Big( e^2 \left(\left(3 \left(5681 r^8 
   +30852 r^6+67813 r^4+77507 r^2 \right. \right. \right. \\ & \left. \left. \left. +39246\right)
   r^2+29647\right) r^2+1104\right)+2772 r^2 \left(r^2+1\right) \left(\left(5 e^2-104\right) r^{10}
   \right. \\ & \left.+8
   \left(3 e^2-56\right) r^8+5 \left(9 e^2-136\right) r^6+40 \left(e^2   
   -8\right) r^4
 \right. \\ & \left. +8 \left(3 e^2-56\right)
   r^2+256\right) \left(2 \log (r)-\log \left(r^2+1\right)\right)
   \\ &  -24 \left(19673 r^{12}+103716 r^{10}+217325
   r^8+230143 r^6\right. \\ & \left.+123462 r^4+10777 r^2-424\right) \Big)\\
 \end{split}
\end{equation}

\begin{equation}\label{uttorph1}
 \begin{split}
 \phi^{out}_{(1,0)}(r)=& \frac{3 r^2-2}{3 \left(r^2+1\right)^3}\\
\phi^{out}_{(1,2)}(r) =& \frac{1}{2 e^2
   \left(r^2+1\right)^4} \left. \Big(3 \left(72-5 e^2\right) r^2
   +6 \left(e^2-72\right) \left(3 r^4+r^2-2\right) \log (r)
  \right. \\ & \left. -3 \left(e^2-72\right) \left(3 r^4+r^2-2\right) \log \left(r^2+1\right)
  -5 e^2+456 \right. \Big)\\
\phi^{out}_{(3,0)}(r)=& -\frac{1}{33264 \left(r^2+1\right)^9} \Big(e^2 \left(r^2+1\right) 
   \left(1308 r^{10} \right. \\  & \left. +5376 r^8+8004 r^6+4633 r^4 +827 r^2-710\right) \\ & +4212 r^2-4
   \left(7632 r^8+41946 r^6+90252 r^4 \right. \\  & \left. +87853 r^2+41412\right) r^4+23720 \Big)
 \end{split}
\end{equation}

In the above formulae $\mathcal{C}_1$ and $\mathcal{C}_2$ are constants which will only be determined 
by regularity and normalisability of the scalar field at one higher order.

\section{Thermodynamics in the Canonical Ensemble}\label{app:Can}
In the following subsections, we will describe the canonical phase diagrams
that result from a competition between small RNAdS black holes, small hairy black holes and small solitons. Everywhere in this section we completely ignore large black holes
and large hairy black holes. 

In the microcanonical ensemble this was logically justified; 
large black holes never have small mass and charge. However large black holes can (and do)
have temperatures (and or chemical potentials) comparable to their small counterparts. 
Consequently the phase diagrams we will draw in this appendix do not, in general, 
represent the true thermo dynamical equilibrium of our system at finite temperature 
and chemical potential. The phase diagrams of this appendix should be regarded as 
formal; their purpose is to help us better understand the formal interrelationship
 between the phases constructed in this paper, ignoring all other phases that might exist
in the system.

In this section we will study the interelationship between small black hole, soliton
and excited black hole phases at fixed charge and temperature. We find it convenient 
to work with the rescaled inverse temperature variable
$$\beta = \frac{1}{4\pi T}.$$
In this section we will assume that $\beta$ and $q$ are small. We 
also assume that $q$ and $\beta^2$ are of the same order, and present 
all formulae only to leading order in $q$ and $\beta^2$.

\subsection{ RNAdS Black Hole}

At any fixed charge, the temperature of a small RNAdS black hole is given, 
as a function of its chemical potential $\mu =\frac{q}{R^2}$ by 
\begin{equation}\label{beta}
\frac{\sqrt{q}}{\beta}= 2\sqrt{\mu}(1-\frac{2}{3 } \mu^2)
\end{equation}
\begin{figure}
\begin{center}
 \includegraphics{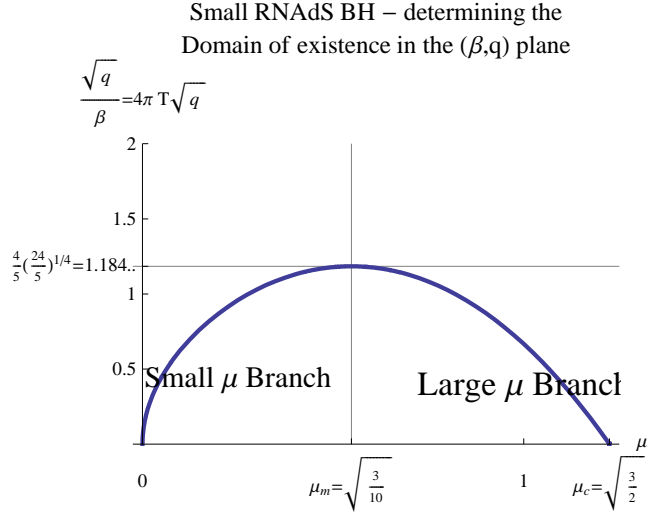}
\end{center}
\caption{\label{figp1} Canonical Ensemble : Small  RNAdS BH -  For any given
charge and temperature there are two possible small black holes with two 
different chemical potentials (Large $\mu$ branch has a lower free-energy).
And for a given temperature,  there is a maximum possible charge 
which is  attained by the black hole with $\mu=\mu_m=\sqrt{3/10}$. }
\end{figure}
In Fig.\ref{figp1}, we present a plot of $\frac{\sqrt{q}}{\beta}$ versus 
$\mu$. Note that $$\mu^2 \leq \frac{3}{2}$$
(the constraint follows directly from  the requirement of positivity 
of the temperature in \eqref{beta}).
Note also that 
\begin{equation}\label{betamin}
\frac{\sqrt{q}}{\beta} \leq \frac{4}{5}\left[ \frac{24}{5}\right]^{1/4}
\end{equation}
This inequality is saturated at at $\mu= \mu_m\equiv \sqrt{\frac{3}{10}}$.

Note that there exist two small black holes (with 
different values of $\mu$) for any given $\beta$ that obeys \eqref{betamin} 
The free energy is given as a function on $q$ and $\mu$ by 
\begin{equation} \label{fen}
F_{sbh}= q \frac{\pi}{24}\left( 10 \mu + \frac{3}{\mu} \right)
\end{equation}
\begin{figure}
 \begin{center}
 \includegraphics{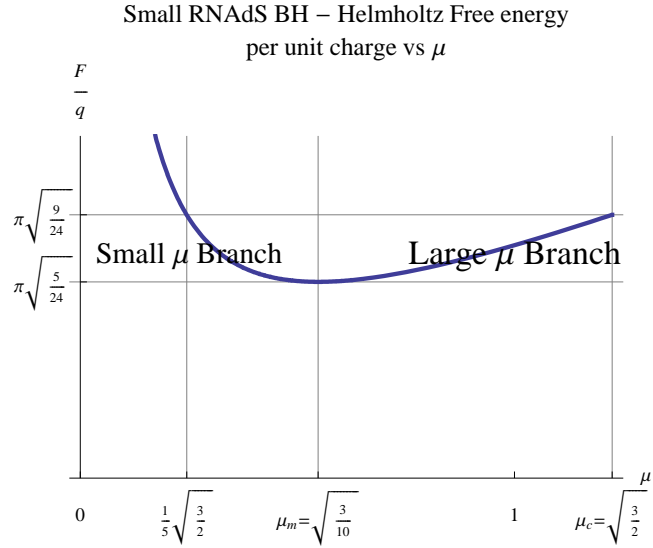}
\end{center}
\caption{\label{figp2} Canonical ensemble : small RNAdS BH - Free energy of the small BHs are
positive and for the large $\mu$ branch, the free-energy varies over a bounded domain.  }
\end{figure}
\begin{figure}
\begin{center}
 \includegraphics{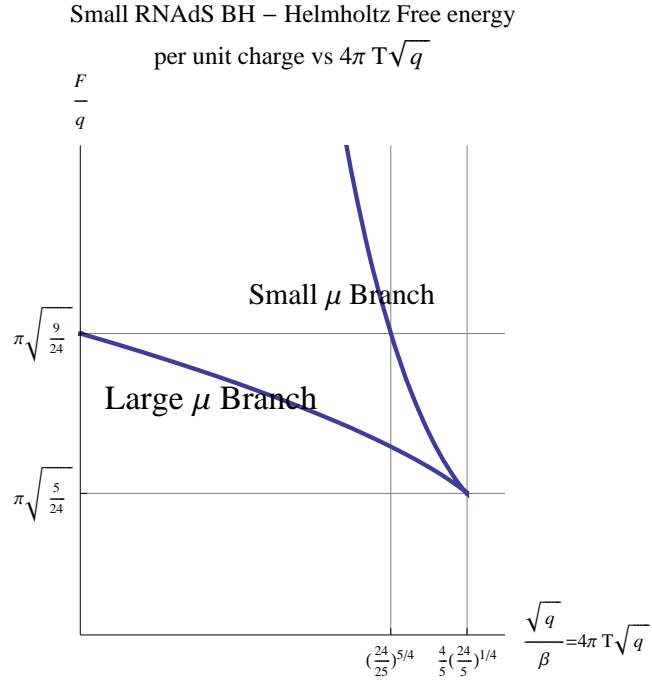}
\end{center}
\caption{\label{figp3} Canonical ensemble : small RNAdS BH - The large $\mu$ branch
always has a lower free-energy. }
\end{figure}
In Fig. \ref{figp2} we present a plot of $\frac{F_{sbh}}{q}$ versus $\mu$.
Note that 
 $$\frac{F_{sbh}}{q} \geq \pi \sqrt{\frac{5}{24}} $$ and the minimum value 
occurs at $\mu=\mu_m\equiv\sqrt{\frac{3}{10}}$ (this is the same value of $\mu$ 
at which the temperature curve has a maximum). As is visually apparent from these 
graphs, the RNAdS black hole with the larger value of $\mu$ has lower free energy 
at any fixed $\beta$ and $q$(see fig.\ref{figp3}). In this appendix we will refer to this solution as the small RNAdS black hole. We will completely ignore the 
free energetically subdominant RNAdS black hole in the rest of this 
Appendix. 

Let us briefly summarise. RNAdS black holes exist whenever 
the inequality \eqref{betamin} is obeyed. Their free energy is 
given as a function of $\beta$ and $q$ by \eqref{fen} and 
\eqref{beta}, where we are instructed always to choose the larger 
of the two roots when inverting \eqref{beta}. The chemical potential 
of these black holes obey 
$$\mu_m\equiv\sqrt{\frac{3}{10}} \geq \mu \geq \sqrt{\frac{3}{2}}\equiv \mu_c.$$

\subsection{ Soliton}

Solitonic solutions exist at all values of $\beta$ and $q$. 
At leading order the Free energy and chemical potential of the 
soliton are given by 
\begin{equation}\label{fancp} \begin{split}
\frac{F_{sol}}{q}&=\frac{2\pi}{e}\\
\mu&=\frac{4}{e}\\ 
\end{split}
\end{equation}

Note that $\frac{2\pi }{e} \leq \sqrt{\frac{5}{24}}$ whenever 
$$e^2 \geq \frac{32}{3} \times \frac{9}{5}=e_c^2 \times \frac{9}{5}
\equiv e_1^2.$$ It follows that the soliton free energetically dominates that RNAdS 
black hole whenever this inequality is obeyed. At smaller values of 
$e$, on the other hand, the RNAdS black hole free energetically 
dominates the soliton at large enough temperatures (but temperatures 
that are small enough to be allowed by \eqref{betamin}, i.e. 
whenever
\begin{equation} \label{bhdom} \begin{split}
&\frac{3 }{2\sqrt{\mu_m}(3-2 \mu_m^2)} \leq \frac{\beta}{\sqrt{q}} \leq  \frac{3 }{2\sqrt{\mu_*}(3-2 \mu_*^2)}\\
\mu_*&\equiv \frac{4}{e} \times  \frac{3}{5} \left( 1+ \sqrt{1-\frac{5}{9}
\times \frac{3 e^2}{32} }\right) = \frac{4}{e} \times  \frac{3}{5} \left( 1+ \sqrt{1-\frac{5}{9}
\times \frac{e^2}{e_c^2} }\right) \\
&\mu_m= \sqrt{\frac{3}{10}}
\end{split}
\end{equation}
We will return to a more detailed comparison of phases below.

\subsection{ Hairy Black hole}

Hairy black holes exist if and only if
\begin{equation} \label{hairyexist}
\frac{\beta}{\sqrt{q}}\leq \sqrt{ \frac{e^5}{16 (e^2-\frac{32}{3})^2} }
\end{equation}
In this regime their chemical potential, mass and free energy are given by 
\begin{equation}\label{canhs}
 \begin{split}
  m =& \left(\frac{4 \left(3 e^2-32\right)^3}{27 e^6} \right)\beta^2 +\frac{16}{3 e} q + {\cal O}(m^2,mq,q^2)\\
\mu =& \frac{4}{e} + \left(\frac{8 \left(32-3 e^2\right)^2 \left(e^2-32\right)}{21 e^7}\right)\beta^2 +\left(\frac{9}{7}-\frac{64}{7 e^2}\right)q  + {\cal O}(m^2,mq,q^2)\\
F(\beta,q) =& \frac{3 \pi}{8}\bigg[\left(\frac{16 q}{3e} +\frac{4 \left(3 e^2-32\right)^3 \beta ^2}{81 e^6}\right) +\frac{16 \left(32-3 e^2\right)^4 \left(21 e^4-384 e^2+5120\right)}{1701
   e^{12}}\beta^4\\
&+\frac{32 \left(32-3 e^2\right)^2 \left(e^2-32\right)}{63 e^7}\beta^2 q +\frac{2 \left(9 e^2-64\right) }{21 e^2}q^2 + {\cal O}\left(\beta^6,\beta^4 q,\beta^2 q^2, q^3\right)\bigg]
 \end{split}
\end{equation}
(where we have listed perturbative corrections, but will only use leading order 
results in what follows). Note that the difference between the free energy of a 
hairy black hole and the soliton is a positive number times $\beta^2$, so that 
hairy black holes are free energetically always subdominant compared to the soliton. 

It is also interesting to perform a comparison between RNAdS and hairy black holes. 
First notice that the existence ranges \eqref{hairyexist} and \eqref{betamin} overlap
only when 
$$e^2  \leq \frac{32}{3} \times 5\equiv e_2^2$$
Moreover hairy black holes only exist for $e^2\geq \frac{32}{3}\equiv e_c^2$. 
Within this range of $e$ hairy and RNAdS black holes in an overlapping region
in the $\beta$, $q$ plane. It turns out that RNAdS black holes always dominate 
over hairy black holes in these overlapping regions. 

\subsection{ Plots of Phase Existence and Dominance}

As we have explained above, the phase diagram of our system depends qualitatively 
on the value of $e^2$. In particular, there are three special values of $e^2$ 
$$ e_c^2=\frac{32}{3}, ~~~e_1^2= e_c^2 \frac{9}{5}, ~~~e_2^2= 5 e_c^2$$
Recall that RNAdS black holes are always stable - and no hairy black holes exist - 
for $e^2\leq e_c^2$. In this regime the only phases of the system are the RNAdS 
black hole and the soliton. It turns out that the RNAdS black hole is 
free energetically dominant whenever it exists. The phase diagram is depicted
in Fig.\ref{figp4} below. 

\begin{figure}
\begin{center}
 \includegraphics{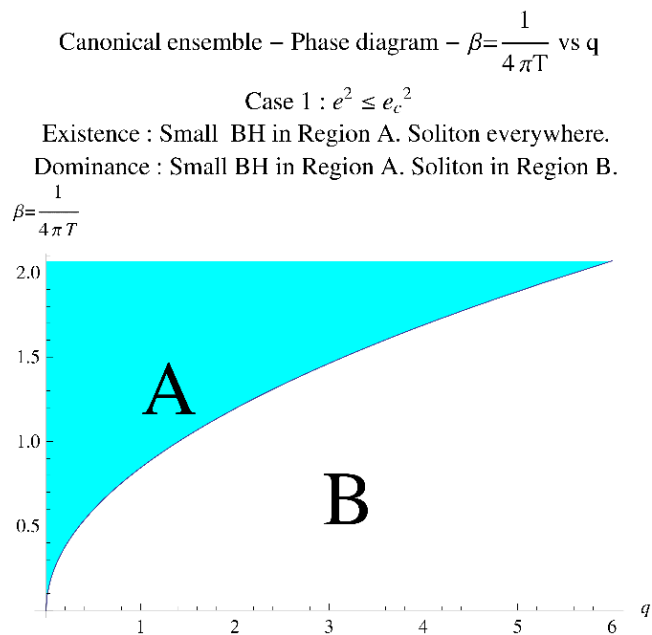}
\end{center}
\caption{\label{figp4} Canonical ensemble : Case 1 : $e^2\leq e_c^2$ . No Hairy black hole. Soliton and small RNAdS dominate two different regions.}
\end{figure}

RNAdS black holes exist only above the line drawn in Fig \ref{figp4}, and give the dominant
phase when they exist. The soliton dominates the phase diagram elsewhere.
 
\begin{figure}
\begin{center}
 \includegraphics{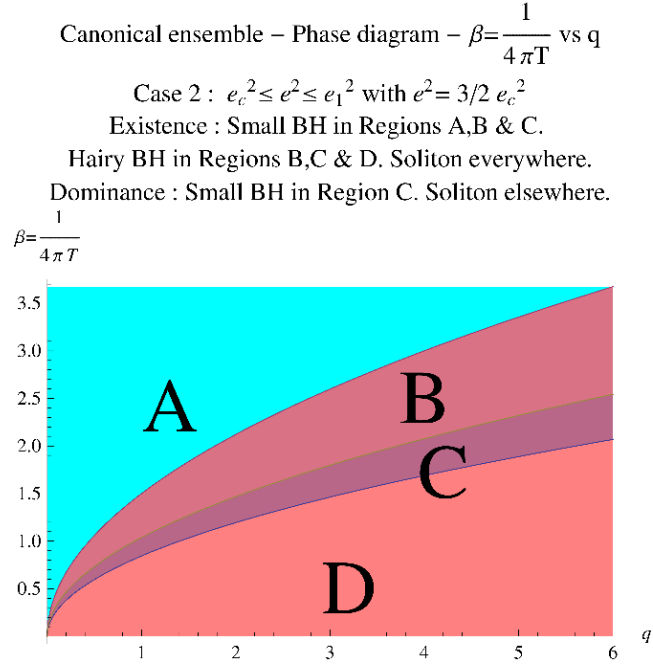}
\end{center}
\caption{\label{figp5} Canonical ensemble : Case 2 : $e_c^2\leq e^2 \leq e_1^2$. Hairy black holes
appear and over regions B and C overlap with the domain of existence of the plain black
holes. But, soliton is dominant over most of the diagram except for the region marked C
where small BHS are preferred. }
\end{figure}

In the regime $e_c^2 \leq e^2 \leq e_1^2$ hairy black holes exist as a phase 
(below the topmost line in Fig. \ref{figp5}) but are never dominant. In this regime the 
RNAdS black  holes exist above the bottom most line but free energetically dominate 
the soliton only in the region between the bottom most line and the intermediate line.
The soliton is free energetically dominant elsewhere. 
The RNAdS black hole is free energetically dominant over the hairy black hole 
over their common region of existence.

\begin{figure}
\begin{center}
 \includegraphics{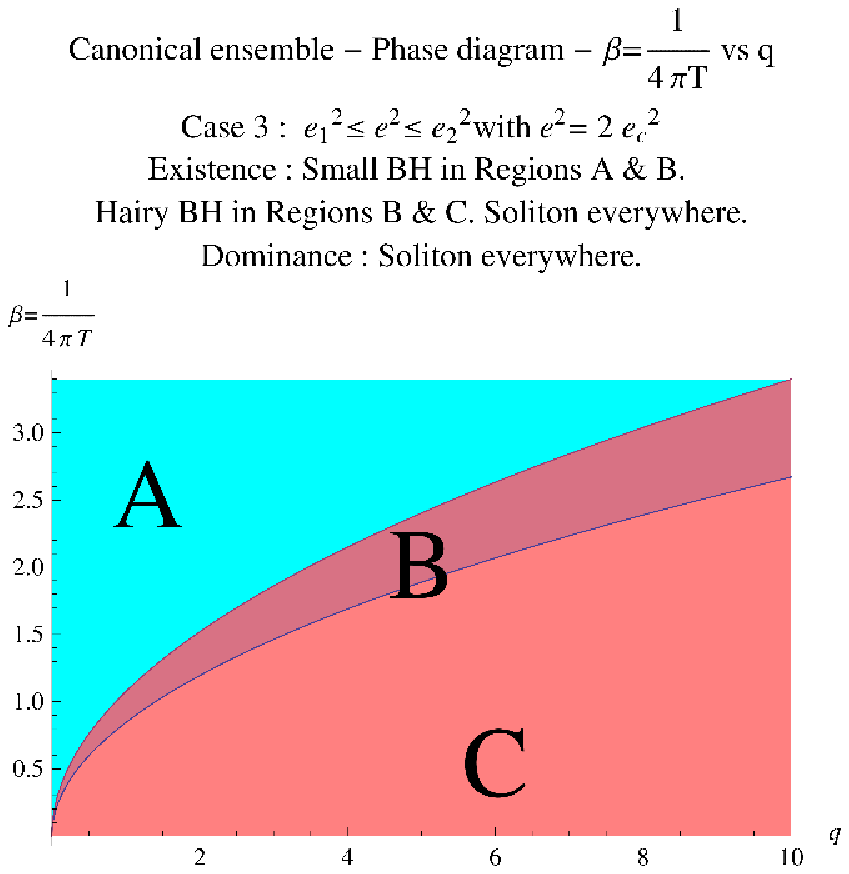}
\end{center}
\caption{\label{figp6}Canonical ensemble : Case 3 : $e_1^2\leq e^2 \leq e_2^2$. Soliton is
preferred everywhere over the RNAdS or the hairy black hole.}
\end{figure}

In the regime $e_1^2 \leq e^2 \leq e_2^2$ the phase diagram (shown in Fig.\ref{figp6}) is 
very similar to that in Fig.\ref{figp5}, except that there is no intermediate region; the soliton is the thermodynamically dominant solution everywhere. Note that in this 
regime the RNAdS and hairy black hole solutions continue to exist as phases; the 
RNAdS black hole is free energetically dominant over the hairy black hole over 
this region of overlap (i.e., region B in \ref{figp6}).

\begin{figure}
\begin{center}
 \includegraphics{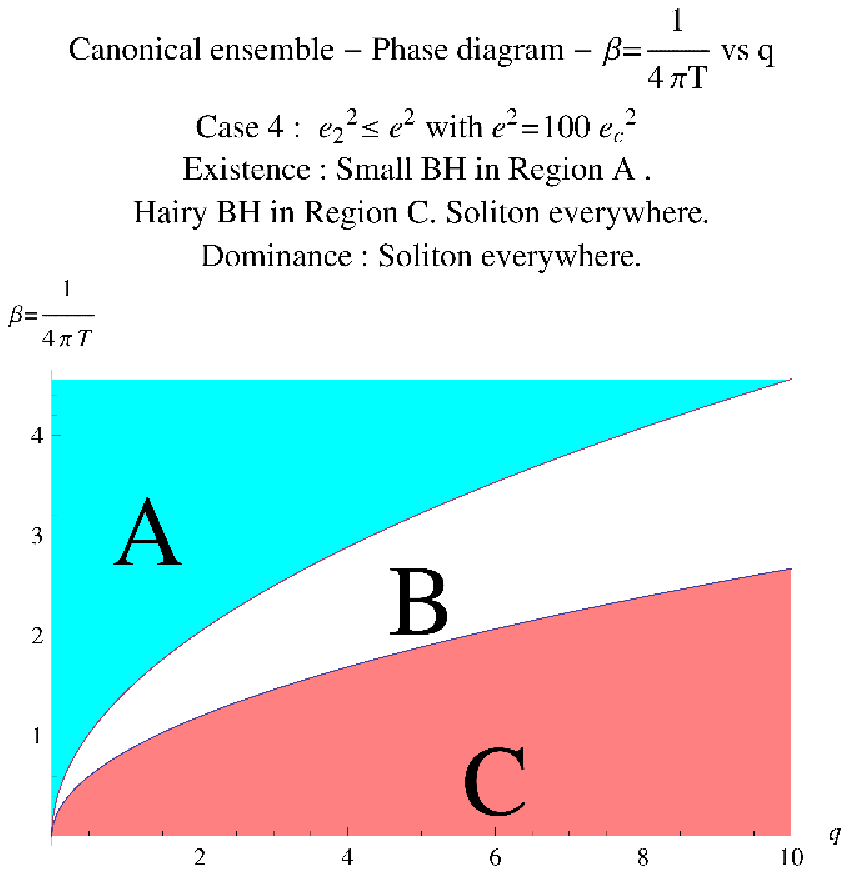}
\end{center}
\caption{\label{figp7}Canonical ensemble : Case 4 : $e_2^2\leq e^2$. The region of
overlap between hairy black holes and RNAdS black holes disappears. Soliton is
still preferred everywhere.}
\end{figure}

We turn, finally to the range $e^2 \geq e_2^2$. Note that in this range 
$\frac{4}{e} \leq \sqrt{\frac{3}{10}}=\mu_m$. Recall that the chemical 
potential of a hairy black hole is equal to $\frac{4}{e}$ at leading order. 
It follows that the RNAdS black hole component of a hairy black hole, 
in this regime, has $\mu \leq \mu_m$. In other words, in this regime, the RNAdS 
black hole that lies in the centre of a hairy black hole is of negative specific 
heat. In this regime, also, there is no overlap in the existence regimes of 
RNAdS black holes \eqref{betamin} and hairy black holes \eqref{hairyexist}. 
At any given value of $\beta$ and $q$ we have at most two phases (soliton and 
black hole or soliton and hairy black hole) and the soliton is always the free 
energetically dominant phase. The phase diagram is displayed in Fig. \ref{figp7}

\section{Thermodynamics in the Grand Canonical Ensemble}\label{app:GC}

In the following subsections, we will describe the  grand canonical phase diagrams that result from a competition between small RNAdS black holes, small hairy black holes and small solitons. Everywhere in this section we completely ignore large black holes
and large hairy black holes.  Like our discussion on canonical ensemble
given in the previous appendix, this is not entirely justified if we are 
interested in the actual phase diagram of the system. However, the formal
diagrams which we present in this appendix are still helpful in 
contrasting the various phases constructed in this paper. 

In this section we will study the thermodynamics of our system as a function of 
$$\beta = \frac{1}{4\pi T},~~\delta\mu = \mu - \frac{4}{e}$$

\subsection{ RNAdS Black Holes}

As we have seen above, small RNAdS black holes exist only for 
$$ \mu \leq \sqrt{\frac{3}{2}}.$$
Moreover these black holes also satisfy the inequality 
$$ \beta^2 \leq \frac{1}{32 (1-\frac{2}{3} \mu^2)}$$
This last inequality is automatically obeyed for parametrically small 
values of $\beta$, of prime interest to us in this paper, and so will 
play no important role in the analysis below. Whenever these inequalities 
are satisfied, we have a unique small black hole. 

The various important thermodynamical quantities of small RNAdS black holes, in the grand 
canonical ensemble, are given by 
\begin{equation}\label{masbh}
\begin{split}
 m =& \frac{4 \left(32-3 e^2\right)^2 \left(3 e^2+32\right) \beta ^2}{27 e^6}+ \left[\frac{16 \left(3 e^2-32\right)^3 \left(5 e^2+32\right)}{27 e^8}\right]\beta^4 \\
&-\left[\frac{64 \left(e^2+32\right) \left(3 e^2-32\right)  }{9 e^5} \right]\beta ^2 \delta \mu + {\cal O}\left(\beta^5,\beta^3\delta\mu,\delta\mu^2\beta\right)\\ 
q =& \frac{16 \left(32-3 e^2\right)^2 \beta ^2}{9 e^5} +\left[ \frac{256 \left(3 e^2-32\right)^3}{27 e^7}\right] \beta ^4\\
 &+\left[\frac{4 \left(9 \left(e^2-64\right) e^2+5120\right)}{9 e^4}\right] \beta ^2 \delta \mu + {\cal O}\left(\beta^5,\beta^3\delta\mu,\delta\mu^2\beta\right)\\ 
\end{split}
\end{equation}
\begin{equation}\label{ansa}
\begin{split}
G(\beta,\delta\mu) =&\frac{3\pi}{8}\bigg[\frac{4 \left(3 e^2-32\right)^3 \beta ^2}{81 e^6} + \left(\frac{16 \left(32-3 e^2\right)^4}{81 e^8}\right)\beta^4 -\left(\frac{64 \left(32-3 e^2\right)^2}{27 e^5}\right)\beta^2\delta\mu\\
&+ {\cal O}\left(\beta^5,\beta^3\delta\mu,\delta\mu^2\beta\right)\bigg]
\end{split}
\end{equation}

\subsection{ Soliton}

The soliton exists for all temperatures but for $\mu \geq \frac{4}{e}$. Its 
thermodynamical quantities are given by 
\begin{equation}\label{soleq}
 \begin{split}
  m = &\left.\frac{112 e  }{3 \left(9 e^2-64\right)}\right.\delta \mu+\left.\frac{e^2 \left(2364219
   e^4-47285088 e^2+244052992\right) }{1485 \left(9
   e^2-64\right)^3}\right.\delta \mu ^2+{\cal O}\left(\delta \mu ^3\right)\\
q =& \left.\frac{7 e^2 }{9 e^2-64}\right.\delta \mu +\left.\frac{e^3 \left(1802889 e^4-39301728
   e^2+215667712\right) }{7920 \left(9
   e^2-64\right)^3}\right.\delta \mu ^2+{\cal O}\left(\delta \mu ^3\right)\\
G(\mu,T)=& \frac{3\pi}{8}\bigg[\left(\frac{14 e^2 }{192-27 e^2}\right)\delta \mu ^2+\left(\frac{\left(-1802889 e^7+39301728
   e^5-215667712 e^3\right)}{17820 \left(9
   e^2-64\right)^3}\right) \delta \mu ^3\\
&+{\cal O}\left(\delta \mu ^4\right)\bigg]
 \end{split}
\end{equation}

It is easy to verify that the solitonic solution always has lower grand free energy 
(it is negative at leading order) than the RNAdS black hole 
(the free energy is positive at leading order ) within the validity of perturbation 
theory. Consequently the system undergoes a first order phase transition from 
the RNAdS black hole to the solitonic phase as  $\mu$ is raised above $\frac{4}{e}$

\subsection{Hairy Black hole}

Hairy black holes exist whenever
\begin{equation}
 \frac{\delta\mu}{\beta^2}\geq \frac{8 \left(3 e^2-32\right)^3}{9 e^7}
\end{equation}
Their thermodynamical quantities are given by 
\begin{equation}\label{masbhs}
\begin{split}
 m =& \frac{4 \left(252 e^7\delta \mu  +\left(32-3 e^2\right)^2 \left(27 e^4-576
   e^2+5120\right) \beta ^2\right)}{27 e^6 \left(9 e^2-64\right)} + \mathcal{O}(\beta^4,\beta^2\delta \mu ,\delta\mu^2)\\
q =& \frac{21 e^7 \delta \mu -8 \left(32-3 e^2\right)^2 \left(e^2-32\right)
   \beta ^2}{3 e^5 \left(9 e^2-64\right)} + \mathcal{O}(\beta^4,\beta^2\delta \mu ,\delta\mu^2)\\
G(\beta,\delta\mu) =& \frac{3 \pi}{8}\bigg[\frac{4 \left(3 e^2-32\right)^3 \beta ^2}{81 e^6} +\frac{16 \left(32-3 e^2\right)^4 \left(27 e^6-696 e^4+10752
   e^2-57344\right) \beta ^4}{243 e^{12} \left(9 e^2-64\right)}\\
&-\frac{2 \left(21 e^7 \delta \mu ^2-16 \left(32-3 e^2\right)^2
   \left(e^2-32\right) \beta ^2 \delta \mu \right)}{9 e^5 \left(9
   e^2-64\right)} + {\cal O}\left(\beta^5,\beta^3\delta\mu,\delta\mu^2\beta\right)\bigg]
\end{split}
\end{equation}

It is easily verified that (within perturbation theory) hairy black holes are 
free energetically subdominant compared to solitons in their common domain 
of existence. On the other hand, it may be checked that they are free energetically
dominant compared to RNAdS black holes, where the solutions coexist. 

\subsection{ Phase Diagrams}

The phase diagram of our system is very simple when $e^2 \leq e_c^2$. Hairy black holes
don't exist. The two phases that do exist - RNAdS black holes and the soliton - never 
coexist at the same $\mu$ and $\beta$. The RNAdS black hole dominates when it exists; 
the soliton dominates when it exists. The phase diagram is sketched in Fig.\ref{figp8} below. 
\begin{figure}
\begin{center}
 \includegraphics{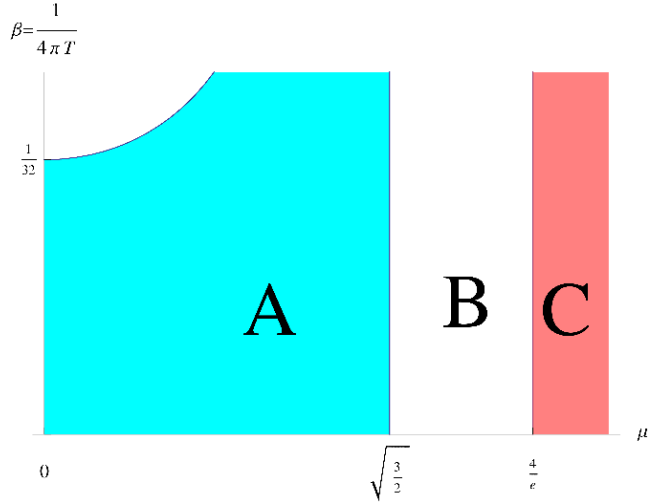}
\end{center}
\caption{\label{figp8} Grand canonical ensemble : Case 1 : $e^2 \leq e_c^2$ . No hairy black holes. Solitons and RNAdS black holes never overlap and they dominate their
respective domains of existence.}
\end{figure}
\begin{figure}
\begin{center}
 \includegraphics{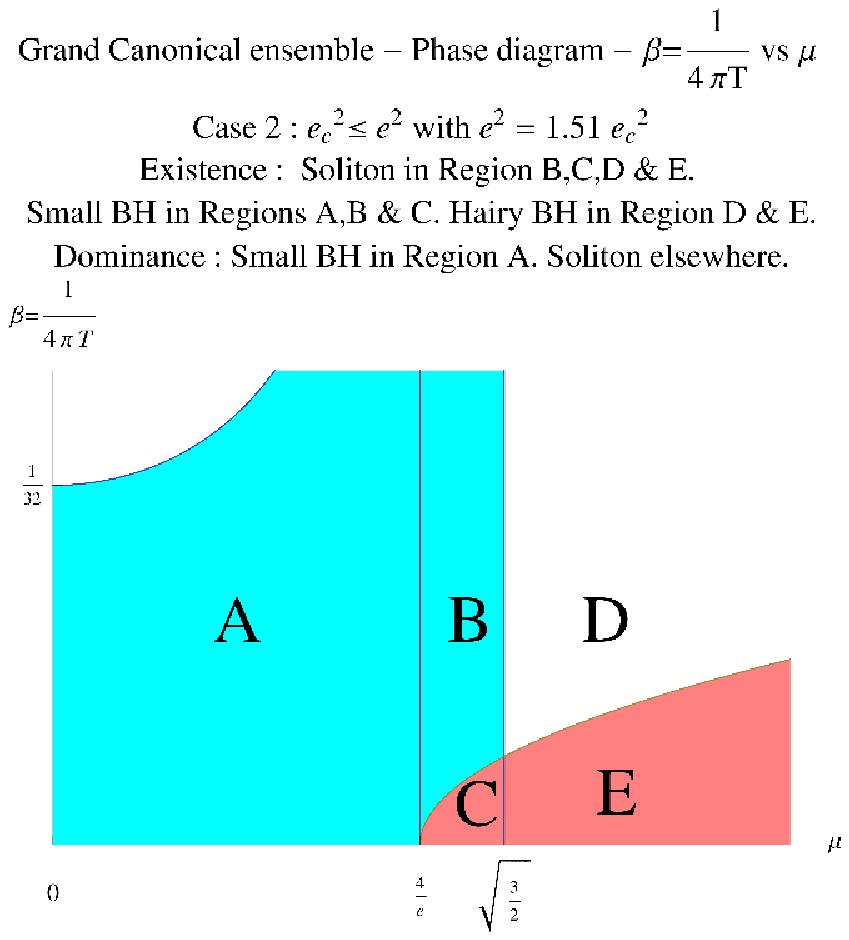}
\end{center}
\caption{\label{figp9}Grand canonical ensemble : Case 2 : $e_c^2 \leq e^2$ . Small BH 
is preferred region A and soliton is preferred elsewhere.}
\end{figure}

The phase diagram is more interesting when $e^2 \geq e_c^2$. As we have mentioned above, 
the system undergoes a first order phase transition from the black hole to the 
solitonic phase as $\mu$ is raised above $\frac{4}{e}$. The hairy black hole phase 
is always subdominant compared to the soliton, but free energetically dominates 
the black hole when both exist. The phase diagram is depicted in Fig. \ref{figp9} below. 
Note that the black hole and hairy black holes phases are identical, where the 
hairy black hole is first created. In the absence of the solitonic solution, 
consequently, our system would have undergone a second order transition from 
the RNAdS black hole to the hairy black hole phase upon raising $\mu$.

\section{Notation} \label{app:notation}

\subsection{Basic Setup}

Throughout this paper, we work in asymptotically (global) 
AdS$_5$ spacetimes with a bulk metric $g$, a bulk 
charged scalar field $\phi$ and a bulk gauge field $A_\mu$
with a Lagrangian
\begin{equation}\begin{split}
S&=\frac{1}{8\pi G_5}\int d^5x  \sqrt{g} \left[ \frac{1}{2} \left(\mathcal{R}[g] +12\right) 
-\frac{1}{4} \mathcal{F}_{\mu\nu}\mathcal{F}^{\mu\nu} 
- |D_\mu \phi|^2 - m_\phi^2|\phi|^2\right]\\
\mathcal{F}_{\mu\nu} &\equiv \nabla_\mu A_\nu-\nabla_\nu A_\mu \qquad\text{and}\qquad
D_\mu \phi \equiv \nabla_\mu\phi -i e A_\mu \phi 
\end{split} 
\end{equation}
where $G_5$ is the Newtons's constant and the radius of AdS$_5$ is set to unity.
This implies that the bulk cosmological constant is taken to be $\Lambda_5=-6$.

The radial co-ordinate of AdS$_5$ is denoted by $r$ with 
the boundary of AdS$_5$ being at $r=\infty$. For solutions
with the horizon, the outer horizon is taken to be at $r=R$.
The Schwarzschild-like temporal co-ordinate is denoted by $t$.
Sometimes, we find it convenient to work with rescaled
co-ordinates $y\equiv r/R$ and $\tau\equiv t/R$, especially in the 
near-field expansion at small radius ($r<<1$). In 
Appendix \ref{app:qnm}, we shift to Eddington-Finkelstein like
co-ordinates, with an Eddington-Finkelstein(EF) time
co-ordinate denoted by $v$.

Since throughout this paper we work with spherically symmetric
solutions, we will leave the co-ordinates parametrising 
S$^3$ implicit. Mostly, we work in a gauge where the bulk fields
take a form
\begin{equation} \begin{split}
ds^2 &=-f(r) dt^2+ g(r) dr^2+ r^2 d \Omega_3^2\\
A_t &= A(r), \qquad A_r = A_i =0\\
\phi &= \phi(r)\\
\end{split} 
\end{equation}
In Appendix \ref{app:qnm}, we work with the EF co-ordinate
metric for a small charged black hole (see below).

The charge of the scalar field $\phi$ is denoted by $e$ and its
mass by $m_\phi$. We take $m_\phi=0$ for most of the paper
except in section \ref{ssec:mphi} . By the standard rules of AdS/CFT,
the dual boundary operator $\mathcal{O}_\phi$ has a scaling dimension
\[ \Delta_0 = \left[\frac{d}{2} + \sqrt{ \left(\frac{d}{2}\right)^2 +m^2_\phi }\
 \right]_{d=4} =  2 + \sqrt{ 4 +m^2_\phi}  \]
This is also the energy of the lowest $\phi$ mode in vacuum AdS$_5$.
For the case $m_\phi=0$, $\Delta_0=4$. The other spherically 
symmetric modes of $\phi$ (dual to the descendants 
$\partial^{2n}\mathcal{O}_\phi$ ) have an energy 
\[ \Delta_n \equiv \Delta_0 + 2n =  2 + \sqrt{ 4 +m^2_\phi}  + 2n \]
For the case $m_\phi=0$, $\Delta_n=4+2n$. The covariant derivative
acting on $\phi$ is denoted by $D_\mu\equiv \nabla_\mu -i e A_\mu$.
The symbol $D^2$ is used to denote the covariant Laplacian.

\subsection{Thermodynamic quantities}
We now turn to notations involving thermodynamic quantities. First, we
omit a factor of $G_5^{-1}$ from all our extensive quantities 
in order to simplify our expressions. With this understanding,
we will denote the ADM mass by $M$, the charge of a solution by $Q$
and its entropy by $S$. We often find it convenient to work with a rescaled mass
$m$ and a rescaled charge $q$ which are related to the actual mass 
$M$ and charge $Q$ via the relations 
\begin{equation} \begin{split}
Q&\equiv\frac{\pi }{2} q\quad\text{and} \quad M\equiv\frac{3 \pi}{8} m
\end{split}\end{equation}
We use $F\equiv M-TS$ to denote Helmholtz free-energy appropriate to 
the canonical ensemble and $G\equiv M-TS-\mu Q$ to denote the grand potential
appropriate to the grand-canonical ensemble. Coming to the intensive
variables $\mu$ represents the chemical potential and $T$ represents
the temperature both of which are defined by the first law
\[ dM = T dS +\mu dQ \]
We sometimes find it convenient to work with the `rationalised' inverse
temperature $\beta \equiv (4\pi T)^{-1}$ and the chemical potential
excess over the super-radiant bound $ \delta\mu \equiv \mu - \Delta_n/e $.

\subsection{Double expansions}

Quantities in this paper are often expressed as double expansions
about two parameters - first is the parameter $\epsilon$ which is the 
amplitude of the leading normalisable mode in $\phi$. Under
AdS/CFT, roughly  $\epsilon\sim\ <\mathcal{O}_\phi>$, the 
boundary expectation value of $\mathcal{O}_\phi$, the operator dual to
$\phi$. The second parameter is the outer horizon radius $R$ of the small
charged blackhole at the core of the hairy blackhole. Further,
our solutions are often expressed in terms of a matched
asymptotic expansion with a near-field expansion at small radius
($r<<1$) and a far field expansion far away from the horizon 
($r>>R$) matched at their common domain of validity. We use 
the superscripts $in$ and $out$ to denote these two expansions
respectively. Many formulae in this paper involve the 
coefficients in this expansion which are defined via 
\begin{equation}\begin{split}
f(r) &=  \left\{ \begin{array}{c} \text{Near field} (r<<1) : f^{in}=\sum_{n=0}^{\infty} \epsilon^{2n} f_{2n}^{in} = \sum_{n=0}^{\infty} \epsilon^{2n} \sum_{k=0}^{\infty} R^{2k}f_{2n,2k}^{in} \\ \text{Far field} (r>>R) : f^{out}=\sum_{n=0}^{\infty} \epsilon^{2n} f_{2n}^{out} = \sum_{n=0}^{\infty} \epsilon^{2n} \sum_{k=0}^{\infty} R^{2k}f_{2n,2k}^{out}  \end{array} \right.\\
g(r) &=  \left\{ \begin{array}{c} \text{Near field} (r<<1) : g^{in}=\sum_{n=0}^{\infty} \epsilon^{2n} g_{2n}^{in} = \sum_{n=0}^{\infty} \epsilon^{2n} \sum_{k=0}^{\infty} R^{2k}g_{2n,2k}^{in} \\ \text{Far field} (r>>R) : g^{out}=\sum_{n=0}^{\infty} \epsilon^{2n} g_{2n}^{out} = \sum_{n=0}^{\infty} \epsilon^{2n} \sum_{k=0}^{\infty} R^{2k}g_{2n,2k}^{out}  \end{array} \right.\\
A(r) &=  \left\{ \begin{array}{c} \text{Near field} (r<<1) : A^{in}=\sum_{n=0}^{\infty} \epsilon^{2n} A_{2n}^{in} = \sum_{n=0}^{\infty} \epsilon^{2n} \sum_{k=0}^{\infty} R^{2k}A_{2n,2k}^{in} \\ \text{Far field} (r>>R) : A^{out}=\sum_{n=0}^{\infty} \epsilon^{2n} A_{2n}^{out} = \sum_{n=0}^{\infty} \epsilon^{2n} \sum_{k=0}^{\infty} R^{2k}A_{2n,2k}^{out}  \end{array} \right.\\
\phi(r) &=  \left\{ \begin{array}{c} \text{Near field} (r<<1) : \phi^{in}=\sum_{n=0}^{\infty} \epsilon^{2n+1} \phi_{2n}^{in} = \sum_{n=0}^{\infty} \epsilon^{2n+1} \sum_{k=0}^{\infty} R^{2k}\phi_{2n,2k}^{in} \\ \text{Far field} (r>>R) : \phi^{out}=\sum_{n=0}^{\infty} \epsilon^{2n+1} \phi_{2n}^{out} = \sum_{n=0}^{\infty} \epsilon^{2n+1} \sum_{k=0}^{\infty} R^{2k}\phi_{2n+,2k}^{out}  \end{array} \right.\\
\end{split}\end{equation}
In a similar vein, one can define an expansion of the covariant Laplacian
\[ D^2 =  \left\{\begin{array}{c}\text{Near field} (r<<1) : (D^2)^{in}=\sum_{n=0}^{\infty} \epsilon^{2n} (D^2)_{2n}^{in} = \sum_{n=0}^{\infty} \epsilon^{2n} \sum_{k=0}^{\infty} R^{2k}(D^2)_{2n,2k}^{in} \\ \text{Far field} (r>>R) : (D^2)^{out}=\sum_{n=0}^{\infty} \epsilon^{2n} (D^2)_{2n}^{out} = \sum_{n=0}^{\infty} \epsilon^{2n} \sum_{k=0}^{\infty} R^{2k-2}(D^2)_{2n,2k}^{out}  \end{array} \right.\]
Further, the boundary value of the gauge field or the chemical potential 
has the double expansion 
\[\mu\equiv \lim_{r\to\infty} A(r) = \sum_{n=0}^{\infty} \epsilon^{2n} \mu_{2n}(R) = \sum_{n=0}^{\infty} \epsilon^{2n} \sum_{k=0}^{\infty} R^{2k}\mu_{2n,2k}\]

\nocite{*}
\bibliographystyle{JHEP}
\bibliography{hairybh}

\end{document}